\documentclass[fleqn]{aa}

\usepackage{natbib}
\bibpunct{(}{)}{;}{a}{}{,}
\usepackage{graphicx} 
\usepackage{amsmath}
\usepackage{amsfonts}
\usepackage{amssymb}

\newlength{\zero}
\newcommand{\wz}{\hspace{\zero}}
\newcommand{\mb}[1]{\mbox{\boldmath $#1$}}
\newcommand{\mbsm}[1]{\mbox{\scriptsize \boldmath $#1$}}

\DeclareMathOperator{\fpart}{FP}
\DeclareMathOperator{\diag}{diag}

\begin{document}

\title{CFC+: Improved dynamics and gravitational waveforms from relativistic 
  core collapse simulations}

\author{Pablo Cerd\'a-Dur\'an \inst{1}
  \and Guillaume Faye \inst{2}
  \and Harald Dimmelmeier \inst{3} 
  \and Jos\'e A. Font \inst{1} 
  \and Jos\'e M$^{\underline{\mbox{a}}}$. Ib\'a\~{n}ez \inst{1}
  \and \\ Ewald M\"uller \inst{3}
  \and Gerhard Sch\"afer \inst{2} }

\offprints{Pablo Cerd\'a-Dur\'an, \\ \email{pablo.cerda@uv.es}}

\institute{
  Departamento de Astronom\'{\i}a y Astrof\'{\i}sica,
  Universidad de Valencia, Dr. Moliner, 50, 46100 Burjassot (Valencia), Spain
  \and
  Theoretisch-Physikalisches Institut, Friedrich-Schiller-Universit\"at Jena,
  Max-Wien-Platz 1, 07743 Jena, Germany
  \and
  Max-Planck-Institut f\"ur Astrophysik,
  Karl-Schwarzschild-Str. 1, 85741 Garching, Germany
}

\date{Received date / Accepted date}


\abstract{Core collapse supernovae are a promising source of
  detectable gravitational waves. Most of the existing
  (multidimensional) numerical simulations of core collapse in general
  relativity have been done using approximations of the Einstein field
  equations. As recently shown by \citet{dimmelmeier_02_a,
    dimmelmeier_02_b}, one of the most interesting such approximation
  is the so-called conformal flatness condition (CFC) of Isenberg,
  Wilson and Mathews. Building on this previous work we present here
  new results from numerical simulations of relativistic rotational
  core collapse in axisymmetry, aiming at improving the dynamics and
  the gravitational waveforms. The computer code used for these
  simulations evolves the coupled system of metric and fluid equations
  using the $ 3 + 1 $ formalism, specialized to a new framework for
  the gravitational field equations which we call CFC+. In this
  approach we add new degrees of freedom to the original CFC
  equations, which extend them by terms of second post-Newtonian
  order. The resulting metric equations are still of elliptic type but
  the number of equations is significantly augmented in comparison to
  the original CFC approach. The hydrodynamics evolution and the CFC
  spacetime metric are calculated using the code developed by
  \citet{dimmelmeier_02_a}, which has been conveniently extended to
  account for the additional CFC+ equations. The corrections for CFC+
  are computed solving a system of elliptic linear equations. The new
  formalism is assessed with time evolutions of both rotating neutron
  stars in equilibrium and gravitational core collapse of
  rotating polytropes. Gravitational wave signals for a comprehensive
  sample of collapse models are extracted using either the quadrupole
  formula or directly from the metric. We discuss our results on the
  dynamics and the gravitational wave emission through a detailed
  comparison between CFC and CFC+ simulations. The main conclusion is
  that, for the neutron star spacetimes analyzed in the present work,
  no significant differences are found among CFC, CFC+, and full
  general relativity, which highlights the suitability of the
  former.

  }

\authorrunning{P.~Cerd\'a-Dur\'an et al.}
\titlerunning{CFC+}

\maketitle


\section{Introduction}
\label{sec:introduction}

One of the most interesting topics in relativistic astrophysics is the
formation and evolution of compact objects, such as those produced in
the rotational core collapse of massive stars, or in the merging of a
compact binary consisting of two white dwarfs, two neutron stars, or a
white dwarf and a neutron star. The catastrophic events in which the
new-born compact objects are formed are a promising source of
gravitational waves in the kHz frequency range, hence optimal to be
detected at galactic distances with the new ground-based
interferometer detectors such as LIGO or VIRGO.

Gravitational radiation is produced by accelerating matter whose
dynamics is not spherically symmetric. Among the list of astrophysical
sources of gravitational radiation, axisymmetric sources such as
rotational core collapse are particularly amenable to numerical
investigations, as present-day computational resources allow for more
accurate calculations than in full three dimensions. The early phase
of the merging of two compact objects, when the two bodies start to
plunge from the innermost stable circular orbit, is a
three-dimensional situation, although after the merging itself the
scenario is again suitable for investigations in axisymmetry. On the
other hand, Newtonian gravity does not describe correctly the dynamics
of these processes, and one has to use general relativity for a proper
description, as well as to compute waveforms, or at least some
approximation better than the Newtonian one. Post-Newtonian
approximations of the metric in the near zone match quite well with
the features described before, and probably most of these scenarios
can still be handled without fully general relativistic calculations:
Let $ d $ be the typical dimension of the source, $ v $ the typical
velocity of the system, and $ t $ the typical timescale of variation
of the gravitational field; hence, the typical frequency of the
gravitational waves in the wave zone is $ 1 / t $, and the wavelength
is $ \lambda \sim c t $, where $ c $ is the speed of light. From the
virial theorem it follows that $ t $ is also the typical dynamical
timescale, hence $ d \sim v t $ and
$ v / c \sim d / c t \sim d / \lambda $. Therefore, a post-Newtonian
expansion around $ v / c \to 0$ is equivalent to an  expansion around
$ d / \lambda \to 0 $. For the typical scales involved in core
collapse and neutron stars we get $ d^2 / \lambda^2 \sim 0.05 $. That
estimate yields first and second order post-Newtonian corrections of
about 5\% and 0.3\%, respectively. We note, however, that although
these corrections may seem to be small, the non-linearity of the
equations makes it difficult to predict beforehand the
effects in a numerical simulation.

During the last two decades calculations of gravitational waveforms
from core collapse have been done using different approaches for
gravity~\citep{mueller_82_a, finn_89_a, moenchmeyer_91_a,
  bonazzola_93_a, yamada_95_a, zwerger_97_a, rampp_98_a,
  dimmelmeier_01_a, dimmelmeier_02_a, dimmelmeier_02_b, fryer_02_a,
  fryer_03_a, imamura_03_a, kotake_03_a, shibata_04_a,
  ott_04_a}. First attempts were made in Newtonian gravity, using
Eulerian codes with artificial viscosity \citep{mueller_82_a,
  finn_89_a, moenchmeyer_91_a} to calculate the
hydrodynamics. Pseudo-spectral methods were used by
\citet{bonazzola_93_a}, who followed the collapse without
including the bounce phase due to the appearance of numerical
instabilities associated with the presence of a shock wave. The need
for correctly treating the shock wave, which forms after core bounce,
gradually led to the use of high-resolution shock-capturing (HRSC)
schemes. \citet{zwerger_97_a} first used HRSC methods to simulate a
sequence of collapsing rotating polytropes in axisymmetry, providing a
comprehensive waveform catalog. Extensions to 3D were carried out
by \citet{rampp_98_a}. In both works a simple equation of state
\citep{janka_93_a} was used. Recently there have been attempts to
include more realistic physics in Newtonian simulations:
\citet{ott_04_a} used the equation of state of \citet{lattimer_91_a},
while \citet{mueller_04_a} also incorporated 2D neutrino transport and
computed the gravitational wave emission produced by neutrino-driven
convection.

A step forward to improve the treatment of the gravitational field was
taken by \citet{dimmelmeier_01_a, dimmelmeier_02_a, dimmelmeier_02_b},
who used the conformal flatness condition (CFC hereafter) of Isenberg,
Wilson, and Mathews (see \citet{isenberg_78_a} and
\citet{wilson_96_a}). Note that in the spherical case CFC is not an
approximation to general relativity but rather is exact. In a more
general context it has been shown that conformally flat gravity agrees
with full general relativity at the level of the first post-Newtonian
expansion \citep{kley_99_a}. Therefore the CFC approximation reproduces
quite well the main features of the models studied (quasi-spherical
sources tractable by the post-Newtonian formalism). Numerically, the
major advantage of the CFC approximation is that the $ 3 + 1$ metric
equations are reduced to elliptic equations, which, in principle,
makes their implementation more stable than in the standard
formulations of the field equations commonly used in numerical
relativity. Concerning the core collapse waveforms,
\citet{dimmelmeier_02_b} found quantitative and also qualitative
differences between Newtonian and CFC calculations. This demonstrates
that a relativistic calculation of the spacetime dynamics
is necessary for properly studying core collapse and the
gravitational waves emitted. Recently this work in axisymmetry has
been extended to 3D by \citet{dimmelmeier_04_a}, combining HRSC methods
for the hydrodynamic equations and spectral methods for the metric
equations.

First multidimensional core collapse simulations in full general
relativity were reported by \citet{siebel_03_a} who used the
characteristic formulation of the Einstein equations. In these
simulations the gravitational waveform extraction using the Bondi news
function at future null infinity was hampered by gauge effects
and numerical inaccuracies. Recently, \citet{shibata_04_a} have
performed simulations in axisymmetry using the so-called Cartoon
method to solve the Einstein equations, finding remarkable agreement
with the CFC results of \citet{dimmelmeier_02_b}. Extensions of this
work to full 3D have been reported in~\citet{shibata_04_b}. Except for
\citet{siebel_03_a}, in all existing simulations thus far the
gravitational wave signals have been computed using the Newtonian
quadrupole formula.

In this paper we propose a new approximation scheme for the metric
equations that is an extension of the CFC approach. Our approximation
consists of adding second post-Newtonian terms to the conformal
flet metric. With the new approach, which we call CFC+, the resulting
metric equations are still elliptic, and reduce to full general
relativity in the spherical case. Therefore, we conserve the main
advantages of the CFC approximation at small extra computational cost
(the additional elliptic equations are linear), while improving the
computation of the dynamics and the metric. The aim of this paper is
to compare the quality of the CFC+ approximation in different
scenarios with that of the CFC approach. In particular we
focus on the study of two scenarios of current interest in
relativistic astrophysics, the gravitational collapse of rotating
stellar cores to neutron stars (the relevant mechanism responsible for
core collapse supernovae of type II/Ib/Ic), and the stability and
oscillations of rotating neutron stars.

This paper is organized as follows: In
Sect.~\ref{sec:general_framework} we present the mathematical
framework and introduce the general relativistic equations for the
hydrodynamics and the spacetime metric that we implement in the
numerical code. Particular emphasis is given to the derivation of the
CFC+ equations (for technical details see
Appendices~\ref{app:calc_htt}, \ref{app:inversion_htt},
and~\ref{app:multipole_expansion}).
Sect.~\ref{sec:numerical_implementation} is devoted to describing the
numerical methods used to solve the hydrodynamics and metric
equations. We further explain in this section the procedure we follow
to extract gravitational waves (see also Appendix~\ref{app:GW}). In
Sect.~\ref{sec:equilibrium_rs} we specify the models used for the
simulations of rotating relativistic stars. The results from the
simulations of evolutions of rotating neutron stars and rotational
stellar core collapse to a neutron star are discussed in
Sects.~\ref{sec:oscillations_rns} and~\ref{sec:core_collapse},
respectively. The conclusions of this work are summarized in
Sect.~\ref{sec:conclusions}.

We use a spacelike signature $ (-, +, +, +) $ and units in which
$ c = G = 1 $ (except in some passages, particularly in the
Appendices, where $ c $ and $ G $ are retained for a better
understanding of the post-Newtonian expansion). Greek indices run from
0 to 3, Latin indices from 1 to 3, and we adopt the standard Einstein
convention for the summation over repeated indices. The indices of
3-vectors and 3-tensors are raised and lowered by means of the
3-metric.


\section{Mathematical framework}
\label{sec:general_framework}

We adopt the $ 3 + 1 $ formalism \citep{arnowitt_62_a} to
foliate spacetime with a metric $ g_{\mu\nu} $ into a set of
non-intersecting spacelike hypersurfaces. The line element reads
\begin{equation}
  ds^2 = - \alpha^2 dt^2 + \gamma_{ij} (dx^i + \beta^i dt)
  (dx^j + \beta^j dt),
  \label{eq:line_element}
\end{equation}
with $ \alpha $ being the lapse function, which describes the rate
of advance of time along a timelike unit vector $ n^{\mu} $ normal
to a hypersurface, $ \beta^i $ being the spacelike shift 3-vector,
which describes the motion of coordinates within a surface, and
$ \gamma_{ij} $ being the spatial 3-metric.

\subsection{Hydrodynamic equations}
\label{subsec:hydro_equations}

The hydrodynamic evolution of a relativistic perfect fluid with
rest-mass current $ J^{\mu} = \rho u^{\mu} $ and energy-momentum
tensor $ T^{\mu\nu} = \rho h u^\mu u^\nu + P g^{\mu\nu} $ in a
(dynamic) spacetime  is
determined by a system of local conservation equations, which read
\begin{equation}
  \nabla_{\mu} J^{\mu} = 0, \qquad \nabla_{\mu} T^{\mu \nu} = 0,
  \label{eq:gr_equations_of_motion}
\end{equation}
where the covariant derivative is denoted by $ \nabla_\mu $. The fluid
is specified by the rest mass density $ \rho $, the specific enthalpy
$ h = 1 + \epsilon + P / \rho $, the pressure $ P $, the specific
internal energy $ \epsilon $, and the 4-velocity $ u^\mu $. The
pressure is determined by an equation of state (EOS)
$ P = P (\rho, \epsilon) $.

The 3-velocity of the fluid, as measured by an Eulerian observer at
rest in a spacelike hypersurface $ \Sigma_{t} $, is
\begin{equation}
  v^i = \frac{u^i}{\alpha u^t} + \frac{\beta^i}{\alpha}.
  \label{eq:three_velocity}
\end{equation}
Following the work of \citet{banyuls_97_a} we introduce a set of
variables in terms of the primitive (physical) hydrodynamic variables
$ (\rho, v^i, \epsilon) $, whose associated densities are conserved:
\begin{align}
  D & = \rho W,
  \label{eq:conserved_quantities_d} \\
  S^i & = \rho h W^2 v^i,
  \label{eq:conserved_quantities_s} \\
  \tau & = \rho h W^2 - P - D.
  \label{eq:conserved_quantities_tau}
\end{align}%
In the above expressions $ W $ is the Lorentz factor defined as
$ W = \alpha u^t $, which satisfies the relation
$ W = 1/\sqrt{1 - v_i v^i} $.

The previous choice allows us to write the conservation
laws~(\ref{eq:gr_equations_of_motion}) as a first-order, flux-conservative
hyperbolic system of equations
\citep{banyuls_97_a}:
\begin{equation}
  \frac{1}{\sqrt{- g}} \left[
  \frac{\partial \sqrt{\gamma} \mb{U}}{\partial t} +
  \frac{\partial \sqrt{- g} \mb{F}^i}{\partial x^i} \right] = \mb{Q},
  \label{eq:hydro_conservation_equation}
\end{equation}
with the state vector, flux vector, and source vector given by
\begin{align}
  \mb{U} & = [D, S_j, \tau], \\
  \mb{F}^i & = \left[
  D \hat{v}^i, S_j \hat{v}^i + \delta^i_j P, \tau \hat{v}^i + P v^i \right], \\
  \mb{Q} & = \left[ 0, \frac{1}{2} T^{\mu \nu}
  \frac{\partial g_{\mu \nu}}{\partial x^j},
  \alpha \! \left( \! T^{\mu 0} \frac{\partial \ln \alpha}{\partial x^\mu} -
  T^{\mu \nu} {\it \Gamma}^0_{\mu \nu} \! \right) \! \right] \!.
  \label{eq:hydro_conservation_equation_vectors}
\end{align}%
Here $ \hat{v}^i = v^i - \beta^i / \alpha $, and
$ \sqrt{-g} = \alpha \sqrt{\gamma} $, with
$ g = \det (g_{\mu \nu}) $ and $ \gamma = \det (\gamma_{ij}) $ being
the determinant of the 4-metric and 3-metric, respectively;
$ {\it \Gamma}^\lambda_{\mu \nu} $ are the Christoffel
symbols.

For convenience, we also define the following modified conserved
quantities:
\begin{equation}
  D^* = \sqrt{\bar{\gamma}} D,
  \qquad
  S^{*}_i = \sqrt{\bar{\gamma}} S_i.
\end{equation}

The determinant $ \bar{\gamma} $ is actually the ratio of $ \gamma $
evaluated on some coordinate grid $ \{x^\mu\} $ and the determinant
$ \hat{\gamma} $ of the flat metric $ \hat{\gamma}_{ij} $ on the same
grid: $ \bar{\gamma} = \gamma / \hat{\gamma} $.

\subsection{3+1 metric equations}
\label{subsec:metric_equations}

In the 3\,+\,1 formalism, the Einstein equations split into (i) evolution
equations for the 3-metric $ \gamma_{ij} $ and the extrinsic curvature $
K_{ij} $,
\begin{align}
  \partial_t \gamma_{ij} & = - 2 \alpha K_{ij} + 2 \nabla_{(i} \beta_{j)},
  \label{eq:adm_metric_equation_1} \\
  \partial_t K_{ij} & = - \nabla_i \nabla_j \alpha + 
  \alpha (R_{ij} + K K_{ij} - 2 K_{ik} K_j^k) 
  \nonumber \\ 
  & + \beta^k \nabla_k K_{ij} + 2 K_{k(i} \nabla_{j)} \beta^k
  \nonumber \\ & -
  8 \pi \alpha \! \left( \! S_{ij} - \frac{\gamma_{ij}}{2}
  (S_k^k - E) \right)\!,
  \label{eq:adm_metric_equation_2}
\end{align}%
and (ii) constraint equations,
\begin{align}
  R + K^2 - K_{ij} K^{ij} - 16 \pi E & = 0,
  \label{eq:adm_metric_equation_3} \\
  \nabla_i (K^{ij} - \gamma^{ij} K) - 8 \pi S^j & = 0,
  \label{eq:adm_metric_equation_4}
\end{align}%
which must be fulfilled at every spacelike hypersurface.

In these equations $ \nabla_i $ is the covariant derivative with
respect to the 3-metric $ \gamma_{ij} $, $ K_{ij} $ is the
extrinsic curvature, $ R_{ij} $ is the Ricci tensor on $ \Sigma_t $, $ R $
the scalar curvature and $ K = K^i_i $ the trace of the extrinsic
curvature. The brackets around indices indicate symmetrization. The
projection of the energy-momentum tensor onto the spatial hypersurface
is defined as $ S_{ij} = \rho h W^2 v_i v_j + \gamma_{ij} P $, while
$ E = \rho h W^2 - P $ is the total energy per volume unit as measured by an
observer at rest in $ \Sigma_{t} $.

The ADM gauge, which we will be using in the derivation of the metric
equations in Sects.~\ref{subsec:cfc} and~\ref{subsec:cfc+}, is
defined as that gauge for which the 3-metric can be decomposed into a
conformally flat term plus a transverse traceless term,
\begin{equation}
  \gamma_{ij} = \phi^4 \hat{\gamma}_{ij} + h^\mathrm{TT}_{ij},
  \label{eq:generic_gamma}
\end{equation}
with
\begin{equation}
  h^\mathrm{TT}_{ij} \hat{\gamma}^{ij} = 0,
  \qquad
  \hat{\gamma}^{ik} \hat{\nabla}_k h^\mathrm{TT}_{ij} = 0,
  \label{eq:condition_gamma}
\end{equation}
where $ \hat{\gamma}_{ij} $ is the flat metric (with inverse
$ \hat{\gamma}^{ij} $;
$ \hat{\gamma}_{ij} = \hat{\gamma}^{ij} = \delta_{ij} $ in Cartesian
coordinates), $ \phi $ is the conformal factor, and
$ h^\mathrm{TT}_{ij} $ is transverse and traceless. The conjugate
momentum $ \pi^{ij} $ of $ \gamma_{ij} $ is traceless as well:
\begin{equation}
  \pi^{ij} \hat{\gamma}_{ij} = 0.
  \label{eq:condition_pi}
\end{equation}

\subsection{CFC metric equations}
\label{subsec:cfc}

In spherical symmetry $ h^\mathrm{TT}_{ij} = 0 $, i.e.\ the
3-metric is conformally flat (CF). A first reasonable approximation in
quasi-spherical scenarios is imposing a vanishing
$ h^\mathrm{TT}_{ij} $ in Eq.~(\ref{eq:generic_gamma}):
\begin{equation}
  \gamma_{ij} = \phi^4 \hat{\gamma}_{ij}.
  \label{eq:cfc_three_metric}
\end{equation}
This is the conformal flatness condition or Isenberg--Wilson--Mathews
approximation~\citep{isenberg_78_a, wilson_96_a}. In the ADM gauge a
conformally flat 3-metric ensures the maximal slicing condition,
$ K = 0 $. Hence, it can be shown that the $ 3 + 1 $ metric
equations~(\ref{eq:adm_metric_equation_1}--\ref{eq:adm_metric_equation_4})
reduce to a set of five coupled elliptic (Poisson-like) equations for
the metric components,
\begin{align}
  \hat{\Delta} \phi & = - 2 \pi \phi^5
  \left( \rho h W^2 - P + \frac{K_{ij} K^{ij}}{16 \pi} \right),
  \label{eq:metric_equation_1} \\
  \hat{\Delta} (\alpha \phi) & = 2 \pi \alpha \phi^5
  \left( \rho h (3 W^2 - 2) + 5 P + \frac{7 K_{ij} K^{ij}}{16 \pi} \right),
  \label{eq:metric_equation_2} \\
  \hat{\Delta} \beta^i & = 16 \pi \alpha \phi^4 S^i + 2 \phi^{10} K^{ij}
  \hat{\nabla}_j \! \left( \! \frac{\alpha}{\phi^6} \! \right) \! -
  \frac{1}{3} \hat{\nabla}^i \hat{\nabla}_k \beta^k,
  \label{eq:metric_equation_3}
\end{align}%
where $ \hat{\nabla} $ and $ \hat{\Delta} = \hat{\gamma}^{ij} \hat{\nabla}_i
\hat{\nabla}_j $ are the flat space Nabla and Laplace operators, respectively.

Applying the CFC to the traceless part of
Eq.~(\ref{eq:adm_metric_equation_1}) yields the following
relation between the conformal factor and the shift vector components:
\begin{equation}
  \partial_t \phi = \frac{\phi}{6} \nabla_k \beta^k.
  \label{eq:phi_t_relation}
\end{equation}
With this, Eq.~(\ref{eq:adm_metric_equation_1}) transforms into an
expression for the extrinsic curvature components:
\begin{equation}
  K_{ij} = \frac {1}{2 \alpha} \left( \nabla_i \beta_j +
    \nabla_j \beta_i - \frac{2}{3} \gamma_{ij} \nabla_k \beta^k \right).
  \label{eq:definition_of_extrinsic_curvature}
\end{equation}

As a consequence of setting the off-diagonal elements of
$ \gamma_{ij} $ to zero, the degrees of freedom representing
gravitational waves are removed from the spacetime. Therefore, we
calculate gravitational waveforms in a post-processing step
using the quadrupole formula. Note that at that approximation level
the ADM gauge is equivalent to the maximal slicing quasi-isotropic
(MSQI) gauge, i.e.\ maximal slicing ($ K = 0 $) plus quasi-isotropic
coordinates ($ g_{r\theta} = 0 $ and
$ g_{\theta\theta} = g_{rr} $ in spherical coordinates in an orthonormal 
basis).

\subsection{CFC+ metric equations}
\label{subsec:cfc+}

We improve the metric approximation by adding to the CFC metric the
second post-Newtonian deviation from isotropy. We call this
approximation CFC+. Compared to the CFC
metric~(\ref{eq:cfc_three_metric}) the CFC+ metric
\begin{equation}
  \gamma_{ij} = \phi^4 \hat{\gamma}_{ij} + h^\mathrm{2PN}_{ij}
\end{equation}
now includes a new traceless and transverse term $ h^\mathrm{2PN}_{ij} $,
which is identical to $h^\mathrm{TT}_{ij}$ in the
decomposition~(\ref{eq:generic_gamma}) up to the second post-Newtonian
expansion. As described in Appendix~\ref{app:calc_htt} this higher order
correction $h^\mathrm{2PN}_{ij}$ is the solution of the tensor Poisson
equation~\citep{schaefer_90_a}
\begin{equation}
  \hat{\Delta} h^\mathrm{2PN}_{ij} =
  \hat{\gamma}_{ij}^{\mathrm{TT}kl} F_{kl},
  \label{eq:equation_hTT}
\end{equation}
where the source $ F_{kl} $ is given by
\begin{equation}
  F_{kl} = - 4 \hat{\nabla}_k U \hat{\nabla}_l U -
  16 \pi \frac{S^*_k S^*_l}{D^*},
  \label{eq:Fij_definition}
\end{equation}
with $ U $ being the ``Newtonian'' potential, which is the solution of the
Poisson equation
\begin{equation}
  \hat{\Delta} U = - 4 \pi D^*.
  \label{eq:newtonian_potential}
\end{equation}
The operator $ \hat{\gamma}_{ij}^{\mathrm{TT}kl} $ is the transverse, 
trace-free (TT) projector as defined in Eq.~(\ref{eq:tt_projector}). 

Of the original CFC metric
equations~(\ref{eq:metric_equation_1}--\ref{eq:metric_equation_3}) for $ \phi
$, $ \alpha $, and $ \beta^i $, only the equation for the lapse function,
Eq.~(\ref{eq:metric_equation_2}), has to be modified. It now depends
explicitly on the second post-Newtonian correction $ h^\mathrm{2PN}_{ij} $:
\begin{equation}
  \hat{\Delta} (\alpha \phi) =
  \left[ \hat{\Delta} \left( \alpha \phi \right)
  \right]_{\!h_{ij}^\mathrm{2PN} = 0} \! - 
  \hat{\gamma}^{ik} \hat{\gamma}^{jl} h^\mathrm{2PN}_{ij}\hat{\nabla}_k 
  \hat{\nabla}_l U.
  \label{eq:alpha_phi}
\end{equation}
The metric equations for the conformal factor $ \phi $ and the shift
vector $ \beta^i $ remain unaltered. However, the CFC+ corrections
couple implicitly as the source terms on their right-hand sides depend
on $ \alpha $.

The computation of $ h^\mathrm{2PN}_{ij} $, i.e.\ the inversion of the
tensor Poisson equation~(\ref{eq:equation_hTT}) by means of the Poisson
integral operator $\hat{\Delta}^{-1}$, is simplified by the
introduction of intermediate potentials $ \mathcal{S} $,
$ \mathcal{S}_i $, $ \mathcal{T}_i $, $ \mathcal{R}_i $, and
$ \mathcal{S}_{ij} $, which are solutions of the following
scalar/vector/tensor-Poisson equations:
\begin{align}
  \hat{\Delta} \mathcal{S} & =
  - 4 \pi \frac{S^*_i S^*_j}{D^*} x^i x^j,
  \label{eq:lap_s} \\
  \hat{\Delta} \mathcal{S}_i & =
  \left[ - 4 \pi \frac{S^*_i S^*_j}{D^*} -
  \hat{\nabla}_i U \hat{\nabla}_j U \right] x^j,
  \label{eq:lap_si} \\
  \hat{\Delta} \mathcal{T}_i & =
  \left[ - 4 \pi \frac{S^*_j S^*_k}{D^*} -
  \hat{\nabla}_j U \hat{\nabla}_k U \right] \hat{\gamma}^{jk} 
  \hat{\gamma}_{il} x^l,
  \label{eq:lap_ti} \\
  \hat{\Delta} \mathcal{R}_i & =
  \hat{\nabla}_i (\hat{\nabla}_j U \hat{\nabla}_k U x^j x^k),
  \label{eq:lap_ri} \\
  \hat{\Delta} \mathcal{S}_{ij} & =
  - 4 \pi \frac{S^*_i S^*_j}{D^*} - \hat{\nabla}_i U \hat{\nabla}_j U.
  \label{eq:lap_sij}
\end{align}%
These equations are designed such that their source terms approach
zero like $ r^{-3} $ when $ r = |\mb{x}| $ tends towards infinity,
which ensures the existence of the corresponding Poisson
integrals. With the help of these potentials $ h^\mathrm{2PN}_{ij} $
can simply be expressed as
\begin{align}
  h^\mathrm{2PN}_{ij} & = \hat{\Delta}^{-1}
  \left( \hat{\gamma}_{ij}^{\mathrm{TT}kl} F_{kl} \right)
  \nonumber \\
  & = \frac{1}{2} \mathcal{S}_{ij} -
  3 x^k \hat{\nabla}_{(i} \mathcal{S}_{j)k} + \frac{5}{4} \hat{\gamma}_{jm}
  x^m \hat{\nabla}_i 
  \left( \hat{\gamma}^{kl} \mathcal{S}_{kl} \right) 
  \nonumber \\
  & +
  \frac{1}{4} x^k x^l \hat{\nabla}_i \hat{\nabla}_j
  \mathcal{S}_{kl} + 3 \hat{\nabla}_{(i} \mathcal{S}_{j)} - \frac{1}{2} x^k
  \hat{\nabla}_i \hat{\nabla}_j \mathcal{S}_k 
  \nonumber \\
  & + \frac{1}{4} \hat{\nabla}_i \hat{\nabla}_j \mathcal{S} -
  \frac{5}{4} \hat{\nabla}_i \mathcal{T}_j -
  \frac{1}{4} \hat{\nabla}_i \mathcal{R}_j \nonumber \\
  & + \hat{\gamma}_{ij} \left[ \frac{1}{4} \hat{\gamma}^{kl} \mathcal{S}_{kl} 
  + x^k \hat{\gamma}^{lm} \hat{\nabla}_m \mathcal{S}_{kl} - \hat{\gamma}^{kl}
  \hat{\nabla}_k \mathcal{S}_l  \right] ,
  \label{eq:hTT}
\end{align}%
as shown in Appendix~\ref{app:inversion_htt}.

Due to the specific type of elliptic solvers employed in our 
computer code (see Sect.~\ref{subsec:metric_solver} below), it is not
possible to use the inverse image method to evaluate the
intermediate potentials up to spatial infinity. We instead solve
Eqs.~(\ref{eq:lap_s}--\ref{eq:lap_sij}) assuming specific boundary
conditions. These are determined from the multipole expansion
$ \mathcal{M} $ of the intermediate potentials $ \mathcal{S} $,
$ \mathcal{S}_i $, $ \mathcal{T}_i $, $ \mathcal{R}_i $, and
$ \mathcal{S}_{ij} $. As detailed in
Appendix~\ref{app:multipole_expansion}, these multipole expansions are
given by
\begin{align}
  \mathcal{M}(\mathcal{S}) & = \frac{1}{r} \int \mathrm{d}^3\mb{x} \,
  \sqrt{\hat{\gamma}} 
  \left( \! \frac{S_k^* S_l^*}{D^*} x^k x^l \! \right) \!,
  \label{eq:mult_expansion_S}
  \\
  \mathcal{M}(\mathcal{S}_i) & = \frac{1}{r} \int \mathrm{d}^3\mb{x} \,
  \sqrt{\hat{\gamma}} D^* \!\! \left( \! \frac{S_i^* S_k^*}{D^{*\,2}}
  x^k + \hat{\gamma}_{ij} x^j (U + x^k\hat{\nabla}_k U) \!\! \right)
  \nonumber \\
  & + \frac{M^2}{2 r} \hat{\gamma}_{ij} n^j,
  \label{eq:mult_expansion_Si}
  \\
  \mathcal{M}(\mathcal{T}_i) & = \frac{1}{r} \int \mathrm{d}^3\mb{x} \,
  \sqrt{\hat{\gamma}} D^*
  \! \left( \! \frac{\hat{\gamma}^{kl} S_k^* S_l^*}{D^{*\,2}} + U \!
  \right) \! \hat{\gamma}_{ij} x^j
  \nonumber \\
  & + \frac{M^2}{2 r} \hat{\gamma}_{ij} n^j,
  \label{eq:mult_expansion_Ti}
  \\
  \mathcal{M}(\mathcal{R}_i) & =
  \frac{M^2 \hat{\gamma}_{ij} n^j}{r},
  \label{eq:mult_expansion_Ri}
  \\
  \mathcal{M}(\mathcal{S}_{ij}) & = \frac{1}{r} \int \mathrm{d}^3\mb{x} \,
  \sqrt{\hat{\gamma}} D^* \! \left( \! \frac{S_i^* S_j^*}{D^{*\,2}} +
  \frac{1}{2} \hat{\gamma}_{ij} U + \hat{\gamma}_{ik} x^k
  \partial_j U \! \right) \!,
  \label{eq:mult_expansion_Sij}
\end{align}%
modulo $\mathcal{O}(1/r^2)$ corrections. 
$ M = \int \mathrm{d}^3\mb{x} \sqrt{\hat{\gamma}}D^* $ is the baryonic rest mass.
By evaluating the above expansions at a finite radius outside
the star we obtain the necessary boundary conditions for
$ \mathcal{S} $, $ \mathcal{S}_i $, $ \mathcal{T}_i $,
$ \mathcal{R}_i $, and $ \mathcal{S}_{ij} $.


\section{Numerical implementation}
\label{sec:numerical_implementation}

The code used for the simulations presented in this paper is an
extension of the one described in \citet{dimmelmeier_02_a,
  dimmelmeier_02_b}, incorporating  the new CFC+ metric
equations. Hence, the interested reader is addressed to those
references for details additional to those discussed below. 

We use spherical polar coordinates $ \{ t, r, \theta, \varphi \} $ and
assume axial symmetry with respect to the rotation axis and symmetry
with respect to the equatorial plane at $ \theta = \pi / 2 $. The
computational grid is composed of $ n_r $ radial zones and
$ n_\theta $ equidistant angular zones, whose specific values for the
simulations reported here are given below. For convenience the radial
cell-spacing is chosen equidistant for evolutions of equilibrium
neutron stars and logarithmically spaced when simulating core
collapse. As in the original code \citep{dimmelmeier_02_a} the part of
the grid outside the star is filled with an artificial atmosphere (as
customary in finite difference codes similar to ours; see
\citet{font_02_a, duez_02_a, baiotti_04_a}). This atmosphere obeys a
polytropic equation of state and has a very low density such that its
presence does not affect the dynamics of the star (see
\citet{dimmelmeier_02_a} for details). Moreover, an extended grid
containing no matter is used beyond the atmosphere for the CFC+ metric
calculations, namely to properly capture the radial fall-off behavior
of the metric potentials and to extract the gravitational radiation
using components of $ h^\mathrm{TT}_{ij} $ in the wave zone (see
Sect.~\ref{subsec:waveforms}).


\subsection{Hydrodynamics solver}
\label{subsec:hydro_solver}

The hydrodynamics solver performs the numerical integration of the
system of hydrodynamic conservation
equations~(\ref{eq:hydro_conservation_equation}) using a
high-resolution shock-capturing (HRSC) scheme. This method ensures
numerical conservation of physically  conserved quantities and a
correct treatment of discontinuities such as shocks. In HRSC methods a
Riemann problem has to be solved at each cell interface, which
requires the reconstruction of the primitive variables
$ (\rho, v^i, \epsilon) $ at this interfaces. We use the PPM method of
\citet{colella_84_a} for the reconstruction, which yields third order
accuracy in space. The solution of the Riemann problems then provides
the numerical fluxes at cell interfaces. To obtain this solution, the
characteristic structure of the hydrodynamics equations is explicitely
needed \citep{banyuls_97_a}.

Contrary to CFC in the CFC+ approximation the metric does not have any
zero off-diagonal elements, and thus we have to use the most general
eigenvalues and eigenvectors in general relativity as reported in
\citet{font_03_a}. Once the spectral information is known, the
numerical fluxes are computed by means of Marquina's approximate flux
formula \citep{donat_96_a}. The time update of the conserved vector
$ \mb U $ is done using the method of lines in combination with a
Runge--Kutta method with second order accuracy in time. Once the
conserved quantities $ (D, S_i, \tau) $ are updated in time, the
primitive variables need to be recovered. This is done through an
iterative Newton--Raphson method, as these variables can not be
obtained in closed form from the conserved variables.

We note that the sources $ \mb Q $ of the hydrodynamic equations have
been implemented in the code using a general form for the metric,
although they can be simplified for a metric with vanishing
nondiagonal terms in the 3-metric, as, for example, in CFC.


\subsection{Metric solver}
\label{subsec:metric_solver}

The main feature of the approximations we are using for the metric is
that only elliptic equations have to be solved to know the metric at
each time step. In our approximate scheme we solve the equations
hierarchically. First a solution of the Poisson
equation~(\ref{eq:newtonian_potential}) for $ U $ is obtained. Then we
solve the equations~(\ref{eq:lap_s}--\ref{eq:lap_sij}) for the
intermediate potentials $ \mathcal{S} $, $ \mathcal{S}_i $,
$ \mathcal{T}_i $, $ \mathcal{R}_i $, and $ \mathcal{S}_{ij} $, which
we need to calculate $ h^\mathrm{TT}_{ij} $ in
Eq.~(\ref{eq:hTT}). Finally, we use $ h^\mathrm{TT}_{ij} $ to solve
the modified equations for $ \phi $, $ \beta^i $, and
$ \alpha$, Eqs.~(\ref{eq:metric_equation_1}, \ref{eq:metric_equation_3},
\ref{eq:alpha_phi}). For each step we use different methods according
to the mathematical characteristics of each equation.

The equation for the Newtonian potential $ U $ is a linear Poisson
equation with compact support sources, and hence standard methods for
Poisson equations like integral solvers can be used to obtain a
numerical solution. For an equation of the form
\begin{equation}
  \hat{\Delta} u (\mb{x}) = f (\mb{x}),
  \label{eq:standard_poisson}
\end{equation}
the solution for the potential $ u $ can be expanded in axisymmetry as
\begin{align}
  u (\mb{x}) & = - \frac{1}{4 \pi} \int \mathrm{d}^3\mb{x}' \,
  \sqrt{\hat{\gamma}} \,
  \frac{f (\mb{x})}{|\mb{x} - \mb{x}'|}
  \nonumber \\ 
  & = - \frac{1}{2} \sum^{\infty}_{l=0} P_l (\mu)
  \left( u^{(l)}_{\mathrm{in}} (\mb{x}) +
  u^{(l)}_{\mathrm{out}} (\mb{x}) \right),
  \label{eq:expansion}
\end{align}%
where
\begin{align}
  u^{(l)}_{\mathrm{in}} (\mb{x}) & =
   \frac{1}{r^{l+1}} \int_{|\mbsm{x}'|<R} \!\!\!\!\!\!\!
  \mathrm{d}r' \, \mathrm{d}\mu' \, f(\mb{x}') r'^{l+2} P_l (\mu'),
  \label{eq:expansion_1}
  \\
  u^{(l)}_{\mathrm{out}} (\mb{x}) & =
   r^l \int_{|\mbsm{x}'|>R} \!\!\!\!\!\!\!
  \mathrm{d}r' \, \mathrm{d}\mu' \,
  f(\mb{x}') \frac{1}{r'^{l-1}} P_l (\mu').
  \label{eq:expansion_2}
\end{align}%
Here $ P_l $ are the Legendre polynomials, and $ \mu = \cos \theta $.
We numerically integrate Eqs.~(\ref{eq:expansion_1},
\ref{eq:expansion_2}) by assuming $ f (\mb{x}') $ to be constant
inside each computational cell, integrating over $ r' $ and $ \theta' $
analytically within each cell, and then adding up the partial integrals to
obtain the solution at every grid point of the computational
domain. This method has been described and tested in
\citet{mueller_95_a} and \citet{zwerger_95_a}, and successfully used
by \citet{zwerger_97_a} to calculate the Newtonian potential in
axisymmetric core collapse simulations.

The equations~(\ref{eq:lap_s}--\ref{eq:lap_sij}) for the intermediate
potentials, which are needed to compute the second post-Newtonian
corrections to CFC, can in general be solved separately as a scalar
Poisson equation for $ \mathcal{S} $, three vector Poisson equations
for $ \mathcal{S}_i $, $ \mathcal{T}_i $, and $ \mathcal{R}_i $, and a
tensor Poisson equation for $ \mathcal{S}_{ij} $. In the axisymmetric
case we can take advantage of some additional simplifications: The
$ \varphi $-component of the vector-Poisson equations decouples from
the $ r $- and $ \theta $-components, and the $ r \varphi $- and
$ \theta \varphi $-components of the tensor-Poisson equation decouple
from all other components, even though e.g.\
$ \hat{\Delta} \mathcal{S}_i \ne (\hat{\Delta} \mathcal{S})_i $
(similar for the other vectors and tensor), which means that the various
components in general couple to each other. Therefore, the equations
can be grouped into 9 sets of linear elliptic equations: four sets of
one equation, four sets of two equations, and one set of four
equations, with coupling only within the respective set.

The discretization of each of the equations on the $ \{r, \theta \} $-grid 
leads to 9 sparse linear matrix equations of the type
\begin{equation}
  \mb{\mathcal{A}} \mb{u} = \mb{f}, 
  \label{eq:matrix_eq}
\end{equation}
where $ \mb{u} = u^k_{ij} $ is the vector of unknowns with $ i,j $
labeling the grid points and $ k $ ranging from 1 to 1, 2, or 4
depending on the number of coupled components. The vector of sources
is respectively denoted as $ \mb{f} = f^k_{ij} $, and
$ \mb{\mathcal{A}} $ is a matrix which does not depend on $ \mb{u} $,
as the original system is linear. The linearity of the equations allows
us to avoid an iteration scheme and to use instead the LU
decomposition method to invert $ \mb{\mathcal{A}} $. The main advantage
of the LU method is that the decomposition itself (which is the most time
consuming step) only has to be done once at the beginning. Then it
can be used to calculate the solution for different source vectors
$ \mb{f} $ during the metric computations, which is computationally
very fast and efficient. The LU decomposition is performed using
standard LAPACK libraries for banded matrices. We use the monopole
solution of
Eqs.~(\ref{eq:mult_expansion_S}--\ref{eq:mult_expansion_Sij})
as explicit outer boundary condition for the intermediate
potentials. This procedure is only accurate far from the sources, and
the matching of the metric to a monopole solution deteriorates if it
is performed too close to a strongly gravitating nonspherical
star. As a consequence when calculating the components of the
spacetime metric we use a grid which extends well beyond both the
boundary of the star and also the atmosphere, and apply the boundary
condition at the outer boundary of this extended vacuum grid.
An example of the influence of the location where the boundary
condition is applied on the accuracy of the metric solution is
presented in Sect.~\ref{subsec:boundary_influence}. Although the
nonlinear metric equations for $ \phi $, $ \alpha $, and $ \beta^i $
have source terms with noncompact support, the location
of the outer grid boundary has much less influence on the accuracy of
the numerical solution (see \citet{dimmelmeier_04_a}).

At this stage we have a numerical solution for $ h^\mathrm{TT}_{ij} $
and are ready to solve the CFC+ metric equations for the conformal
factor $ \phi $, the shift vector $ \beta^i $, and the lapse function
$ \alpha $, Eqs.~(\ref{eq:metric_equation_1},
\ref{eq:metric_equation_3}, \ref{eq:alpha_phi}). For the comparison
between CFC and CFC+ presented in Sects.~\ref{sec:oscillations_rns}
and~\ref{sec:core_collapse}, we also need to solve the CFC metric
equations, which, we recall, are equivalent to the CFC+ equations up
to corrections in the metric equation for
$ \alpha $~(\ref{eq:metric_equation_2}). As both systems of equations
are 5 nonlinear elliptic coupled Poisson-like equations, we can apply
the same methods to solve them. In either case, we can write them in
compact form as
\begin{equation}
  \hat{\Delta} \mb{u} (\mb{x}) = \mb{f} (\mb{x}; \mb{u} (\mb{x})),
\end{equation}
where $ \mb{u} = u^k = (\phi, \alpha \phi, \beta^j) $, and
$ \mb{f} = f^k $ is the vector of the respective sources. These
five scalar equations for each component of $ \mb{u} $ couple to each
other via the source terms which in general depend on the various
components of $ \mb{u} $.

We use a fix-point iteration scheme in combination with the linear
Poisson solver~(\ref{eq:expansion}) described above to solve these
equations. The value of $ \mb{u} $ at each iteration $ s $, denoted by
$ \mb{u}^s $, is set constant in the sources $ \mb{f} $ to calculate a
new value $ \mb{u}^{s + 1} $,
\begin{equation} 
  \mb{u}^{s + 1} (\mb{x}) = \hat{\Delta}^{-1} \mb{f}
  \left( \mb{x}; \mb{u}^s (\mb{x}) \right).
  \label{eq:metric_iteration}
\end{equation}
As a result the 5 previously coupled nonlinear equations reduce to a
decoupled set of 5 linear scalar Poisson equations. The solution
vector $ \mb{u}^{s + 1} $ at each iteration step is obtained by
solving the associated 5 Poisson equations of the
type~(\ref{eq:standard_poisson}) separately. After the computation of
$ \mb{u}^{s + 1} $ the right-hand side of
Eq.~(\ref{eq:metric_iteration}) is updated by replacing
$ \mb{u}^s \to \mb{u}^{s + 1} $, which is used as a new starting value
for the next iteration. Convergence to the desired numerical solution
$ \mb{u} $ is achieved when the relative variation
$ |\mb{u}^{s + 1} / \mb{u}^s - 1| $ of the numerical solution of
$ \mb{u} $ between two successive iterations is smaller than a certain
threshold, which we set to $ 10^{-6} $. In the simulations reported
here the metric solver successfully converges when using the flat
metric as initial guess in each metric computation. However, to
accelerate convergence, during the hydrodynamic evolution we take the
metric from the previous metric computation as starting value for the
subsequent one. Furthermore, we cut the Legendre polynomial expansion
in Eq.~(\ref{eq:expansion}) at $ l = 10 $. For the CFC metric
equations in axisymmetry the above method has been recently discussed
in detail in~\citet{dimmelmeier_04_a}.


\subsection{Gravitational wave extraction}
\label{subsec:gravitational_wave_extraction}

For an isolated source of gravitational radiation the metric can be
considered as asymptotically flat. In the wave zone, defined as a
neighborhood of future radiative infinity $ \mathcal{I}^+ $ for which
$ r \gg \lambda / (2 \pi) $, the metric can be decomposed into the
Minkowski metric $ \eta_{\mu\nu} $ plus a small linear perturbation
$ h_{\mu\nu}^\mathrm{rad} $. By an appropriate choice of gauge it
is always possible to make the latter quantity algebraically
transverse and traceless, up to quadratic corrections. At the linear
order we have then
\begin{equation}
  h_{ij}^\mathrm{rad} =
  P_{ij}^{\mathrm{TT}kl} (\gamma_{kl} - \hat{\gamma}_{kl}),
\end{equation}
where
$ P_{ij}^{\mathrm{TT}kl} = \hat{\gamma}_{i(p} \hat{\gamma}_{q)j}
\bigl[ \left( n^p n^k - \hat{\gamma}^{pk} \right)
\left( n^q n^l - \hat{\gamma}^{ql} \right)
- \left( n^p n^q - \hat{\gamma}^{pq} \right)
\left( n^k n^l - \hat{\gamma}^{kl} \right)/2 \bigr] $
stands for the algebraic transverse traceless projector. We note
that the above expression stays invariant with respect to
asymptotically Minkowskian gauge transformations of the 3-metric. It
involves only two independent components, $ h_{+} $ and $ h_{\times} $,
representing the two degrees of freedom of the gravitational waves.

At linear order in the wave zone, $ h_{ij}^\mathrm{rad} $ can be
approximated by means of the Newtonian quadrupole formula in some
Bondi-like coordinate system. In this representation, the waveform
depends only on the second time derivative (denoted by a double dot) of the
mass quadrupole moment $ Q_{ij} $ of the source:
\begin{equation}
  h^\mathrm{rad}_{ij} (\mb{x}, t) \approx
  h^\mathrm{quad}_{ij} (\mb{x}, t) =
  P^{\mathrm{TT}kl}_{ij}
  \frac{2 \ddot{Q}_{kl}(t_\mathrm{ret})}{r}.
\end{equation}
with the retarded time $ t_\mathrm{ret} = t - r/c $.
The argument of $ \ddot{Q}_{kl} $ in the ADM gauge is not
simply equal to $ t - r / c $. It also contains a logarithmic
correction \citep{schaefer_90_a}
$ - 2 \, (M + \mathcal{O}(1/c^2)) \, G \ln [r / (c \eta)] / c^3 + 
\mathcal{O}(G^2) $,  
where $ \eta $ denotes an arbitrary constant with the
dimension of time that cancels a term entering the first tail
contribution. For large distances near future null infinity
$ \mathcal{I}^+ $, this correction cannot be neglected, but it may be
omitted in investigations focusing on time dependences as it depends
on $ r $ only. Following \citet{finn_89_a}, the time derivatives in
the {\it standard quadrupole formula} can be replaced by spatial
derivatives making use of the equations of motion. This yields the
so-called {\it stress formula},
\begin{equation}
  \ddot{Q}_{ij} = 2 \! \int \! \mathrm{d}^3\mb{x} \, \sqrt{\hat{\gamma}} D^*
  \left[ \hat{\gamma}_{k<i} \hat{\gamma}_{j>l} v^{k} v^{l} + x^k
  \hat{\gamma}_{k<i} \hat{\nabla}_{j>} U \right],
\end{equation}
where the brackets $ <> $ denote the symmetric and trace-free sets of
indices, e.g.\ $ \hat{\gamma}_{k<i} \hat{\nabla}_{j>} U =
(\hat{\gamma}_{ki} \hat{\nabla}_{j} U +
\hat{\gamma}_{kj} \hat{\nabla}_{i} U)/2 - 
\hat{\gamma}_{ij} \hat{\nabla}_k U/3 $.

In axisymmetry, $ h^\mathrm{rad}_{ij} $ has only one degree of
freedom, $ h_+ $. In the quadrupole approximation and in
spherical coordinates, it takes the form
\begin{equation}
  h^\mathrm{quad}_+ (\mb{x}, t) =
  \frac{1}{8} \sqrt{\frac{15}{\pi}} \sin^2{\theta}
  \frac{A^\mathrm{E2}_{20} (t-r/c)}{r},
\end{equation}
with $ A^\mathrm{E2}_{20} \propto \ddot{Q}_{\theta \theta} $ being the
amplitude of the quadrupole signal. The above formula allows us to
calculate waveforms directly from the sources, with no need of knowing
the dynamical part of the metric. It may be used with
profit in the CFC approximation, for which the metric is isotropic and
has therefore no gravitational wave content.

In the CFC+ approach, the non-conformal part of the 3-metric in the
near zone at the dominant (second) post-Newtonian order,
$ h^\mathrm{2PN}_{ij} $, is the solution of the pseudo-Poisson
equation~(\ref{eq:equation_hTT}), where the remainder
$ \mathcal{O}(1 / c^6) $ has been removed. This equation has an
elliptic character, so that the information it carries propagates
instantaneously. As a result, $ h^\mathrm{2PN}_{ij} $ is neither
wave-like nor associated to any retardation effects, reflecting the
fact that the post-Newtonian expansion is only valid at distances
comparable to the wavelength of the system $ \lambda $. By contrast,
we know that the full non-conformally flat part
$ h^\mathrm{full}_{ij} $ of the 3-metric carries the dynamical degrees
of freedom of the gravitational field. This indicates a relation
between $ h_{ij}^\mathrm{full} $ and $ h_{ij}^\mathrm{rad} $ near
$ \mathcal{I}^+ $ (they are indeed equal). Now, remarkably, it turns
out that there is also a link between $ h_{ij}^\mathrm{2PN} $ and the
radiative metric. As explained in Appendix~\ref{app:GW}, the algebraic
transverse traceless projection of the former quantity evaluated at
retarded time $ t_\mathrm{ret} $ behaves as
$ h_{ij}^\mathrm{quad}/8 $ at future null infinity,
$ P_{ij}^{\mathrm{TT}kl} h^\mathrm{2PN}_{kl} (\mb{x}, t_\mathrm{ret})
\sim h_{ij}^\mathrm{quad}/8 $, or in components,
\begin{align}
  h^\mathrm{2PN}_+ (\mb{x}, t_\mathrm{ret}) & \sim
  \frac{1}{8} h^\mathrm{quad}_+ (\mb{x}, t),
  \label{eq:h+2pn} \\
  h^\mathrm{2PN}_\times (\mb{x}, t_\mathrm{ret}) & \sim
  \frac{1}{8} h^\mathrm{quad}_\times (\mb{x}, t),
  \label{eq:hx2pn}
\end{align}%
where $ h^\mathrm{2PN}_+ $ can be trivially calculated from the combination $
h_+ = (h^\mathrm{TT}_{\theta \theta} - h^\mathrm{TT}_{\varphi \varphi})/2
$, whereas $ h^\mathrm{2PN}_\times $ is
equal to $ h^\mathrm{2PN}_{\theta \varphi} $.

We see that the two polarizations extracted directly from
$ h^\mathrm{2PN}_{ij} $ in the wave zone of the true waves differ from
the quadrupole waveforms by (i) a factor 8, and (ii) the absence of
retardation. The gravitational waveforms can thus be evaluated
directly from $ h_+^\mathrm{2PN} $ and $ h_\times^\mathrm{2PN} $ even
though there are no propagating gravitational waves in the CFC+
spacetime (which just means that there is no radiation back-reaction
due to energy losses caused by gravitational waves). Furthermore, the
wave amplitude can be deduced solely from $ h_+^\mathrm{2PN} $ in the
axisymmetric case.

In our simulations we extract the waveforms at a distance
$ r \sim 50 \lambda / (2 \pi) $ in the equatorial plane. This ensures
that we are indeed in the wave zone of the true waves where the
Newtonian quadrupole formula applies. We have checked that
$ h_+^\mathrm{2PN} \propto \sin^2 \theta / r $ there, so that the
gravitational wave amplitude $ A^\mathrm{E2}_{20} $ is approximately
constant independent of the radius $ r $ or angle $ \theta $ (except
near the rotation axis where $ h_+^\mathrm{2PN} $
vanishes). Due to the smallness of $ h_+^\mathrm{2PN} $ for
$ r \gg \lambda / (2 \pi) $, some numerical error appears in the
cancellations of the different terms in Eq.~(\ref{eq:hTT}). That
yields an offset in the gravitational wave signal that can be
corrected as follows:
\begin{equation}
  h_+^\mathrm{2PN\,corrected} = h_+^\mathrm{2PN} -
  a \, \hat{\gamma}^{ij} h^\mathrm{2PN}_{ij}.
  \label{eq:h+2pn_corrected}
\end{equation}
Although the term $ \hat{\gamma}^{ij} h^\mathrm{2PN}_{ij} $ should be
zero in principle, it is numerically comparable to
$ h^\mathrm{2PN}_{ij} $ in the wave zone; the parameter $ a $ 
corrects the offset in the waveforms.

Notice that the link between $ h^\mathrm{2PN}_{ij} $ and
$ h^\mathrm{full}_{ij} $ is trivial in the near zone, where the
post-Newtonian approximation is valid:
$ h^\mathrm{full}_{ij} \approx h^\mathrm{2PN}_{ij} $. We shall denote
their common value as $ h^\mathrm{TT}_{ij} $ henceforth.


\section{Initial models}
\label{sec:equilibrium_rs}

As initial models to describe either rotating neutron stars or
rotating stellar cores at the onset of gravitational collapse we
construct uniformly or differentially rotating relativistic polytropes
in equilibrium. These are obtained using Hachisu's self-consistent
field (HSCF) method as described in \citet{komatsu_89_a,
  komatsu_89_b}, which solves the general relativistic hydrostatic
equations for self-gravitating rotating matter distributions, whose
pressure obeys the polytropic relation
\begin{equation}
  P = K \rho^\gamma,
  \label{eq:polytropic_eos}
\end{equation}
where $K$ is the polytropic constant and $\gamma$ is the adiabatic index.
The gauge used in the HSCF method is maximal slicing with
quasi-isotropic coordinates (MSQI). The most general metric to
describe these objects in the MSQI gauge is
\begin{align}
  ds^2 & = - e^{2\tilde{\nu}} d\tilde{t}^2 +
  e^{2\tilde{\alpha}} (d\tilde{r}^2 + \tilde{r}^2 d\tilde{\theta}^2) 
  \nonumber \\
  & + e^{2\tilde{\beta}} \tilde{r}^2 \sin{\tilde{\theta}}^2 
  (d\tilde{\varphi} - \tilde{\omega} d\tilde{t})^2,
\end{align}%
where $ \{\tilde{t}, \tilde{r}, \tilde{\theta}, \tilde{\varphi}\} $
are the coordinates in the MSQI gauge, and $\tilde{\nu}$,
$\tilde{\alpha}$, $\tilde{\beta}$, and $\tilde{\omega}$ are metric
potentials. Hereafter quantities with a tilde are in the MSQI gauge,
and all other quantities are in the ADM gauge. The hydrostatic
equilibrium equations can be numerically integrated by prescribing a
value for the central density $ \rho_\mathrm{c} $, the rotation rate
(selected by setting a ratio of polar radius $ r_\mathrm{p} $ to
equatorial radius $ r_\mathrm{e} $), and choosing a rotation law. As
it is commonly done in the literature we use
\begin{equation}
  \Omega = \Omega_\mathrm{c} \frac{A^2}{A^2 + d^2} 
  \label{rotlaw}
\end{equation}
as rotation law, where $ \Omega $ is the angular velocity of the fluid
as measured from infinity, $ \Omega_\mathrm{c} $ its value at the
center, and $ d = r \sin \theta $ the distance to the rotation axis.
The positive constant $ A $ parametrizes the rate of differential
rotation, with $ A \to \infty $ for a rigid rotator and
$ A \lesssim r_\mathrm{e} $ for differentially rotating stars.

For the study of rotating neutron stars, we choose the polytropic
EOS~(\ref{eq:polytropic_eos}) with $ K = 1.456 \times 10^5 $ (in cgs
units) and $ \gamma = 2 $. We have constructed a sequence of uniformly
rotating models with a central density
$ \rho_\mathrm{c} = 7.95 \times 10^{14} \mathrm{\ g\ cm}^{-3} $, and a
ratio $ r_\mathrm{p} / r_\mathrm{e} $ of polar to equatorial
coordinate radius ranging from 1.00 (spherical model; labeled RNS0)
to 0.65 (rapidly rotating model near the mass-shedding limit; labeled
RNS5). A summary of important quantities specifying these models is
given in Table~\ref{tab:RNS_models}. Note also that models with this
choice of parameters are of widespread use in the literature (see
e.g.\ \citet{font_02_a} or \citet{dimmelmeier_04_a} and references
therein) and will be used in this work for comparison with previous
results obtained with independent codes.

\begin{table}
  \centering
  \caption{Equatorial radius $ r_\mathrm{e} $, axis ratio
    $ r_\mathrm{p} / r_\mathrm{e} $, angular velocity $ \Omega $,
    and ADM mass $ M_\mathrm{ADM} $ for a sequence of
    uniformly rotating neutron stars used in this paper. Along the
    sequence the rotation rate increases, from the spherical model
    RNS0 to the most rapidly rotating model RNS5, which rotates near
    the mass shedding limit at
    $ \Omega_\mathrm{K} = 5.363 \mathrm{\ kHz} $.}
  \begin{tabular}{ccccc}
  \hline \hline
   Model & $ r_\mathrm{e} $ [km] & $ r_\mathrm{p} / r_\mathrm{e} $ &
   $ \Omega / \Omega_\mathrm{K} $ & $ M_\mathrm{ADM} $ [$ M_\odot $] \\
   \hline
   RNS0  &  14.1   &   1.00   &   0.00         & 1.40     \\
   RNS1  &  16.2   &   0.95   &   0.42         & 1.44     \\
   RNS2  &  17.3   &   0.85   &   0.70         & 1.51     \\
   RNS3  &  18.7   &   0.75   &   0.87         & 1.59     \\
   RNS4  &  19.6   &   0.70   &   0.93         & 1.63     \\
   RNS5  &  20.7   &   0.65   &   0.98         & 1.67     \\
   \hline \hline  
   \end{tabular}
   \label{tab:RNS_models}
\end{table}

To model a stellar iron core as initial model for simulating rotational 
stellar core collapse to a neutron star we again utilize the HSCF method 
with the EOS parameters $ K = 4.897 \times 10^{14} $ (in cgs units) and 
$ \gamma = 4 / 3 $, chosen to approximate the pressure profile in a 
degenerate electron gas. The initial central density is set to
$ \rho_\mathrm{c\,ini} = 10^{10} \mathrm{\ g\ cm}^{-3} $. Again each initial
model is further determined by its rotation profile parameter $ A $
and the rotation rate, which is specified by the axis ratio
$ r_\mathrm{p} / r_\mathrm{e} $ or equivalently by the ratio of
rotational energy and gravitational binding energy, $ \beta = T / |W| $.

\begin{table}
  \centering
  \caption{Parameters used in the core collapse simulations. The
    initial models are differentially rotating stellar cores specified by
    the parameter $ A $ controling the degree of differential rotation
    (cf.\ Eq.~(\ref{rotlaw})) and the ratio of rotational to potential
    energy, $ \beta = T / |W| $. The collapse is initiated by reducing
    $ \gamma $ from its initial value $ 4/3 $ to $ \gamma_1 $.
    Additionally, the values for the initial equatorial radius
    $ r_\mathrm{e} $ and the initial ADM mass
    $ M_\mathrm{ADM} $ are given. The label AxByGz of each model is a
    combination of the initial rotation parameters $ A $ and
    $ \beta $ (AxBy) and the value of $ \gamma_1 $ during collapse
    (Gz).}
  \begin{tabular}{cccccc}
    \hline \hline \\ [-1 em]
    Model & $ A $                  & $ \beta $ &
    $ r_\mathrm{e} $ & $ M_\mathrm{ADM} $ & $ \gamma_1 $ \\
          & [$10^3 \mathrm{\ km}$] & [\%]      &
    [km]             & [$ M_\odot $]      &              \\
    \hline
    A1B3G3 & 50.0     & 0.89 & 2247 & 1.46 & 1.31 \\
    A1B3G5 & 50.0     & 0.89 & 2247 & 1.46 & 1.28 \\
    A2B4G5 & \wz 1.0    & 1.81 & 1739 & 1.50 & 1.28 \\
    A4B5G5 & \wz 0.5  & 4.03 & 1249 & 1.61 & 1.28 \\
    \hline \hline  \\ [-3 em]
  \end{tabular}
  \label{tab:CC_models}
\end{table}

Following \citet{dimmelmeier_02_a} we use a hybrid EOS in the core collapse 
simulations. This EOS consists of a polytropic part $ P_\mathrm{p} = K 
\rho^\gamma $ plus a thermal part
$ P_\mathrm{th} = (\gamma_\mathrm{th} - 1) \epsilon_\mathrm{th} $,
where $ \gamma_\mathrm{th} = 1.5 $ and
$ \epsilon_\mathrm{th} = \epsilon - \epsilon_\mathrm{p} $. The thermal
contribution is chosen to take into account the rise of thermal energy
due to shock heating. Details of this EOS can be found in
\citet{dimmelmeier_02_a}. Gravitational collapse is initiated by
slightly decreasing $ \gamma $ from its initial value to
$ \gamma_1 < 4 / 3 $.  As $ \rho $ reaches nuclear density
$ \rho_\mathrm{nuc} = 2.0 \times 10^{14} \mathrm{\ g\ cm}^{-3} $,
we raise $ \gamma $ to $ \gamma_2 = 2.5 $ and adjust $ K $ accordingly
to maintain monotonicity of the pressure. Due to this stiffening of
the EOS the core undergoes a bounce. In Table~\ref{tab:CC_models} we
present the main properties of models A1B3G3, A1B3G5, A2B4G5, and
A4B5G5 used in this work to study core collapse. These models are
identical to those with the same labels in the comprehensive core
collapse study performed by \citet{dimmelmeier_02_b}.

\begin{figure}
  \resizebox{\hsize}{!}{\includegraphics{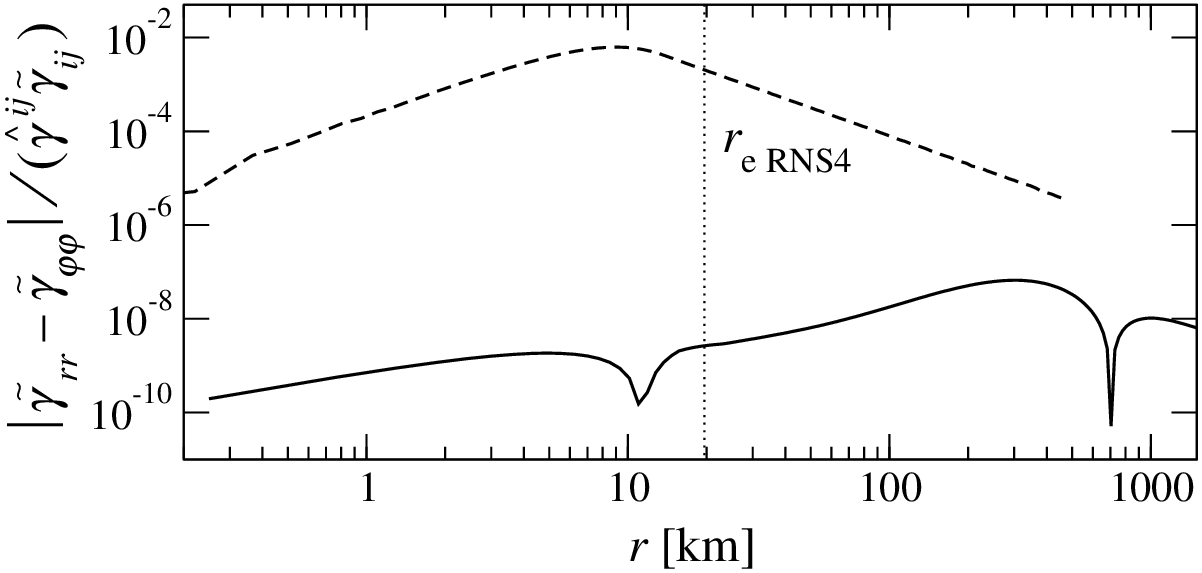}}
  \caption{Deviation from conformal flatness along the equatorial
    plane for a typical rotating stellar core initial model (model
    A1B3; solid line), and for a typical rotating neutron star in
    equilibrium with axis ratio $ r_\mathrm{p} / r_\mathrm{e} = 0.7 $
    (model RNS4; dashed line). The vertical dotted line indicates the
    equatorial stellar radius for the neutron star.}
  \label{fig:delta_g_MSQI}
\end{figure}

We note that when evolving the initial models constructed on the basis
of the HSCF method (which uses the MSQI gauge) with the CFC+ evolution
code (which uses the ADM gauge), we have to consider that in general
the two gauges differ. This could potentially lead to an unsuitable
matching of the data describing the initial models when evolved with
the numerical code. Let us consider the most general metric in a
generic dynamic scenario,
\begin{align}
  \gamma_{ij} & = \phi^4 \hat{\gamma}_{ij} + h^\mathrm{TT}_{ij}
  & \mathrm{(ADM\ gauge)},\:
  \\
  \tilde{\gamma}_{ij} & = \tilde{\phi}^4 \hat{\gamma}_{ij} +
  \tilde{\gamma}_{<ij>}
  & \mathrm{(MSQI\ gauge)},
\end{align}%
In general $ \tilde{\gamma}_{<ij>} $ is not transverse, so that the
ADM gauge and the MSQI gauge differ. To quantify the differences
between both gauges it is relevant to  compare the traceless part
of the 3-metric $ \gamma_{<ij>} $ and the trace
$ \hat{\gamma}^{ij} \gamma_{ij} = 3\phi^4 $ itself. For an equilibrium stellar
model constructed within the MSQI gauge one obtains
\begin{equation}
  \tilde{\gamma}_{<ij>} =
  \left( \tilde{\gamma}_{rr} - \tilde{\gamma}_{\varphi\varphi} \right)
  \diag \left( \frac{1}{3}, \frac{1}{3}, - \frac{2}{3} \right).
\end{equation}
In Fig.~\ref{fig:delta_g_MSQI} we plot equatorial
profiles of $ |\tilde{\gamma}_{rr} - \tilde{\gamma}_{\varphi \varphi}| /
(\hat{\gamma}^{ij} \tilde{\gamma}_{ij}) $ for (i) the equilibrium model
A1B3 used as initial data for core collapse simulations (lower curve; cf.\
Table~\ref{tab:CC_models}), and for (ii) the equilibrium rotating neutron star
model RNS4 (upper curve; cf.\ Table~\ref{tab:RNS_models}). In both cases,
especially in the (only weakly relativistic) collapse initial model,
we observe that deviations from conformal flatness are
negligible. This fact makes the initial models built with HSCF method
suitable for time evolution in the CFC approximation. It also shows
that the differences between both gauges are very small, namely of the
same order of magnitude as $ \tilde{\gamma}_{<ij>} $. Consequently,
the use of initial models computed in the MSQI gauge for evolutions
using the ADM gauge is entirely justified since the differences are
negligible and only appear at the second post-Newtonian order.


\section{Rotating neutron stars}
\label{sec:oscillations_rns}


\subsection{Effects of the boundary on the metric solution}
\label{subsec:boundary_influence}

As mentioned in Sect.~\ref{subsec:metric_solver} the location where
the boundary conditions for the CFC+ intermediate potentials are
imposed can have a significant influence on the accuracy of the metric
solution. Additionally, the extension of the numerical grid is also of
paramount importance for the numerical computation of the CFC
equations for the conformal factor, the shift vector, and the lapse
function. As the sources of these equations have non-compact support,
it is necessary to integrate out far enough in radius to obtain the
desired accuracy of the numerical solution.

\begin{figure}
  \resizebox{\hsize}{!}{\includegraphics{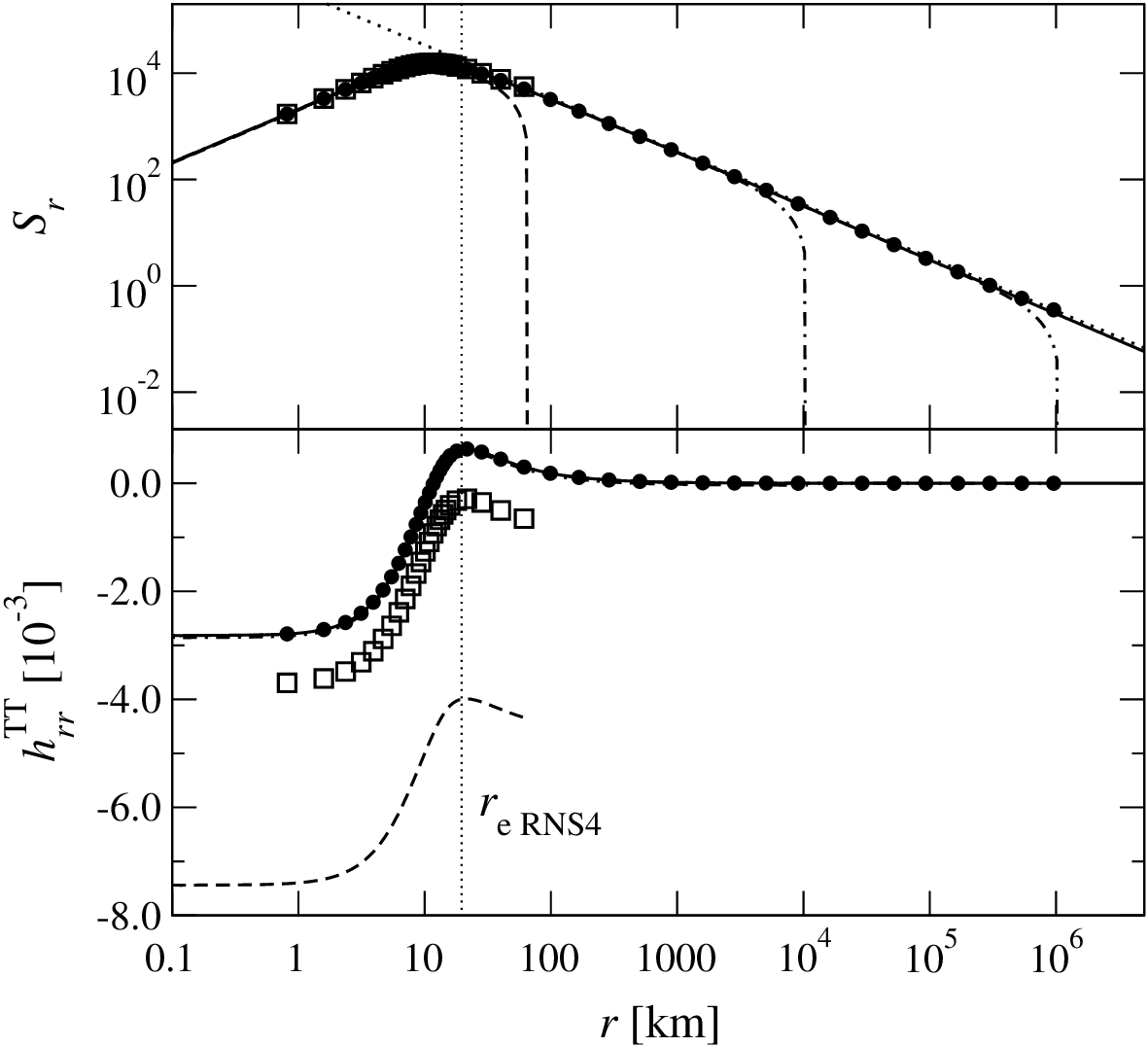}}
  \caption{Effects of the outer boundary conditions on the equatorial
    radial profile of the intermediate potential $ \mathcal{S}_r $
    (top panel) and the metric component $ h^\mathrm{TT}_{rr} $
    (bottom panel) for the rotating neutron star model
    RNS4. Zero-value boundary conditions are imposed at
    $ 62 \mathrm{\ km} $ (dashed line), at
    $ 10^4 \mathrm{\ km} $ (dashed-dotted line), at
    $ 10^6 \mathrm{\ km} $ (dashed-dot-dotted line), and at
    $ 10^{11} \mathrm{\ km} $ (solid line). Alternatively, monopole
    boundary conditions are imposed at $ 62 \mathrm{\ km} $ (open
    boxes) and at $ 10^6 \mathrm{\ km} $ (filled circles). Overplotted
    to $ \mathcal{S}_r $ is the monopole behavior (dotted line).}
  \label{fig:htt_boundary}
\end{figure}

To accomplish this we use an extended (vacuum) grid going far beyond
the actual stellar boundary. Monopole behavior at the outer boundary
of this vacuum domain has been checked by comparing with calculations
done imposing zero values for the potentials as outer boundary
conditions at extremely large distances (see
Fig.~\ref{fig:htt_boundary}). Convergence tests with different
parameters for the outer grid (number of cells, distance of the outer
boundary, amplification factor of the grid) have been performed. The
conclusion of these tests is that for simulations of rotating neutron
stars we need to use a numerical grid consisting of $ n_\theta = 30 $
angular cells and $ n_r = 250 $ radial cells (of which 100 are equally
spaced inside the neutron star and 150 are logarithmically spaced for
the atmosphere) to correctly capture the conformally flat part of the
metric. In order to impose outer boundary conditions for the CFC+
metric potentials, an extra grid of 70 radial cells extending out to
$ 10^6 \mathrm{\ km} $ needs to be added.


\subsection{CFC+ corrections to the 3-metric}
\label{subsec:rns}

\begin{figure}
  \resizebox{\hsize}{!}{\includegraphics{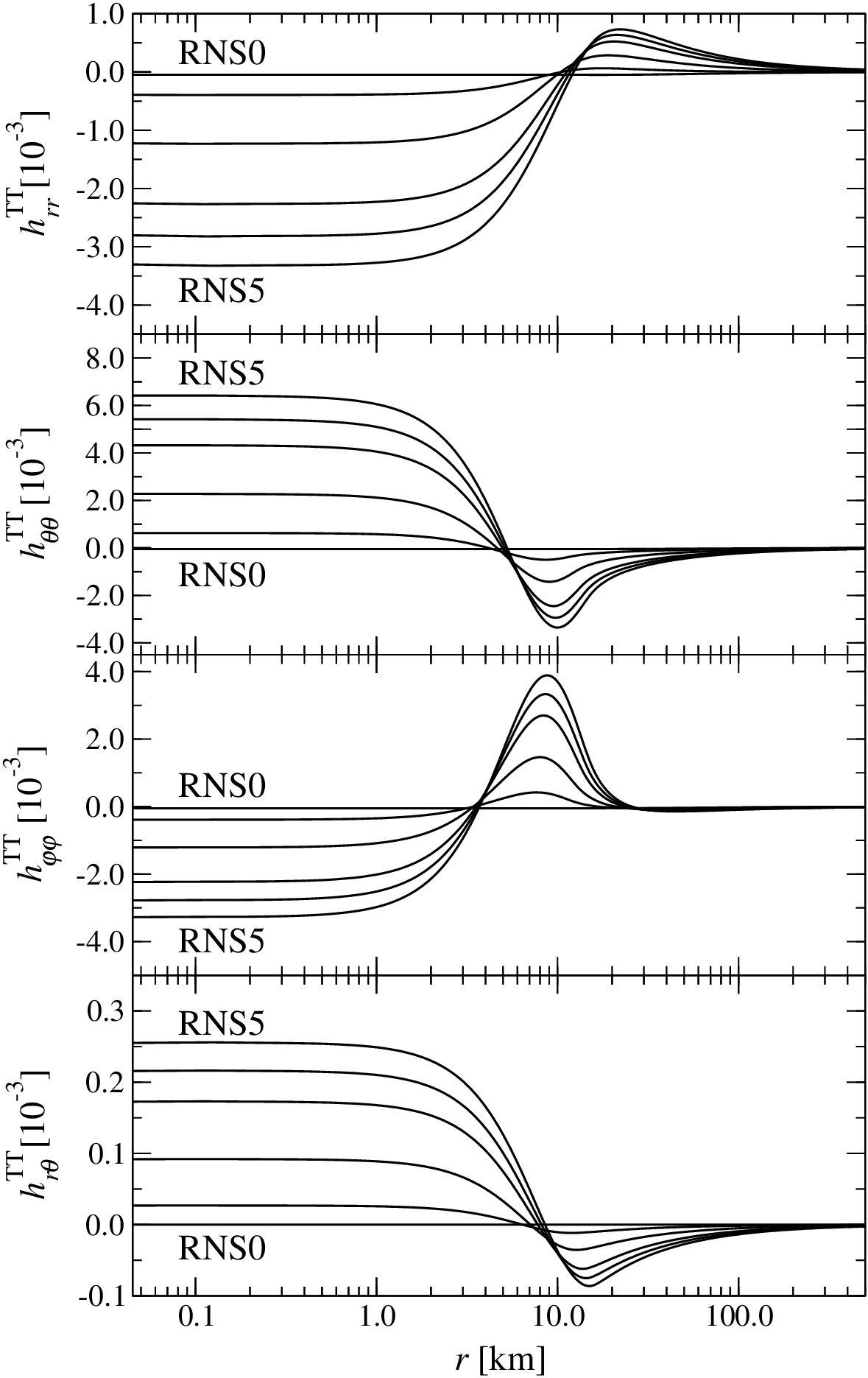}}
  \caption{Radial profiles at the equator of the non-vanishing
    components of $ h^\mathrm{TT}_{ij} $ for the sequence of models
    RNS0 to RNS5. The strength of the correction
    $ h^\mathrm{TT}_{ij} $ increases with the rotation rate. The
    equatorial radius of each model is given in
    Table~\ref{tab:RNS_models}.}
  \label{fig:htt_RNS}
\end{figure}

We turn now to measure the magnitude of the CFC+ corrections to the
metric of our sample of rotating neutron star models (cf.\
Table~\ref{tab:RNS_models}). First, from the distribution of the
hydrodynamic variables in the equilibrium models provided by the
HSCF method we recompute the CFC+ metric components. For equilibrium
configurations, i.e.\ in an axisymmetric stationary spacetime, the
components $ h^\mathrm{TT}_{r \varphi} $ and
$ h^\mathrm{TT}_{\theta \varphi} $ vanish. All other components are
shown in Fig.~\ref{fig:htt_RNS} for different models with increasing
rotation rate. Note that as discussed in Sect.~\ref{sec:equilibrium_rs}
the ADM gauge is not quasi-isotropic in the CFC+ approach, because
$ h^\mathrm{TT}_{r \theta} \ne 0 $ and
$ h^\mathrm{TT}_{rr} \ne h^\mathrm{TT}_{\theta \theta} $.
Therefore, a direct comparison with the full general relativistic metric
calculated in the MSQI gauge is not possible. As expected, $
h^\mathrm{TT}_{ij} $ vanishes for the spherical case RNS0, in which
both CFC and CFC+ agree exactly with general relativity, and there is
no traceless and transverse part. As we increase the rotation rate
from RNS1 to RNS5, the value of $ h^\mathrm{TT}_{ij} $ increases,
resulting in a departure of the metric from conformal flatness. As
$ h^\mathrm{TT}_{ij} $ is a correction to the conformally flat 3-metric
(which is of order unity, $ \phi \sim 1 $), it can be shown that even
in our most extreme case RNS5, which is very near to the mass shedding
limit, the values of $ h^\mathrm{TT}_{ij} $ amount to a correction of
only about 1\%. Therefore, qualitative differences with respect to the
CFC approximation are not expected in the dynamics of such objects.


\subsection{Oscillations of rotating neutron stars}
\label{subsec:rns_oscillations}

A further test of the new approximation for the metric is
provided by the study of pulsations of rotating neutron stars. To this
aim we perturb the neutron star models described in
Sect.~\ref{sec:equilibrium_rs} with a density perturbation of the form
$ \rho_\mathrm{pert} = \rho \, [1 + a \cos (\pi r / (2 r_\mathrm{e}))] $,
where $ a $ is an arbitrary parameter controlling the strength of the
perturbation. The perturbed models are evolved in time in two
different ways, either considering {\it coupled} evolutions for the
hydrodynamics and the metric, or evolving {\it only} the hydrodynamics
in a fixed background metric corresponding to the metric provided by
the elliptic solvers at the first time step (an approximation commonly
referred to in the literature as the Cowling approximation). Both
metric approximations, CFC and CFC+, are used to compare the
respective results. 

The oscillations of the stars can be observed in various hydrodynamic
and metric quantities. In particular we monitor the central density
$ \rho_\mathrm{c} $, the radial velocity $ v_r $ at half the stellar
radius, and the gravitational wave amplitudes $ A^\mathrm{E2}_{20} $
extracted with the quadrupole formula. When Fourier transforming the
time evolution of these quantities, distinctive peaks appear at the
same (discrete) frequencies for any of these variables. Those
frequencies can be identified with the quasi-normal modes of pulsation
of the star, as described in \citet{font_00_b}. To further validate
our approach we present the quasi-normal modes calculated in the
CFC and CFC+ approximation in comparison with those reported
by~\citet{font_02_a}, which are calculated with a 3D code in full
general relativity (without any approximation).

\begin{table}
  \centering
  \caption{Frequencies of small-amplitude quasi-radial pulsations for
    model RNS4. We compare the frequencies obtained from
    simulations with the present code (using either the CFC or the CFC+
    approximation) with those obtained independently from a 3D full general
    relativistic code (GR). The results are extracted from time
    evolutions where the spacetime metric is kept fixed (Cowling
    approximation). The relative differences between the CFC+ and the
    GR code are shown in the last column.}
  \begin{tabular}{ccccc}
    \hline \hline \\ [-1 em]
    Mode & $ f^\mathrm{CFC} $ & $ f^\mathrm{CFC+} $ &
    $ f^\mathrm{GR} $ & Rel.\ diff. \\
    & [kHz] & [kHz] & [kHz] & [\%] \\
    \hline
    F  & 2.48 & 2.48 & 2.468 & 0.5 \\
    H1 & 4.39 & 4.39 & 4.344 & 1.1 \\
    H2 & 6.30 & 6.30 & 6.250 & 0.8 \\
    \hline \hline
  \end{tabular}
  \label{tab:oscillating_RNS4_cowling}
\end{table}

Table~\ref{tab:oscillating_RNS4_cowling} shows the mode-frequencies
for the fundamental mode (F), as well as for the first (H1) and second
(H2) harmonics obtained in evolutions of model RNS4 in which the
spacetime metric is kept fixed (Cowling approximation). The accuracy
in the frequency values depends on the total time of the evolution,
increasing as the evolution is extended. We have evolved all models
for $ 30 \mathrm{\ ms} $, finding no significant deviations in the
hydrodynamic profiles with respect to the original profiles of the
equilibrium models. For such an evolution time the FFT yields a
maximum frequency resolution of $ 0.03 \mathrm{\ kHz} $.
Table~\ref{tab:oscillating_RNS4_cowling} shows that no differences
can be observed between the mode-frequencies computed with CFC and
CFC+, and that, in addition, there is very good agreement with the
general relativistic results since the reported values are within the
affordable resolution in frequency.

\begin{table}
  \centering
  \caption{Fundamental mode frequency $ f_\mathrm{F} $ of
    small-amplitude quasi-radial pulsations for a sequence of rotating
    polytropes with increasing ratio of polar to equatorial radius
    $ r_\mathrm{p} / r_\mathrm{e} $. We compare the frequencies
    obtained from simulations with the present code using either the
    CFC or the CFC+ approximation with those obtained independently
    from a 3D full general relativistic code (GR). The results are
    extracted from coupled spacetime metric and hydrodynamic time
    evolutions. The relative differences between the CFC+ and the
    GR code are shown in the last column.}
  \begin{tabular}{cccccc}
    \hline \hline \\ [-1 em]
    Model & $ r_\mathrm{p} / r_\mathrm{e} $ &
    $ f_\mathrm{F}^\mathrm{CFC} $ & $ f_\mathrm{F}^\mathrm{CFC+} $ &
    $ f_\mathrm{F}^\mathrm{GR} $ & Rel.\ diff. \\
    & & [kHz] & [kHz] & [kHz] & [\%] \\
    \hline
    RNS0 & 1.00 & 1.43 & 1.43 & 1.450 & 1.4 \\
    RNS1 & 0.95 & 1.40 & 1.40 & 1.411 & 0.8 \\
    RNS2 & 0.85 & 1.34 & 1.34 & 1.350 & 0.7 \\
    RNS3 & 0.75 & 1.27 & 1.27 & 1.265 & 0.4 \\
    RNS4 & 0.70 & 1.24 & 1.24 & 1.247 & 0.6 \\
    RNS5 & 0.65 & 1.21 & 1.21 & 1.195 & 1.0 \\
    \hline \hline
  \end{tabular}
  \label{tab:oscillating_RNS_coupled_F}
\end{table}

\begin{figure*}
  \resizebox{\hsize}{!}{\includegraphics{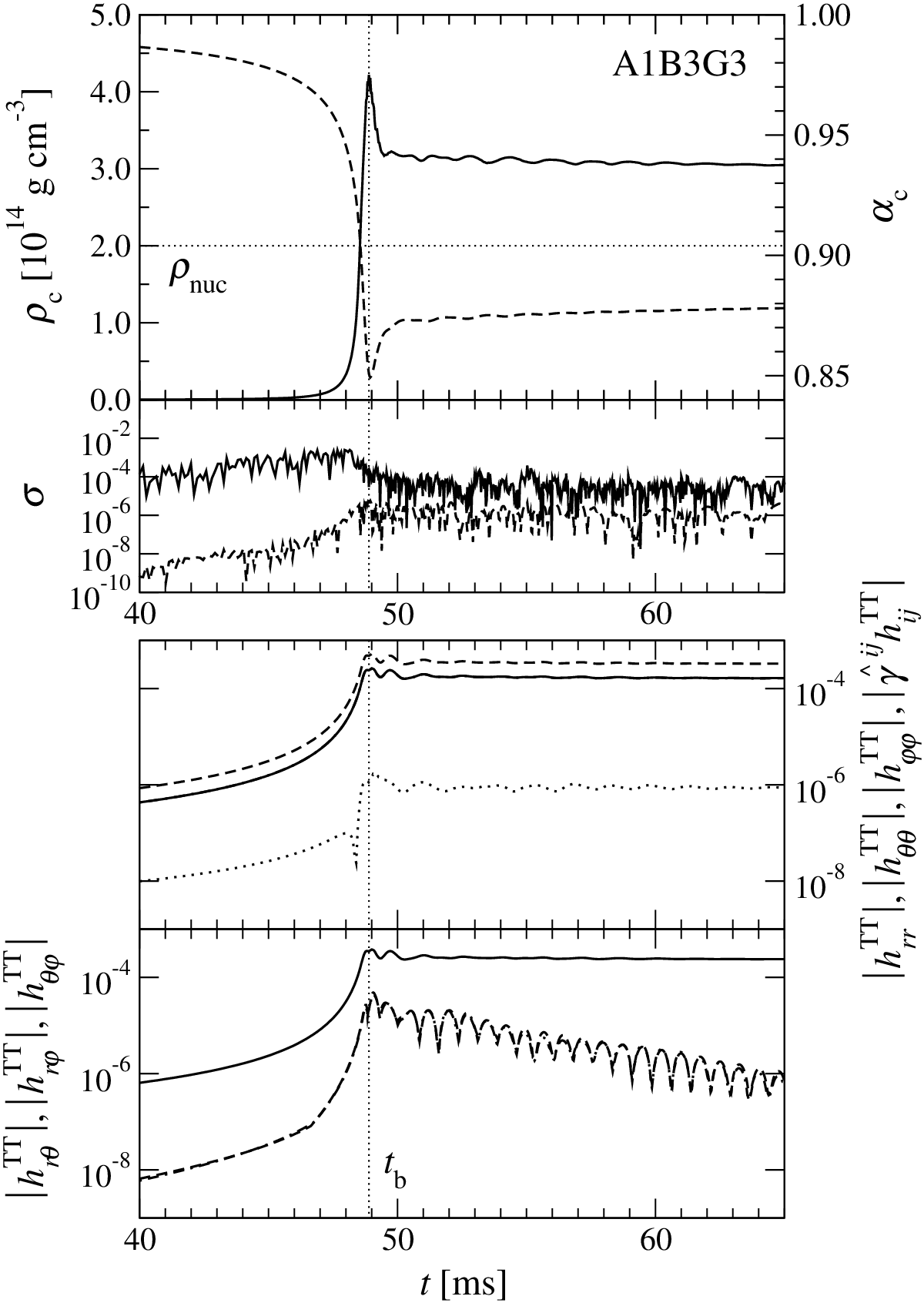}
                        \qquad
                        \includegraphics{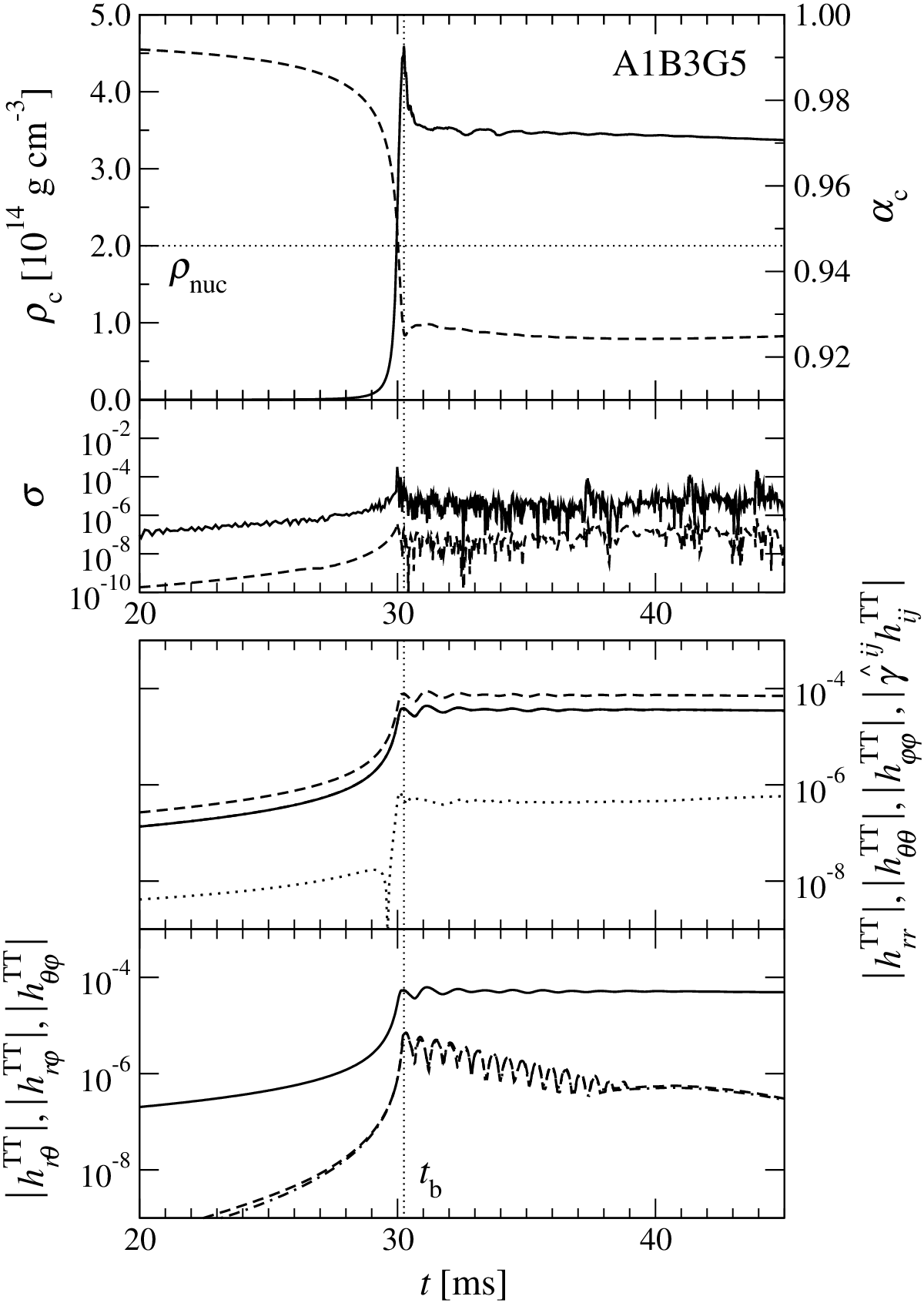}}
  \caption{Time evolution of hydrodynamic and metric quantities for
    the regular collapse model A1B3G3 (left) and the rapid collapse
    model A1B3G5 (right). The top panels show the central
    density $ \rho_\mathrm{c} $ (solid line) and lapse function
    $ \alpha_\mathrm{c} $ (dashed line). Both the CFC and the CFC+
    results overlap. Nuclear matter density $ \rho_\mathrm{nuc} $ is
    indicated by the horizontal dotted line. The second panels from the
    top display the relative difference $ \sigma $ of
    $ \rho_\mathrm{c} $ (solid line) and $ \alpha_\mathrm{c} $ (dashed
    line) between the simulation using CFC and CFC+. In the third
    panels the CFC+ evolution of the absolute value of
    $ h^\mathrm{TT}_{rr} $ (solid line),
    $ h^\mathrm{TT}_{\theta \theta}$ (dashed line),
    $ h^\mathrm{TT}_{\varphi \varphi} $ (dashed-dotted line), as well
    as the trace of $ h^\mathrm{TT}_{ij} $ (dotted line) is
    shown, all measured at the center of the star. Note that
    $ h^\mathrm{TT}_{rr} $ and $ h^\mathrm{TT}_{\varphi \varphi} $
    cannot be discerned, as they practically overlap. The bottom panels
    depict the evolution of the maximum absolute value of
    $ h^\mathrm{TT}_{r\theta} $ (solid line),
    $ h^\mathrm{TT}_{r \varphi} $ (dashed line), and
    $ h^\mathrm{TT}_{\theta \varphi} $ (dashed-dotted line),
    respectively. Again the latter two quantities closely
    coincide. The vertical dotted line in all panels marks the time of
    bounce $ t_\mathrm{b} $.}
  \label{fig:collapse_1_2}
\end{figure*}

\begin{figure*}
  \resizebox{\hsize}{!}{\includegraphics{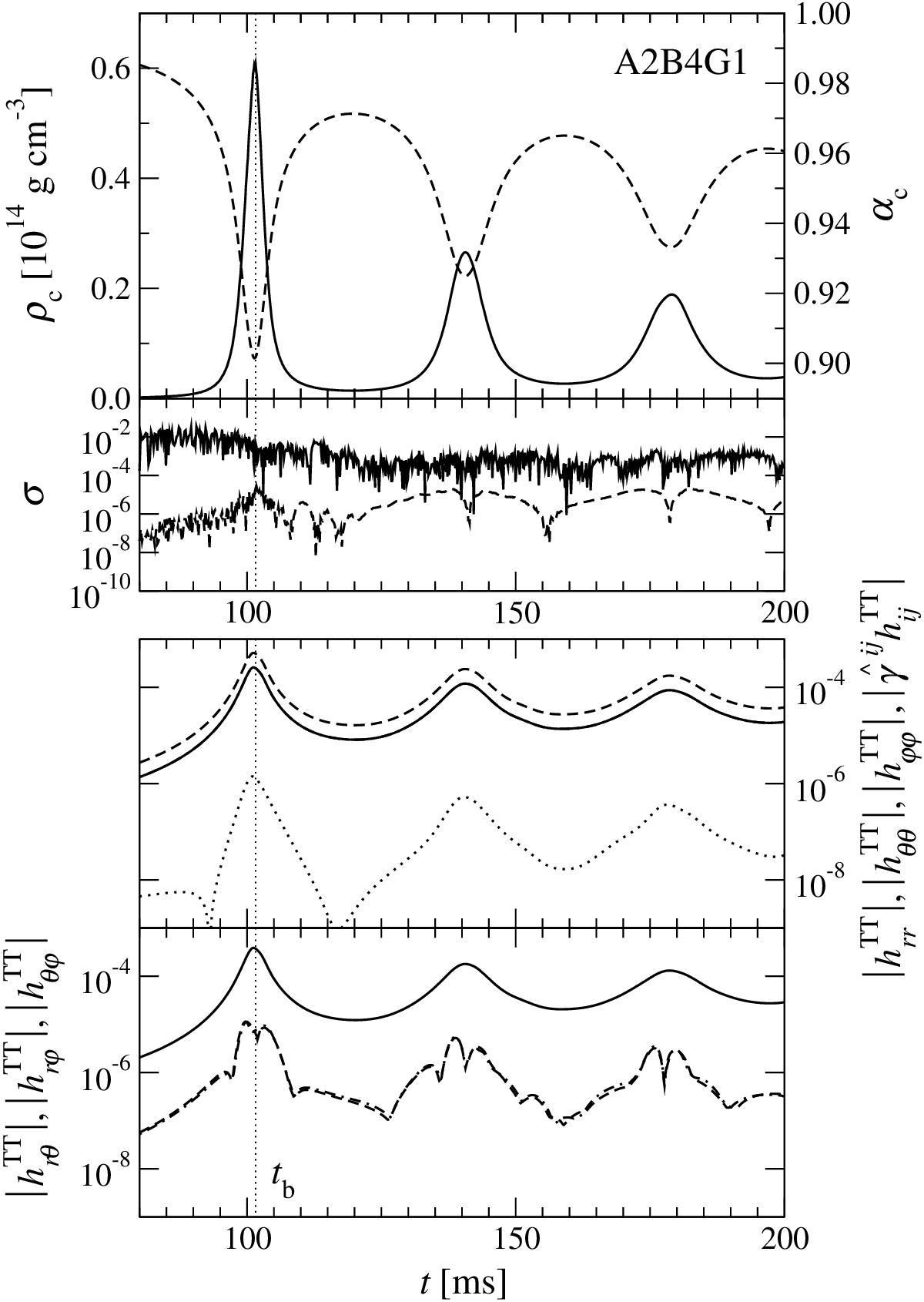}
                        \qquad
                        \includegraphics{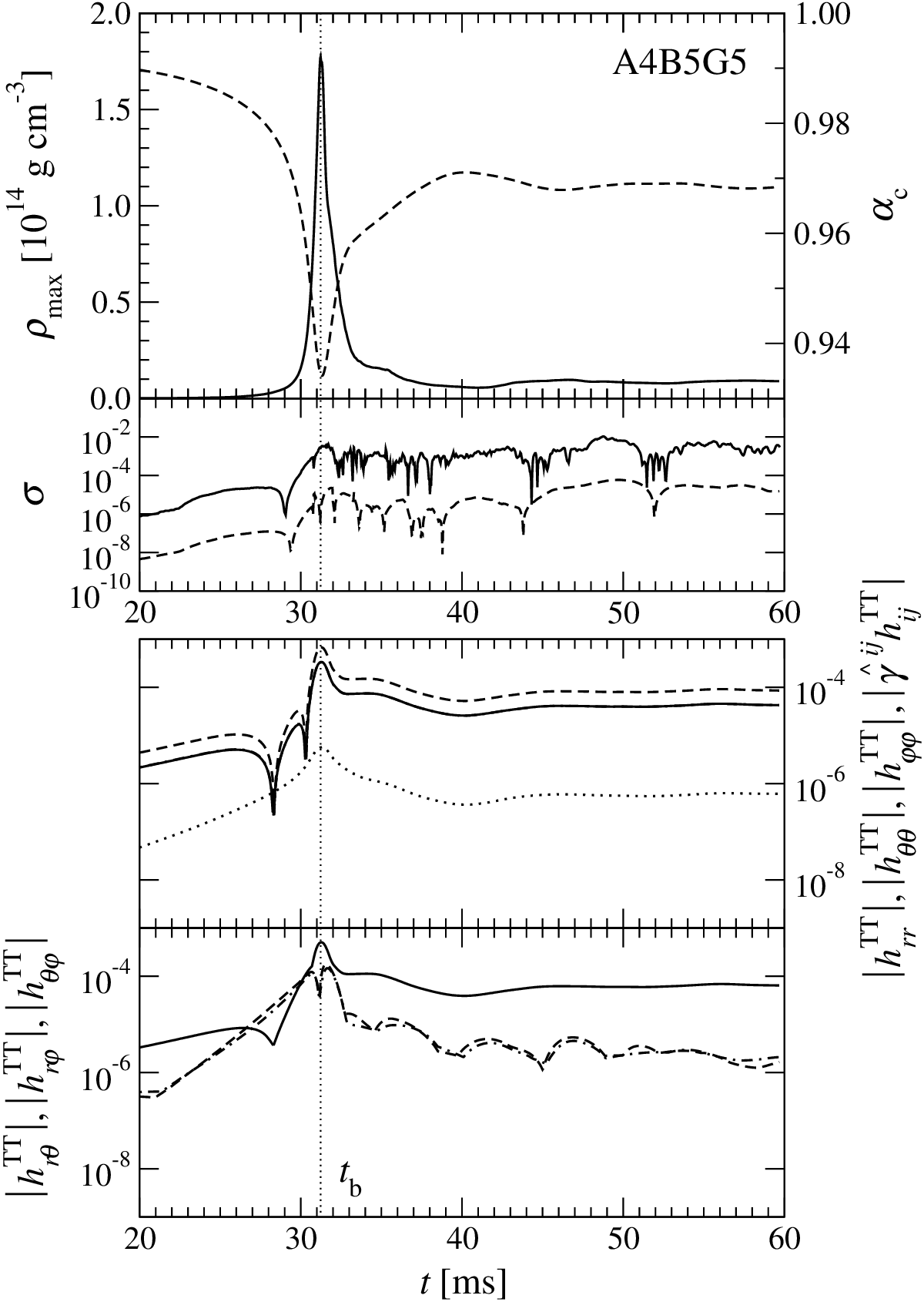}}
  \caption{Same as Fig.~\ref{fig:collapse_1_2} for the multiple
    bounce collapse model A2B4G1 (left) and the extremely rotating
    collapse model A4B5G5 (right). Both models centrifugally bounce
    before reaching $ \rho_\mathrm{nuc} $. Note that for model A4B5G5
    we plot the maximum density $ \rho_\mathrm{max} $ instead of
    $ \rho_\mathrm{c} $, as this model has a toroidal density
    configuration.}
  \label{fig:collapse_3_4}
\end{figure*}

The corresponding results for the case of coupled evolutions of the
spacetime metric and the hydrodynamics are shown in
Tables~\ref{tab:oscillating_RNS_coupled_F}
and~\ref{tab:oscillating_RNS_coupled_H1} for the fundamental mode and
the first harmonic, respectively. In these simulations, for the sake of
computational efficiency and without affecting the dynamics, the
metric is calculated every 100th hydrodynamic time steps and
extrapolated in between, as explained in
\citet{dimmelmeier_02_a}. As in the Cowling simulations, all models
are evolved for $ 30 \mathrm{\ ms} $. Even for the more rapidly
rotating models, no differences in the frequencies from the CFC and
CFC+ simulations can be found. This result can again be explained by
the smallness of the components of $ h^\mathrm{TT}_{ij} $, which does
not modify the dynamics considerably. Furthermore, the results agree
to high precision with the GR results of \citet{font_02_a} within the
limits set by the temporal and spatial resolution.

\begin{table}
  \centering
  \caption{Same as Table~\ref{tab:oscillating_RNS_coupled_F} but
    for the frequency $ f_\mathrm{H1} $ of the first harmonic mode. In
    the model RNS5 this harmonic was not sufficiently excited by the 
    perturbation chosen for a clear identification of its frequency.}
  \begin{tabular}{cccccc}
    \hline \hline \\ [-1 em]
    Model & $ r_\mathrm{p} / r_\mathrm{e} $ &
    $ f_\mathrm{H1}^\mathrm{CFC} $ & $ f_\mathrm{H1}^\mathrm{CFC+} $ &
    $ f_\mathrm{H1}^\mathrm{GR} $ & Rel.\ diff. \\
    & & [kHz] & [kHz] & [kHz] & [\%] \\
    \hline
    RNS0 & 1.00 & 3.97 & 3.97 & 3.958 & 0.3 \\
    RNS1 & 0.95 & 3.87 & 3.87 & 3.852 & 0.5 \\
    RNS2 & 0.85 & 3.95 & 3.95 & 3.867 & 2.0 \\
    RNS3 & 0.75 & 3.98 & 3.98 & 4.031 & 1.3 \\
    RNS4 & 0.70 & 4.02 & 4.02 & 3.887 & 2.0 \\
    RNS5 & 0.65 & ---  & ---  & 3.717 & --- \\
    \hline\hline
  \end{tabular}
  \label{tab:oscillating_RNS_coupled_H1}
\end{table}

We emphasize that, for accurately extracting the oscillation mode
frequencies, the code has to maintain the initial equilibrium
configuration in a hydrodynamical evolution for many rotation
periods (usually several tens of periods). Independently of the
approximation assumed for the metric (either CFC or CFC+) and the
(small) gauge mismatch, we have tested that our code is able to
perform that task successfully.


\section{Rotational core collapse}
\label{sec:core_collapse}

We now present results of simulations of rotational core collapse to
neutron stars. The core collapse models we have selected (see
Table~\ref{tab:CC_models}) are representative for the different types
of collapse dynamics and gravitational radiation waveforms observed in
the CFC simulations of \citet{dimmelmeier_02_a}: A1B3G5 as type I
(regular collapse), A2B4G5 as type II (multiple bounce collapse),
A1B3G5 as type III (rapid collapse), and A4B5G5 as a case with extreme
rotation, i.e.\ a strongly and highly differentially rotating core
with an initial torus-like structure which is strongly enhanced during
collapse. We use a numerical grid consisting of $ n_{\theta} = 30 $
equally spaced angular cells and $ n_r = 300 $ logarithmically spaced
radial cells covering the star. In order to calculate
$ h^\mathrm{TT}_{ij} $ and extract waveforms, an extra grid of 300
radial cells extendig out to $ 10^{11} \mathrm{\ km} $ needs to be
added (600 cells for model A2B4G1).

\subsection{Collapse dynamics}
\label{subsec:dynamics}

In Figs.~\ref{fig:collapse_1_2} and~\ref{fig:collapse_3_4} we compare the
time evolution of selected matter and metric quantities for all four
collapse models considered using both the CFC and the CFC+
approximation. We show the time evolution of the central density
$ \rho_\mathrm{c} $, which is a representative quantity of the
hydrodynamic evolution. We present $ \rho_\mathrm{c} $ for all
models except model A4B5G5 for which the time evolution of the
maximum density $ \rho_\mathrm{max} $ is used instead. In this model
the density maximum is not attained at the center due to the strong
differential rotation. The evolution of the central density in models
A1B3G3 and A1B3G5 (see Fig.~\ref{fig:collapse_1_2}) shows a
distinctive rise during collapse until $ \rho_\mathrm{c} $ reaches
its maximum at the time of bounce
$ t_\mathrm{b} $ (at $ t_\mathrm{b} \sim 49 \mathrm{\ ms} $ and
$ t_\mathrm{b} \sim 30 \mathrm{\ ms} $, respectively). Later on the
density oscillates around the new equilibrium value of the compact
remnant (which can be identified with the new-born proto-neutron
star). These oscillations are highly damped due to the presence of an
extended stellar envelope surrounding the proto-neutron star. Note
that in models A2B4G1 and A4B5G5 (see Fig.~\ref{fig:collapse_3_4}) the
collapse is stopped by rotation before nuclear matter density is
reached, as strong centrifugal forces build up during the
collapse. Hence, the evolution of model A2B4G1 is characterized by
the presence of consecutive multiple bounces, while the centrifugal
hang-up in model A4B5G5 causes a single bounce below nuclear matter
density, leaving a low density proto-neutron star behind.

The top panels of Figs.~\ref{fig:collapse_1_2} and~\ref{fig:collapse_3_4}
also show the central lapse function $ \alpha_\mathrm{c} $ (dashed
line; labels on the right vertical axis). In the CFC+ approach the new
$ h^\mathrm{TT}_{ij} $ terms directly couple to the metric
equation~(\ref{eq:alpha_phi}) for $ \alpha $, while they couple
indirectly to the metric equations for $ \phi $ and $ \beta^i $
through $ \alpha $ itself. The evolution of the lapse closely mirrors
that of the density, decreasing while the density increases, i.e.\
while the star contracts, and vice-versa.

For the four collapse models considered here, there is no direct visual 
evidence of discrepancies between the CFC and CFC+ results. The
corresponding curves for CFC and CFC+ in the top panels of
Figs.~\ref{fig:collapse_1_2} and~\ref{fig:collapse_3_4} coincide
perfectly in the case of both the lapse function and the
density. Therefore, as no appreciable differences are visible, we plot
in the second panels (from the top) of these figures the relative
differences $ \sigma $ of these two quantities between CFC and
CFC+. Maximum differences of the order of $ \sim 1\% $ are found
in the density evolution (solid line) for the strongly differentially
rotating models A2B4G1 and A4B5G5 (see Fig.~\ref{fig:collapse_3_4}). 
In the two other models the differences are two orders of magnitude
smaller. The lapse function (dashed line) shows even smaller
differences between CFC and CFC+, the maximum values of $ \sigma $
being smaller than $ 0.1\% $ even for the rapidly rotating models.

\begin{table*}
  \centering
  \caption{Summary of various quantities that characterize
    the different core collapse models. The table shows
    the time of bounce $ t_\mathrm{b} $, the maximum density at bounce
    $ \rho_\mathrm{b} $, the maximum density reached after the bounce
    $ \rho_\mathrm{f} $, the gravitational wave amplitude at bounce
    $ \left| A^\mathrm{E2}_{20} \right|_\mathrm{b} $ as measured using the
    quadrupole formula, the dominant frequency of oscillations of the
    proto-neutron stars $ f_\mathrm{max} $, the radius of the star after
    bounce and ringdown $ r_\mathrm{e\,f} $ (defined as the radial location
    along the equator where the density first falls below 10\% of the maximum
    density), the size of the near zone
    $ \lambda / (2 \pi) $, and the distance $ r_\mathrm{ex} $ at which
    gravitational waves are extracted from $ h^\mathrm{TT}_{ij} $.}
  \label{tab:CC_results}
  \begin{tabular}{ccccccccc}
    \hline \hline \\ [-1 em]
    Model & $ t_\mathrm{b} $ & $ \rho_\mathrm{b} $ & $ \rho_\mathrm{f} $ &
    $ \left| A^\mathrm{E2}_{20} \right|_\mathrm{b} $ &
    $ f_\mathrm{max} $ & $ r_\mathrm{e\,f} $ &
    $ \lambda / (2 \pi) $ & $ r_\mathrm{ex} $ \\ 
    & [ms] & [$ 10^{14} \mathrm{\ g\ cm}^{-3} $] &
    [$ 10^{14} \mathrm{\ g\ cm}^{-3} $] & [cm] & [Hz] & [km] &
    [km] & [$ 10^3 \mathrm{\ km} $] \\ 
    \hline
    A1B3G3 & \wz 48.89 & 4.23 & 3.22 \wz &    1223 &    674 &    13 & \wz 71 & \wz 2.6 \\
    A1B3G5 & \wz 30.25 & 4.65 & 3.53 \wz & \wz 131 &    890 & \wz 9 & \wz 52 & \wz 2.6 \\
    A2B4G1 &    101.60 & 0.60 & 0.27 \wz & \wz 936 & \wz 54 &    34 &    884 &    40.0 \\
    A4B5G5 & \wz 31.23 & 1.78 & 0.096    &    3757 &    142 &    60 &    334 &    20.0 \\
     \hline\hline  
  \end{tabular}
\end{table*}

The time evolution of the diagonal components of
$ h^\mathrm{TT}_{ij} $ at the center ($ r = 0 $) are also plotted in
Figs.~\ref{fig:collapse_1_2} and~\ref{fig:collapse_3_4} (third panels
from the top), along with the trace of $ h^\mathrm{TT}_{ij}
$. Correspondingly, the maximum absolute values of the off-diagonal
terms of $ h^\mathrm{TT}_{ij} $ are displayed in the panels at the
bottom. As expected, the various components of $ h^\mathrm{TT}_{ij} $
arise and increase when deviations from sphericity occur. The profiles
show that in all collapse simulations $ h^\mathrm{TT}_{ij} $ is quite
small in comparison with the isotropic part which is of order
unity. It can be seen that models with strong gravity but small
asphericities (as A1B3G3) and models with weaker gravity but more
apparent deviations from sphericity (as A2B4G1 or A4B5G5) all reach
values for $ h^\mathrm{TT}_{ij} $ of similar magnitude. Note that the
components $ h^\mathrm{TT}_{r \varphi} $ and
$ h^\mathrm{TT}_{\theta \varphi} $ rapidly decrease after the bounce,
because a quasi-equilibrium configuration is reached in the new-born
proto-neutron star. In all cases considered the trace of
$ h^\mathrm{TT}_{ij} $ is much smaller than the $ h^\mathrm{TT}_{ij} $
components themselves, i.e.\ numerically $ h^\mathrm{TT}_{ij} $ is
traceless to high accuracy, and also remains traceless during the
entire evolution. In addition, we have checked the transverse
character of $ h^\mathrm{TT}_{ij} $, i.e.\
$ \nabla^i h^\mathrm{TT}_{ij} = 0 $. The latter expression is found to
be compatible with zero, as the dimensionless quantity
$ r \, \nabla^i h^\mathrm{TT}_{ij} $ is much smaller than
$ h^\mathrm{TT}_{ij} $.

\begin{figure*}
  \resizebox{6.9in}{!}{\includegraphics{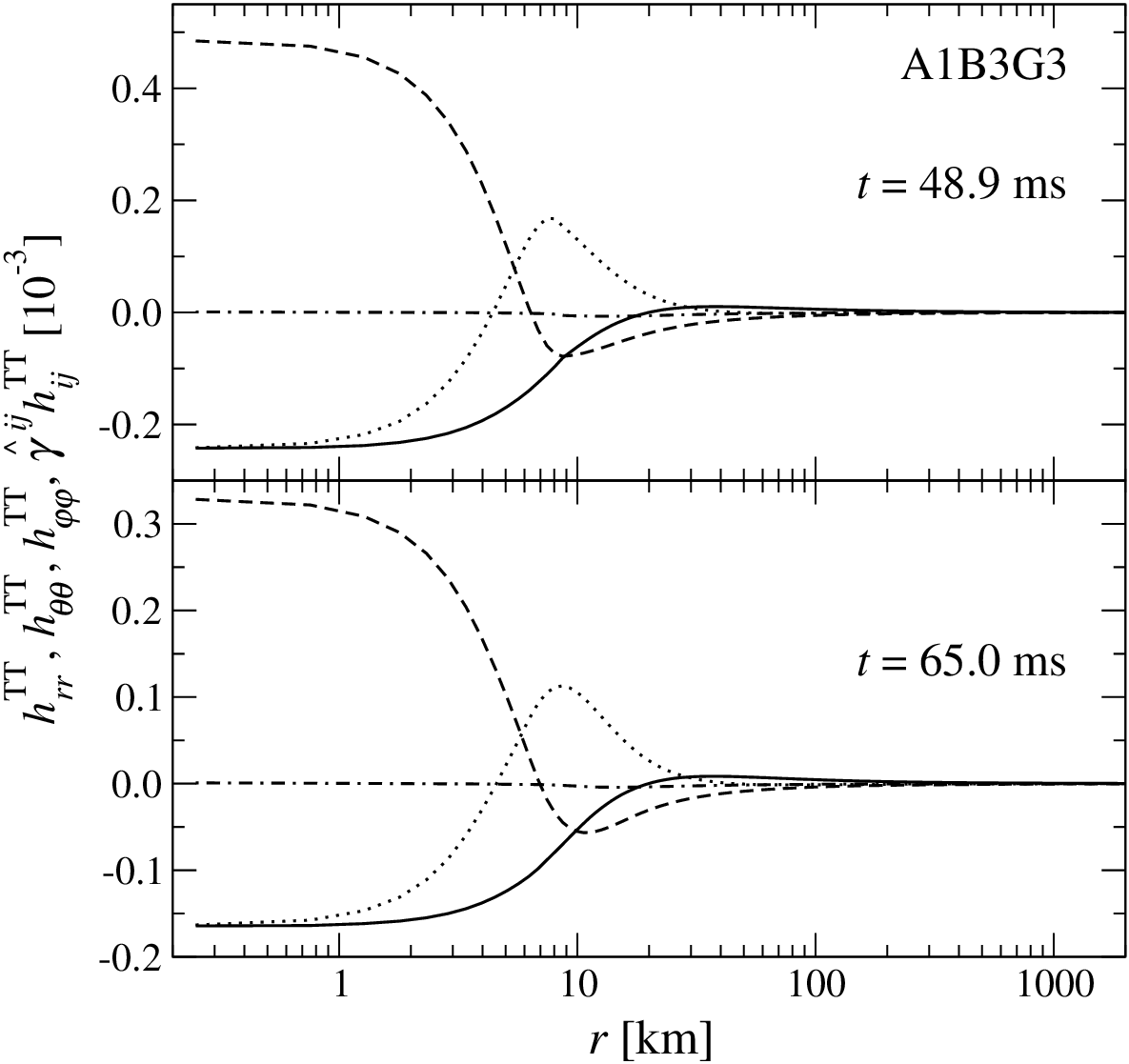}
                        \qquad
                        \includegraphics{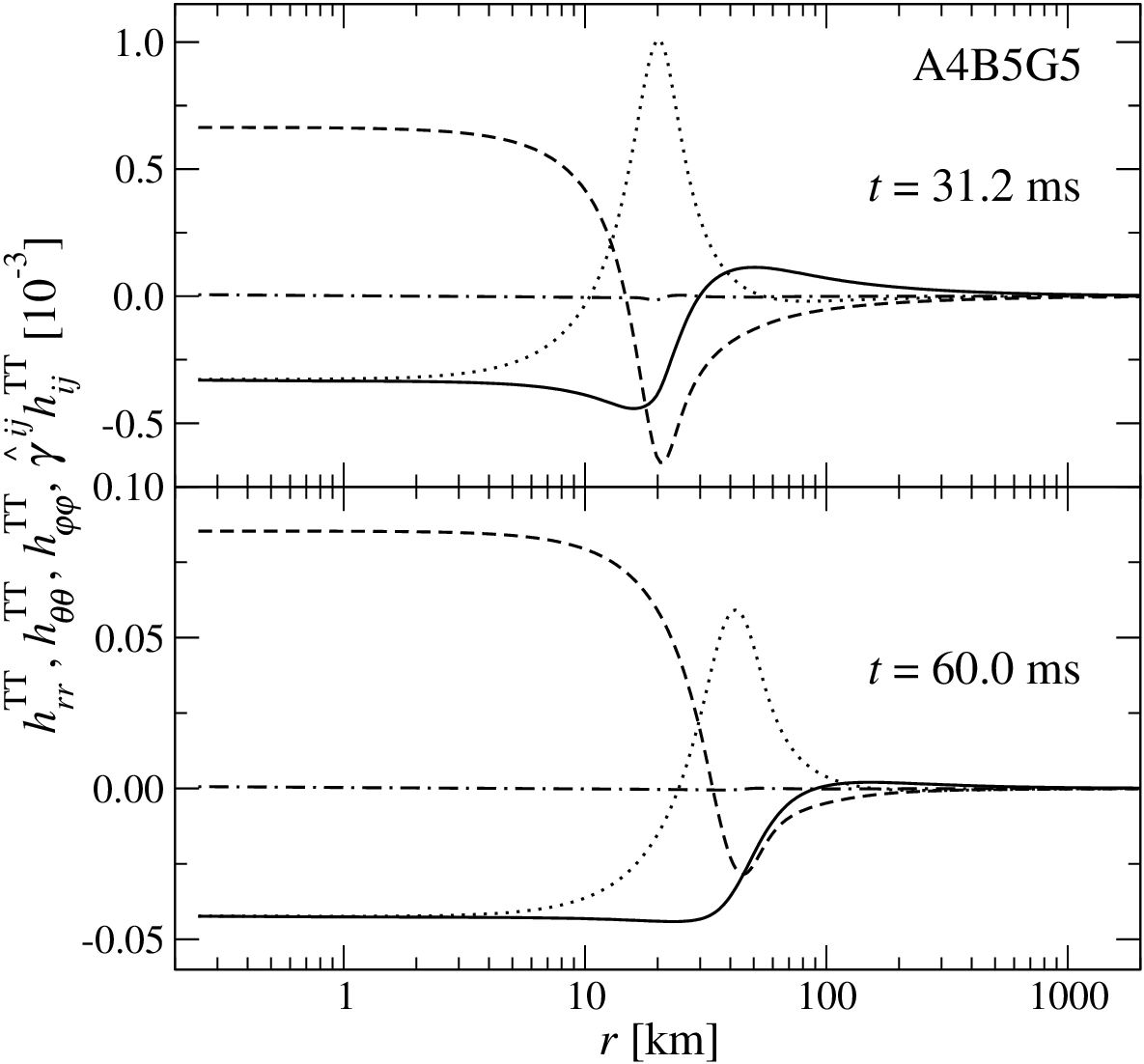}}
  \caption{Radial profiles of $ h^\mathrm{TT}_{ij} $ at the equator
    for model A1B3G3 (left) and model A4B5G5 (right) at the time of
    maximum density (upper panels) and at the final time of the
    simulation (lower panels). The curves plotted
    correspond to $ h^\mathrm{TT}_{rr} $ (solid line),
    $ h^\mathrm{TT}_{\theta\theta} $ (dashed line),
    $ h^\mathrm{TT}_{\varphi \varphi} $ (dotted line), and the trace
    of $ h^\mathrm{TT}_{ij} $ (dashed-dotted line), respectively.}
  \label{fig:htt_profiles}
\end{figure*}

The radial profiles of $ h^\mathrm{TT}_{ij} $ are very similar for all
collapse models we have analyzed except for model A4B5G5 that
collapses off-center, with a torus-like structure. In
Fig.~\ref{fig:htt_profiles} we compare this model with a model in
which the maximum density is reached at the center (A1B3G3). The
profiles are depicted at the instant of maximum density
($ t_\mathrm{b} = 48.9 \mathrm{\ ms} $ for model A1B3G3 and
$ t_\mathrm{b} = 31.2 \mathrm{\ ms} $ for model A4B5G5) and at the
end of the simulation, when the system has reached an equilibrium
state. For the spheroidal model A1B3G3 the maximum values of
$ h^\mathrm{TT}_{ij} $ are reached at the center, and the components
$ h^\mathrm{TT}_{\theta \theta} $ and
$ h^\mathrm{TT}_{\varphi \varphi} $ have local maxima inside the
star. However, in the toroidal model A4B5G5 the maximum values are
off-centered, while the three components exhibit their peak value
inside the torus. Note that the strong deviations from sphericity
generate in model A4B5G5 larger values of $ h^\mathrm{TT}_{ij} $ than
in model A1B3G3 at the time of bounce, but once the torus collapses to
the final oblate star, the values become smaller than for the regular
collapse model.

Table~\ref{tab:CC_results} summarizes the results of all collapse
simulations, including relevant information to calculate the size of
the near zone $ \lambda / (2 \pi) $ needed for the gravitational wave
extraction which we discuss next.

\subsection{Gravitational radiation waveforms}
\label{subsec:waveforms}

Gravitational waves from the collapse simulations discussed in the
preceding section have been calculated for both the CFC and the CFC+
approximation of the field equations, using the quadrupole formula. In
addition, in the case of CFC+ they have also been extracted directly
from $ h^\mathrm{TT}_{ij} $ evaluated in the wave zone. The
radial extension of the near zone $ \lambda / (2 \pi) $ can be
calculated from the approximate size of the source $ r_\mathrm{e} $
and the typical timescale of gravitational wave emission
$ 1 / f_\mathrm{max} $. Results for each collapse model are listed in
Table~\ref{tab:CC_results}, which also gives the distance
$ r_\mathrm{ex} $ at which the waveforms are actually extracted from
$ h^\mathrm{TT}_{ij} $. For all models this distance is much larger
than $ \lambda / (2 \pi) $ (by a factor of about 50), which ensures
that the wave extraction is done in the wave zone far away from the
sources.

\begin{figure*}
  \resizebox{\hsize}{!}{\includegraphics{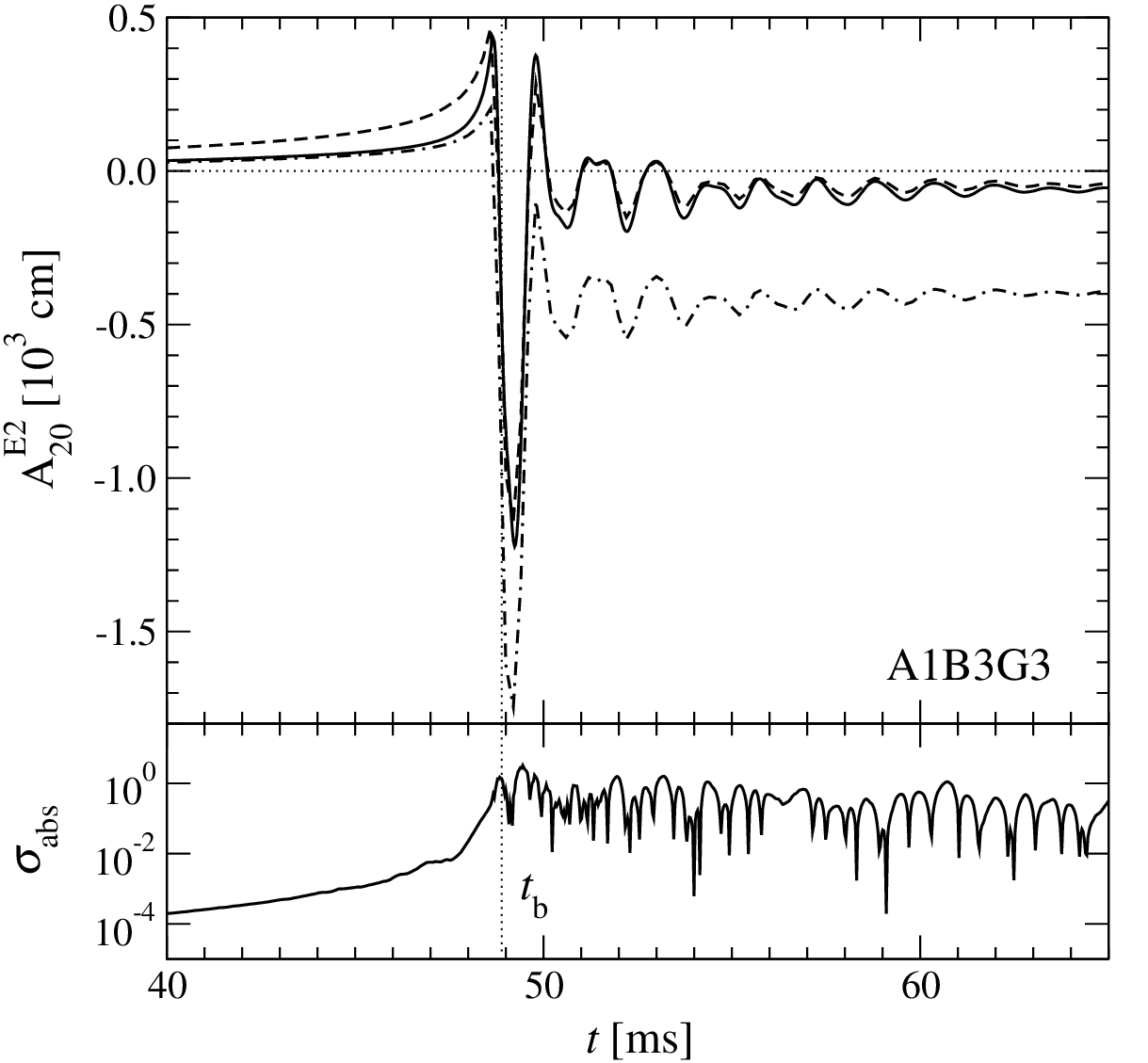}
                        \qquad
                        \includegraphics{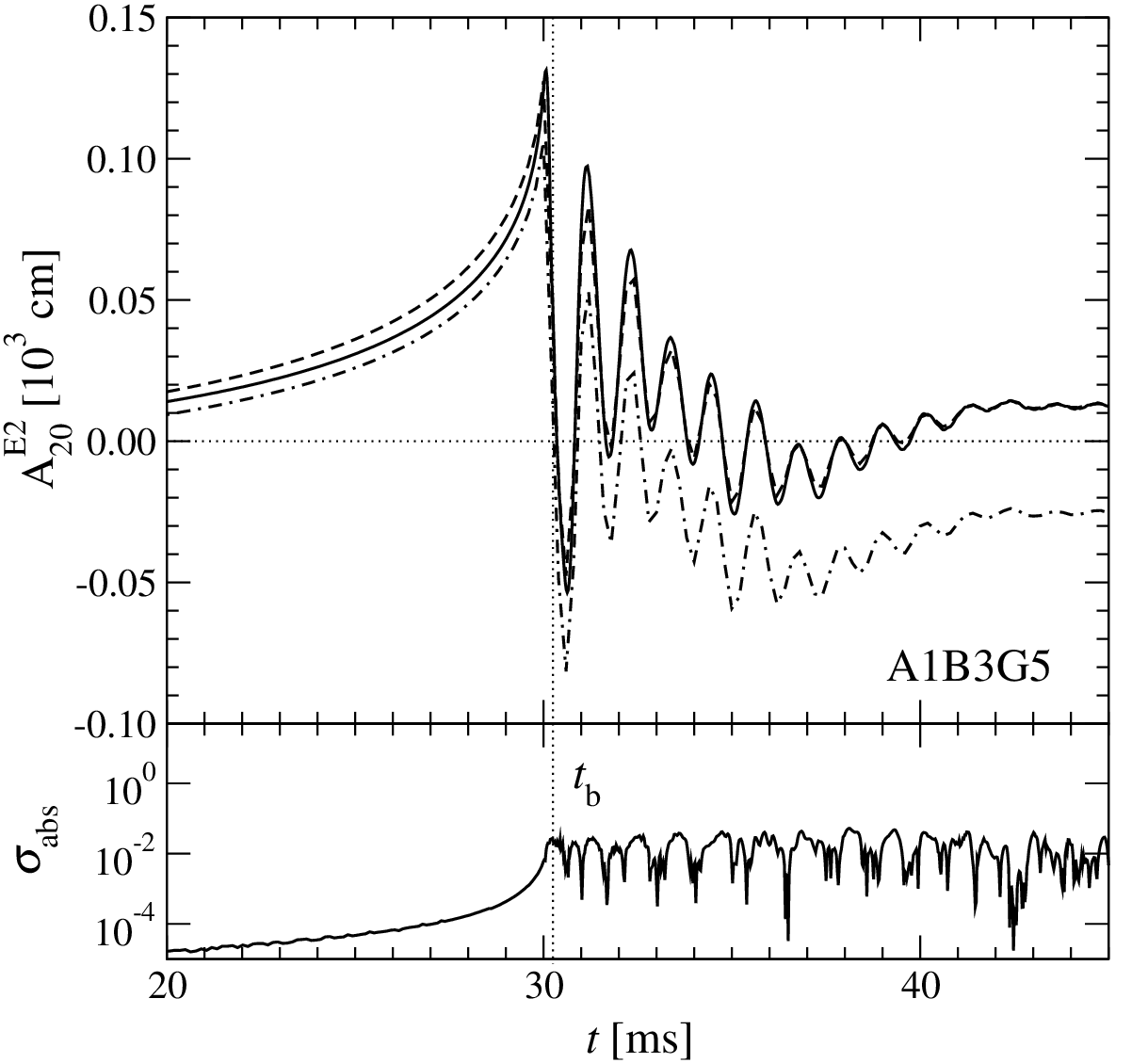}}
  \caption{Gravitational wave amplitude $ A^\mathrm{E2}_{20} $
    computed with the CFC+
    approximation of the spacetime metric for the regular collapse
    model A1B3G3 (left) and the rapid collapse model A1B3G5
    (right). Depicted in the top panels are the waveforms
    extracted using the quadrupole formula (solid line) and extracted
    directly from $ h^\mathrm{TT}_{ij} $ with (dashed line) or without
    (dashed-dotted line) corrections for the offset after core
    bounce. The lower panels show the absolute difference
    $ \sigma_\mathrm{abs} $ of the quadrupole waveforms obtained using
    the CFC+ and CFC approximation  of the spacetime metric. The
    vertical dotted line marks the time of bounce $ t_\mathrm{b} $.}
  \label{fig:GW_1_2}
\end{figure*}

\begin{figure*}
  \resizebox{\hsize}{!}{\includegraphics{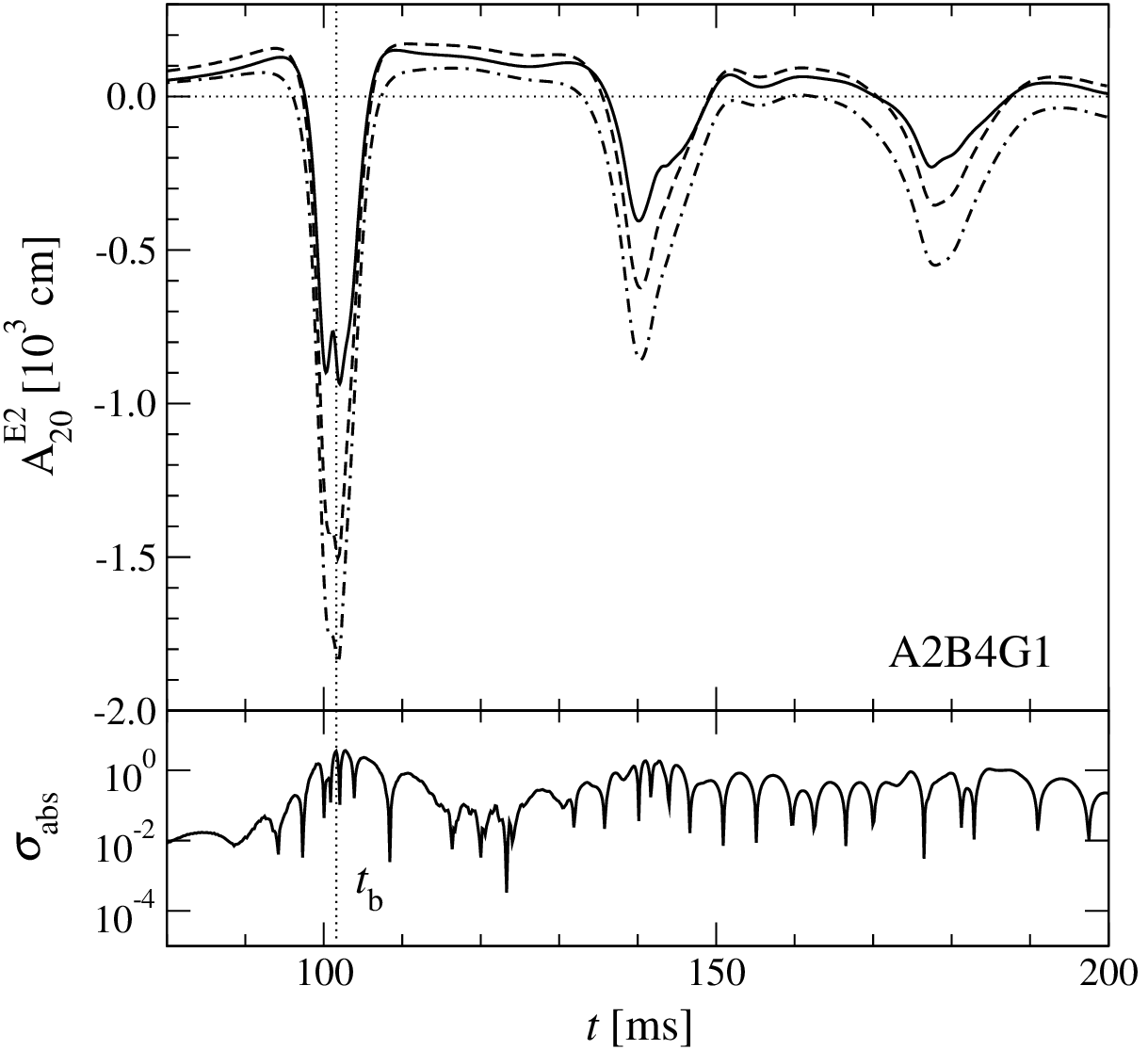}
                        \qquad
                        \includegraphics{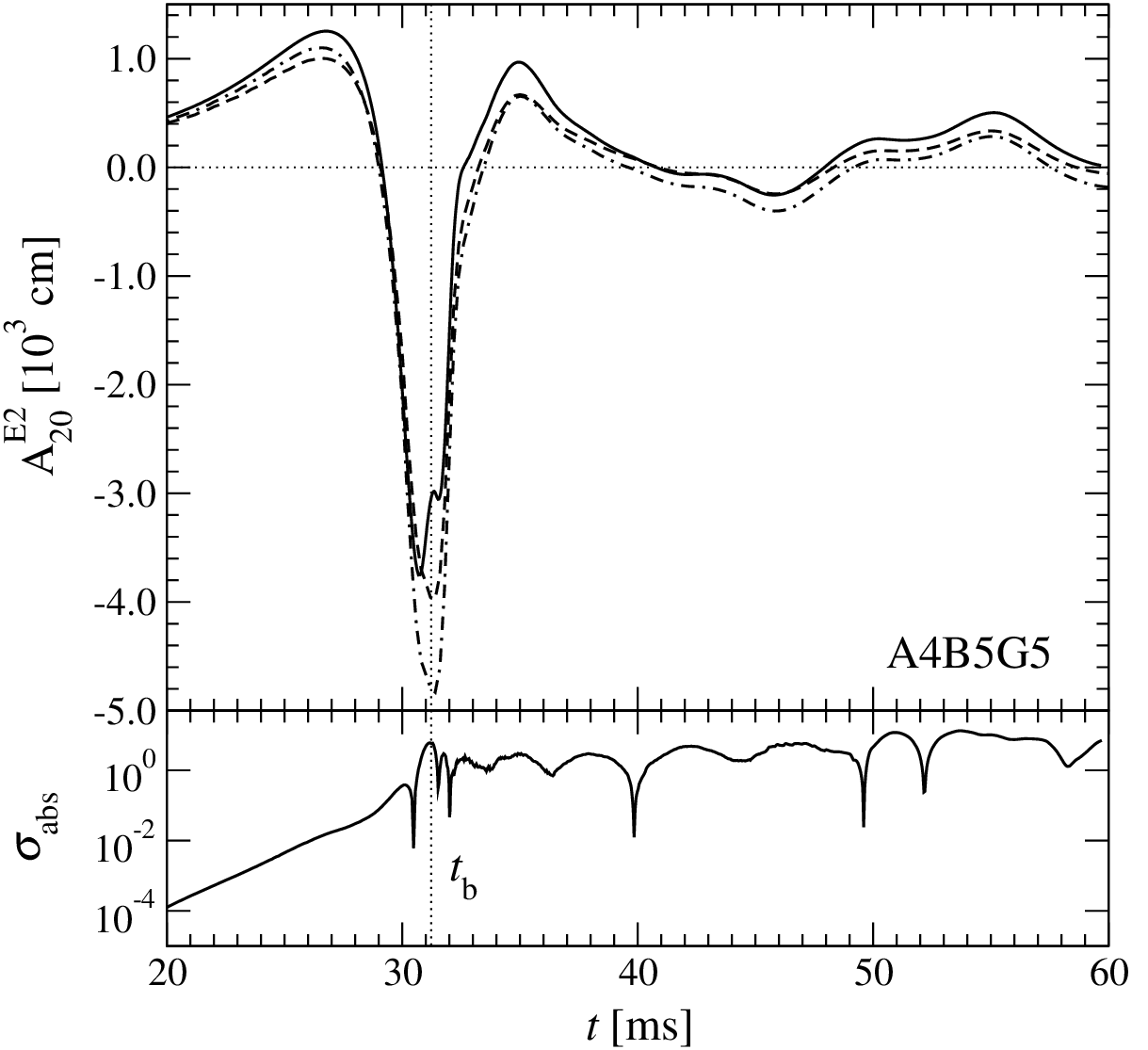}}
  \caption{Same as Fig.~\ref{fig:GW_1_2} for the multiple bounce
    model A2B4G1 (left) and the extremely rotating collapse model
    A4B5G5 (right).}
  \label{fig:GW_3_4}
\end{figure*}

The gravitational waveforms are displayed in the top panels of 
Figs.~\ref{fig:GW_1_2} and~\ref{fig:GW_3_4}, respectively. The waveforms 
extracted using the quadrupole formula (solid lines) are very similar 
for CFC and CFC+ and would not be discernible in the figures. Thus, 
only the CFC+ waveforms are plotted, along with the absolute 
differences $ \sigma_\mathrm{abs} $ with respect to CFC (shown in the 
bottom panels of each figure). For all collapse models considered the
differences are smaller than 0.1\% of the signal maximum. This is
expected from the fact that the quadrupole formula involves an
integral of hydrodynamic quantities, and from the
observation that the modifications in the collapse dynamics
between CFC and CFC+ are not significant, as mentioned in
Sect.~\ref{subsec:dynamics}.

Concerning the waveforms extracted directly from
$ h^\mathrm{TT}_{ij} $ (dashed and dotted lines in
Figs.~\ref{fig:GW_1_2} and~\ref{fig:GW_3_4}), it can be seen that when
directly using Eq.~(\ref{eq:h+2pn}) to calculate the wave amplitude
$ A^\mathrm{E2}_{20} $, the resulting signals are larger at bounce
for all models. After bounce these signals show an offset for the
models with stronger gravity (A1B3G3 and A1B3G5, see
Fig.~\ref{fig:GW_1_2}), where the waveform amplitude should actually
approach zero, because the pulsations of the new-born proto-neutron
star are rapidly damped and the star tends towards an equilibrium
state. If the signals are corrected by means of
Eq.~(\ref{eq:h+2pn_corrected}), the offset disappears and the
gravitational wave amplitude agrees remarkably well with the waveforms
calculated with the quadrupole amplitude. Although the extraction
methods are not really independent, the agreement found between the
two ways of calculating waveforms in the CFC+ approach is a
consistency check for the calculation of the $ h^\mathrm{TT}_{ij} $,
because the asymptotic behavior given by Eq.~(\ref{eq:h+2pn}) can be
assessed numerically this way.

The agreement holds for all cases except for model A2B4G1, where the
amplitudes obtained by the quadrupole formula and from the
$ h^\mathrm{TT}_{ij} $ differ by about 50\% both at the main bounce
and at the subsequent bounces even after the offset correction is
applied. Note that for this particular model the size of the near zone
is very large, $ \lambda / (2 \pi) = 884 \mathrm{\ km} $. Therefore,
an accurate extraction of the waveforms can only be performed at a
radius very far away from the star. We thus set the extraction radius
$ r_\mathrm{ex} = 4 \times 10^4 \mathrm{\ km} $. As a consequence the
extended grid needs to be covered with at least 600 radial zones in
order to avoid too extreme logarithmic cell spacing, which would be
the source of numerical inaccuracies in the fall-off behavior when
solving the elliptic equations with finite difference methods.

\section{Conclusions}
\label{sec:conclusions}

We have presented a new approximation for the Einstein field
equations, which we call CFC+. We have tested its suitability for
simulations of rotating neutron star spacetimes, both for equilibrium
models and for configurations which are formed after gravitational
core collapse. This approach is based on second order post-Newtonian
corrections from conformal flatness, i.e.\ CFC+ represents an
extension of the CFC (or Isenberg--Wilson--Mathews) approximation. The
derivation of the extended system of equations has been presented in
great detail, as well as the boundary conditions to apply when
numerically solving them. All CFC+ field equations have elliptic
character, since at second post-Newtonian order the hyperbolic
character of the Einstein equations disappears. This is a
consequence of the fact that the time derivatives of
$ h^\mathrm{TT}_{ij} $ appear first at the 2.5th post-Newtonian order.

We note in passing that solving elliptic equations ensures the
numerical stability of the solution and avoids the numerical problems
sometimes encountered in long-term evolutions of strongly gravitating
systems when using the $ 3 + 1 $ formulation of general relativity. On
the other hand, the price to pay for using this approximation is that
gravitational radiation reaction on the dynamics, caused by
gravitational waves carrying away energy and angular momentum from the
system, is absent. However, in the case of models where a comparison
of our results to fully general relativistic results is
possible, we have checked that the absence of gravitational
back-reaction does not affect the results significantly. In
scenarios such as the merging of compact binaries (not investigated
here), this effect would indeed be important, but only at late
times. Hence, CFC+ should also be a good approximation to model
phenomena occurring on dynamical timescales such as the final stages
of the plunge and merger.

We have compared the new CFC+ approximation with the CFC approach used
by \citet{dimmelmeier_02_a,dimmelmeier_02_b} in two different
scenarios, oscillating relativistic stars and core collapse to a
neutron star. In the case of pulsating neutron stars, we find that
there are no differences in the calculation of the
quasi-radial normal mode frequencies of those objects, even in the
most extreme situations considered when the star is rotating at the
maximum allowed rate (i.e.\ at the mass-shedding limit). It has been
possible to compare our results directly with fully general
relativistic computations and, again, a very close agreement is
found. Furthermore, our simulations of stellar core collapse to
neutron stars covered well the basic morphology and dynamics of core
collapse types studied by \citet{dimmelmeier_02_b}, including the
extreme case of a core with strong differential rotation and
torus-like structure. Once more, no significant differences between
the two approximations are observed. Therefore, we can
conclude that second post-Newtonian corrections to CFC do not
significantly improve the results when simulating the dynamics of
core collapse to a neutron star as well as when investigating the
evolution of neutron stars in equilibrium.

Regarding the gravitational wave extraction we have not observed any
substantial differences between CFC and CFC+ as well. The comparison
has been carried out using the quadrupole formula, commonly employed
in the literature to extract gravitational waveforms. In addition we
have also calculated the gravitational waves directly from the
$ h^\mathrm{TT}_{ij} $, which permits a straightforward use of
the spacetime metric to study the gravitational wave generation
mechanism from the near zone to the wave zone. Although the waveforms
computed with the latter approach cannot be regarded as an independent
way of calculating gravitational wave signals, it nevertheless
provides a good consistency check of the CFC+ approximation that has
served to validate the numerical scheme we use to calculate
$ h^\mathrm{TT}_{ij} $.

The main conclusion of this work is the assessment of the CFC
approximation as a highly suitable alternative to the full Einstein
equations in axisymmetric scenarios, involving rotating neutron stars
in equilibrium and as end states of core collapse. These findings are
supported by two facts: First, we have demonstated that second
post-Newtonian corrections do not play an important role in neither
the dynamics nor the gravitational radiation waveforms of core
collapse. This suggests that higher order post-Newtonian corrections
will be completely insignificant at least on dynamic
timescales. Secondly, the direct comparison of the CFC approach with
exact fully general relativistic simulations of pulsating neutron stars
yields normal-mode frequencies in excellent agreement. Furthermore,
comparisons of the CFC approach with fully general relativistic
simulations have also recently been reported by \citet{shibata_04_a}
in the context of axisymmetric core collapse simulations. Again, the
differences found in both the dynamics and the waveforms are minute,
which highlights the suitability of CFC (and CFC+) for performing
accurate simulations of those scenarios without the need for solving
the full system of the Einstein equations.

With this investigation we have shown that a numerical code
based on CFC is a very useful tool to investigate core collapse to
neutron stars in general relativity. This approach is suitable
to form the basis of a future core collapse supernova code which can
be gradually extended in various directions to incorporate additional
physics such as realistic equations of state, magnetic fields, and
eventually neutrino transport. In the near future we plan to further
validate the CFC+ equations in other scenarios where higher densities
are present (e.g.\ collapse to a black hole), as well as in the fully
three-dimensional case (namely to investigate the merging of neutron
stars). Such scenarios are in principle beyond the range of
applicability of the CFC approximation, but can possibly still be
handled in a satisfactory way with the new CFC+ approach presented in
this work.

\begin{acknowledgements}
  This work has been supported by the Spanish Ministerio de Ciencia y
  Tecnolog\'{\i}a (grant AYA 2001-3490-C02-01), by the German Science
  Foundation DFG (SFB/Transregio~7
  ``Gravitations\-wellen\-astronomie''), and by the EU Network
  Programme (Research Training Network Contract
  HPRN-CT-2000-00137). All numerical computations were performed with
  the SGI Altix 3700 computer {\tt CERCA} at the Departamento de
  Astronom\'{\i}a y Astrof\'{\i}sica of the University of Valencia.
\end{acknowledgements}

\appendix


\section{Metric equation for the lapse function in CFC+}
\label{app:calc_htt}

The correction to the CFC metric up to second post-Newtonian order may
be obtained in the Hamiltonian framework of ADM within the eponym
gauge (see also \citet{regge_74_a} for an analysis in asymptotically
flat spacetimes and \citet{holm_85_a} for a generalization to isolated
perfect fluids). The original canonical variables are chosen to be the
3-metric $ \gamma_{ij} $ and its conjugate momentum
$ c^3 \pi^{ij} / (16 \pi G) $ with
$ \pi^{ij} = - \sqrt{\bar{\gamma}} (K^{ij} - K \gamma^{ij}) $, but once the
coordinate system has been fixed, four of the six remaining field
degrees of freedom are eliminated by imposing the Hamiltonian
constraint~(\ref{eq:adm_metric_equation_3}) and the three momentum
constraints~(\ref{eq:adm_metric_equation_4}). These four field degrees
of freedom correspond to the conformal factor $\phi$ and to the
symmetric trace-free part  $ \pi^{ij}_\mathrm{LL} $ of the tensor
$ 2 \hat{\Delta}^{-1} (\hat{\gamma}^{ik} \hat{\nabla}_k \hat{\nabla}_l
\pi^{jl}) - \hat{\Delta}^{-2}
(\hat{\gamma}^{im} \hat{\gamma}^{jn} \hat{\nabla}_k
\hat{\nabla}_l \hat{\nabla}_m \hat{\nabla}_n \pi^{kl})/2$, respectively.

Only two transverse trace-free (TT) field variables are left, namely
$ h_{ij}^\mathrm{TT} = \gamma_{ij} - \phi^4 \hat{\gamma}_{ij} $
on the one hand, and
$ \pi^{ij}_\mathrm{TT} = \pi^{ij} - \pi^{ij}_\mathrm{LL} $ on the
other hand. By construction, we have:
\begin{align}
  \hat{\gamma}^{kl} \hat{\nabla}_k h_{il}^\mathrm{TT} & = 0,
  & \hat{\gamma}^{ij} h_{ij}^\mathrm{TT} = 0,
  \\
  \hat{\nabla}_j \pi^{ij}_\mathrm{TT} & = 0,
  & \hat{\gamma}_{ij} \pi^{ij}_\mathrm{TT} = 0.
\end{align}%
The reduced Hamiltonian $ H $ is obtained by substituting $ \phi $
with $ \phi [h_{pq}^\mathrm{TT}, \pi^{pq}_\mathrm{TT}] $ and
$ \pi^{ij} $ with
$ \pi^{ij} [h_{pq}^\mathrm{TT}, \pi^{pq}_\mathrm{TT}] $ in the
Hamiltonian of general relativity for asymptotically flat
spacetimes. The contributions of the super-Hamiltonian and
super-momentum densities vanish, so that
$ H [\mathrm{matter\ variables}, h_{pq}^\mathrm{TT}, \pi^{pq}_\mathrm{TT}] $
is given by the surface integral defining the ADM mass ``on a shell'',
\begin{align}
  H & = \frac{c^4}{16\pi G} \int \mathrm{d}^2\hat{S}^i \,
  \sqrt{\hat{\gamma}} \gamma^{jk}
  \left( \hat{\nabla}_j \gamma_{ik} - \hat{\nabla}_i \gamma_{jk} \right)
  \nonumber \\
  & = - \frac{c^4}{2\pi G} \int \mathrm{d}^3\mb{x} \, \sqrt{\hat{\gamma}}
  \hat{\Delta} \phi,
  \label{eq:H_reduced}
\end{align}%
where the surface element $ d\hat{S}^i \, \sqrt{\hat{\gamma}} $ refers
to the flat metric. The reduced Hamiltonian~(\ref{eq:H_reduced})
contains the full dynamical information about the system. In
particular, as shown by \citet{regge_74_a}, the field evolution is
governed by the equations of motion
\begin{align}
  \partial_t h_{ij}^\mathrm{TT} & =
  \frac{16\pi G}{c^3} \hat{\gamma}_{ij}^{\mathrm{TT}kl}
  \frac{\delta H}{\delta \pi^{kl}_\mathrm{TT}},
\label{eq:h_dot}
  \\
  \partial_t \pi^{ij}_\mathrm{TT} & = -\frac{16\pi G}{c^3} 
  \hat{\gamma}^{ij}_{\mathrm{TT}kl}
  \frac{\delta H}{\delta h_{kl}^\mathrm{TT}},
\label{eq:pi_dot}
\end{align}%
with
\begin{align}
  \hat{\gamma}_{ij}^{\mathrm{TT}kl} & =
  \frac{1}{2} \left( \delta^k_i - \hat{\gamma}^{kp}
  \hat{\Delta}^{-1} \hat{\nabla}_i \hat{\nabla}_p \right)
  \left( \delta^l_j - \hat{\gamma}^{lq}
  \hat{\Delta}^{-1}\hat{\nabla}_j \hat{\nabla}_q \right)
  \nonumber \\ 
  & + \frac{1}{2} \left( \delta^k_j -
  \hat{\gamma}^{kp}\hat{\Delta}^{-1}
  \hat{\nabla}_j \hat{\nabla}_p \right)
  \left( \delta^l_i - \hat{\gamma}^{lq}
  \hat{\Delta}^{-1} \hat{\nabla}_i \hat{\nabla}_q \right)
  \nonumber \\ 
  & - \frac{1}{2} \left( \hat{\gamma}_{ij} -
  \hat{\Delta}^{-1} \hat{\nabla}_i \hat{\nabla}_j \right)
  \left( \hat{\gamma}^{kl} - \hat{\gamma}^{kp}
  \hat{\gamma}^{lq} \hat{\Delta}^{-1} \hat{\nabla}_p
  \hat{\nabla}_q \right),
  \label{eq:tt_projector}
\end{align}%
and a similar formula for $ \hat{\gamma}^{ij}_{\mathrm{TT}kl} $. The
role of these operators is to ensure the transverse trace-free
projection of the Fr{\'e}chet derivative 
$ \delta H / \delta \pi^{kl}_\mathrm{TT} $ and
$ \delta H / \delta h_{kl}^\mathrm{TT} $, respectively. 
The calculation of $ H $ in terms of $ D^* $, $ S^*_i $, $ P $, as well as the
field variables $ h_{ij}^\mathrm{TT} $ and $ \pi^{ij}_\mathrm{TT} $, can be
done essentially by eliminating $ \phi $ in Eq.~(\ref{eq:H_reduced})
with the help of the Hamiltonian constraint
(\ref{eq:adm_metric_equation_3}). This is achievable in perturbative
treatments such as the post-Minkowskian or the post-Newtonian ones,
consisting of the formal expansion of all quantities at play in powers
of the gravitational constant $ G $ or of the inverse of the square of
the speed of light $ 1 / c^2 $. 

In the course of eliminating $ \phi $, we use
Eq.~(\ref{eq:adm_metric_equation_3}) in a more explicit form.
For this we first express the 3-curvature $ R $ as a function of
$ \phi $ and $ h_{ij}^\mathrm{TT} $. By making extensive use of the
relations $ \gamma_{ij} \hat{\nabla}_l \gamma^{jk} = - \gamma^{jk}
\hat{\nabla}_l \gamma_{ij} $ and
$ \hat{\nabla}_k \bar{\gamma} = (\bar{\gamma} \gamma^{ij})
\hat{\nabla}_k \gamma_{ij} $, the combination $ \bar{\gamma}^3 R $
can be written as (cf.~\cite{schwinger_63_a})
\begin{align}
  \bar{\gamma}^3 R & =
  - \bar{\gamma}^2 \hat{\nabla}_i \hat{\nabla}_j
  \left( \bar{\gamma} \gamma^{ij} \right)
  \nonumber \\
  & + \frac{1}{2} \hat{\nabla}_i \bar{\gamma}^2 \hat{\nabla}_j
  \left( \bar{\gamma} \gamma^{ij} \right) -
  \frac{1}{4} \left( \bar{\gamma} \gamma^{ij} \right)
  \hat{\nabla}_i \bar{\gamma} \hat{\nabla}_j \bar{\gamma}
  \nonumber \\
  & - \frac{1}{2} \bar{\gamma} \left(\bar{\gamma} \gamma^{jl} \right)
  \hat{\nabla}_j \left( \bar{\gamma} \gamma^{ik} \right)
  \hat{\nabla}_i \gamma_{kl}
  \nonumber \\
  & + \frac{1}{2} \left( \bar{\gamma} \gamma^{ik} \right)
  \left( \bar{\gamma} \gamma^{jl} \right)
  \hat{\nabla}_j \bar{\gamma} \hat{\nabla}_i \gamma_{kl}
  \nonumber \\
  & + \frac{1}{4} \bar{\gamma} \left( \bar{\gamma} \gamma^{ik} \right)
  \hat{\nabla}_k \left( \bar{\gamma} \gamma^{jl} \right)
  \hat{\nabla}_i \gamma_{jl}
  \nonumber \\
  & - \frac{1}{4} \left( \bar{\gamma} \gamma^{jl} \right)
  \left( \bar{\gamma} \gamma^{ik} \right)
  \hat{\nabla}_k \bar{\gamma} \hat{\nabla}_i \gamma_{jl}.
  \label{eq:g3R}
\end{align}%
By definition, the determinant $ \bar{\gamma} $ it is equal
to the antisymmetric sum of products $ 3! \, \hat{\gamma}^{p [1}
\hat{\gamma}^{2\underline{q}} \hat{\gamma}^{3] r} \gamma_{p1}
\gamma_{q2} \gamma_{r3} = \hat{\gamma}^{p [i}
\hat{\gamma}^{j\underline{q}} \hat{\gamma}^{k] r} \gamma_{pi}
\gamma_{qj} \gamma_{rk} $ on a Cartesian grid, the square brackets denoting
antisymmetrization of non-underlined indices. Its explicit expression
is given, e.g., by \cite{schaefer_85_a}. Similarly, the $ ij $-component
of the comatrix $ \bar{\gamma} \gamma^{ij} $ is known to be
$ \varepsilon^{ikl} (\varepsilon_{rmn} \hat{\gamma}^{jr}
\hat{\gamma}^{mp} \hat{\gamma}^{nq} \gamma_{pk} \gamma_{ql}) / 2 $,
where $ \varepsilon^{ikl} $ is the permutation operator.
From the identity $ \varepsilon^{rkl} \varepsilon_{jmn} = 3! \,
\delta^{[r}_j \delta^k_m \delta^{l]}_n $ it is straightforward to obtain 
$ \hat{\gamma}_{ik} \hat{\gamma}_{jl} \bar{\gamma} \hat{\gamma}^{kl} $
as a function of $ \phi $ and $ h_{ij}^\mathrm{TT} $. After inserting
the relations for $ \bar{\gamma} $ and $ \bar{\gamma} \gamma^{ij} $ in
the right-hand side of Eq.~(\ref{eq:g3R}), it
is expanded in powers of
$ h_{ij}^\mathrm{TT} $ and truncated consistently at the
post-Newtonian level of $ h_{ij}^\mathrm{TT} h_{kl}^\mathrm{TT} $,
denoted as $ \mathcal{O}(h^2) $. At this point we have an expression for
$\bar{\gamma} R$ as a function of $\hat{\Delta}\phi$.

In addition to $ R $, there is a second contribution to the left-hand
side of the Hamiltonian constraint surviving in the absence of matter,
namely $ K_{ij} K^{ij} - K^2 = [\pi^i_j \pi^j_i - \frac{1}{2} (\pi^i_i)^2] / 
\bar{\gamma} $. In the ADM formulation, the momentum $ \pi^{ij} $
decomposes into $ \pi^{ij}_\mathrm{LL} + \pi^{ij}_\mathrm{TT} $. The first
term $ \pi^{ij}_\mathrm{LL} $ is of order $ 1 / c^3 $, being a sum of
derivatives of Poisson inverse operators acting on
$ \hat{\nabla}_j \pi^{ij} = \mathcal{O}(\nabla_j K^{ij}) $, which is
itself $ \mathcal{O}(1 / c^3) $ according to the momentum constraint
equation. The second term, linear in $ h_{ij}^\mathrm{TT} $, is
transverse and trace-free, hence
$ \hat{\gamma}_{ij} \pi^{ij}_\mathrm{TT} = 0 $. Moreover, the ADM
gauge condition implies
$ \hat{\gamma}_{ij} \pi^{ij} = \hat{\gamma}_{ij} \pi^{ij}_\mathrm{LL} = 0 $.

Finally, we consider the matter source term $ E $ in the Hamiltonian
constraint. The corresponding density $ E^* = \sqrt{\bar{\gamma}} E $ 
may be written as
\begin{align}
  \frac{16\pi G}{c^4} E^* &= \frac{16\pi G}{c^2} \Biggl\{\! \left[(D^{*} h)^2
  + \phi^{-4} \hat{\gamma}^{mn}
  \frac{S^*_m S^*_n}{c^2}\right]^{1/2}
  \nonumber \\
  & \qquad \qquad ~~ \times \left(1 -
  \frac{1}{2} \hat{\gamma}^{ik} \hat{\gamma}^{jl}
  h_{kl}^\mathrm{TT} \frac{S^*_i S^*_j}{c^2 D^{*\,2}} \right) -
  \frac{\phi^6 P}{c^2} \!\Biggr\}
  \nonumber \\
  & + \mathcal{O}\left(\frac{h}{c^6}\right) +
  \mathcal{O}\left(\frac{h^2}{c^2}\right).
\end{align}%

This yields an elliptic equation for
$ V = 2 (\phi - 1) = \mathcal{O}(1/c^2) $ up to
$ \mathcal{O}(h / c^6) $ and $ \mathcal{O}(h^2 / c^2) $ corrections:
\begin{align}
  - 4 \hat{\Delta} V & = - 4 \hat{\nabla}_i \phi
  \hat{\nabla}_j \phi \, \hat{\gamma}^{ik}
  \hat{\gamma}^{jl} h_{kl}^\mathrm{TT}
  \nonumber \\
  & + \frac{1}{4} \hat{\gamma}^{kl} \hat{\gamma}^{im}
  \hat{\gamma}^{jn} \hat{\nabla}_k h_{ij}^\mathrm{TT}
  \hat{\nabla}_l h_{mn}^\mathrm{TT}
  \nonumber \\
  & - 4
  \hat{\nabla}_i \left( \phi^{-4} \hat{\nabla}_j \phi \,
  \hat{\gamma}^{ik} \hat{\gamma}^{jl} h_{kl}^\mathrm{TT} \right)
  - \frac{1}{2} \hat{\Delta} \left( \hat{\gamma}^{ij} \hat{\gamma}^{kl} 
  h_{ik}^\mathrm{TT} h_{jl}^\mathrm{TT} \right) \nonumber \\
  &  +
  \frac{1}{2} \hat{\nabla}_i \hat{\nabla}_j \left( \hat{\gamma}^{kl}
  \hat{\gamma}^{mi} \hat{\gamma}^{nj} h_{km}^\mathrm{TT}
  h_{ln}^\mathrm{TT} \right)
  \nonumber \\
  & + \hat{\gamma}_{ik} \hat{\gamma}_{jl}
  \left( \pi^{ij}_\mathrm{LL} \pi^{kl}_\mathrm{LL} +
  2 \pi^{ij}_\mathrm{LL} \pi^{kl}_\mathrm{TT} +
  \pi^{ij}_\mathrm{TT} \pi^{kl}_\mathrm{TT} \right)
  \nonumber \\
  & + \frac{16 \pi G}{c^2} \phi^{-1} \Biggl\{\! \left[ (D^{*} h)^2
  + \phi^{-4}
  \hat{\gamma}^{mn} \frac{S^*_m S^*_n}{c^2}\right]^{1/2}
  \nonumber \\
  & \qquad \qquad \qquad ~ \times \left(\! 1 -
  \frac{1}{2} \hat{\gamma}^{ik}
  \hat{\gamma}^{jl} h_{kl}^\mathrm{TT}
  \frac{S^*_i S^*_j}{c^2 D^{*\,2}} \!\right) -
  \frac{\phi^6 P}{c^2} \!\Biggr\}
  \nonumber \\
  & + \mathcal{O} \left( \frac{h}{c^6} \right) +
  \mathcal{O} \left( \frac{h^2}{c^2} \right).
  \label{eq:DV}
\end{align}%
As the terms containing a factor $ h_{ij}^\mathrm{TT} $ or
$ \pi^{ij}_\mathrm{TT} $ are proportional to the coupling
constant $ G / c^2 $ of general relativity, Eq.~(\ref{eq:DV}) reduces
to $ \hat{\Delta} V = -4 \pi G D^* \! / c^2 + \mathcal{O}(1 / c^4) $
at the lowest post-Newtonian approximation. Thus, if we introduce the
Newtonian potential $ U $ defined as the smooth solution of the Poisson
equation~(\ref{eq:newtonian_potential}) vanishing at spatial infinity,
we have $ V = U / c^2 + \mathcal{O}(1 / c^4) $, plus a possible
harmonic function. Assuming an asymptotically flat space-time, this
function must tend asymptotically towards zero while being regular,
and so it has to be identically zero. Another important piece of
information provided by Eq.~(\ref{eq:DV}) is the value of the lowest
order contribution to the potential $ V $ that depends on the field
variables. It is given by the equation 
\begin{align}
  - 4 V & = \hat{\Delta}^{-1}
  \left( - 4 \hat{\nabla}_i \hat{\nabla}_j \phi
  \hat{\gamma}^{ik} \hat{\gamma}^{jl} h_{kl}^\mathrm{TT} \right) +
  \mathcal{O} \left( \frac{h}{c^4} \right)
  \nonumber \\
  & + \mathrm{pure\ matter\ part},
  \label{eq:phi_equation}
\end{align}%
which shows incidentally that $ \phi = 1 + V/2 $ is not
affected by a non-zero $ h_{ij}^\mathrm{TT} $ at the leading
post-Newtonian approximation in the field. Inserting the resulting
expression for the conformal factor into the Hamiltonian
constraint~(\ref{eq:DV}), we arrive at
\begin{align}
  - 4 \hat{\Delta} V & = - \frac{1}{c^4}
  \left(2 \hat{\nabla}_i U \hat{\nabla}_j U +
  8\pi G \frac{S^*_i S^*_j}{D^*} \right)
  \hat{\gamma}^{ik} \hat{\gamma}^{jl} h_{kl}^\mathrm{TT}
  \nonumber \\
  & + \frac{1}{4} \hat{\gamma}^{kl} \hat{\gamma}^{im}
  \hat{\gamma}^{jn} \hat{\nabla}_k h_{ij}^\mathrm{TT}
  \hat{\nabla}_l h_{mn}^\mathrm{TT}
  \nonumber \\
  & + \hat{\gamma}_{ik} \hat{\gamma}_{jl}
  \left( 2 \pi^{ij}_\mathrm{LL} \pi^{kl}_\mathrm{TT} +
  \pi^{ij}_\mathrm{TT} \pi^{kl}_\mathrm{TT} \right)
  \nonumber \\
  & + \mathcal{O}\left(\frac{h}{c^6}\right) +
  \mathcal{O}\left(\frac{h^2}{c^2}\right)
  \nonumber \\
  & + \mathrm{total\ derivative} + \mathrm{pure\ matter\ part}.
\end{align}%
The pure matter part has been kept aside because it does not enter the
computation of $ h_{ij}^\mathrm{TT} $. The total derivative terms are
irrelevant for the investigation of the field evolution. Indeed, by
virtue of relation~(\ref{eq:H_reduced}), the Hamiltonian is given by
the spatial integral of $ - c^4 \hat{\Delta} V / (4 \pi G) $. Thus the
dynamics of the gravitational interaction is described by
\begin{align}
  H_{\genfrac{}{}{0pt}{}{\mathrm{field}}{\mathrm{+int.}}} & =
  \frac{c^4}{16 \pi G} \int \mathrm{d}^3\mb{x} \, \sqrt{\hat{\gamma}}
  \nonumber \\
  & \times \Biggl[ - \frac{1}{c^4}
  \left( 2 \hat{\nabla}_i U \hat{\nabla}_j U +
  8\pi G \frac{S^*_i S^*_j}{D^*} \right)
  \hat{\gamma}^{ik} \hat{\gamma}^{jl} h_{kl}^\mathrm{TT}
  \nonumber \\
  & \quad ~~ - \frac{1}{4} \hat{\Delta}
  \left( \hat{\gamma}^{im} \hat{\gamma}^{jn}
  h_{mn}^\mathrm{TT} \right) h_{ij}^\mathrm{TT}
  \nonumber \\
  & \quad ~~ +  \hat{\gamma}_{ik} \hat{\gamma}_{jl} 
  \left( 2 \pi^{ij}_\mathrm{LL} \pi^{kl}_\mathrm{TT} +
  \pi^{ij}_\mathrm{TT} \pi^{kl}_\mathrm{TT} \right) \Biggr]
  \nonumber \\
  & + \mathcal{O} \left( \frac{h}{c^6} \right) +
  \mathcal{O} \left( \frac{h^2}{c^2} \right),
  \label{eq:hamiltonian}
\end{align}%
in agreement with \citet{schaefer_90_a}. The Hamilton equations
provide the evolution of the field. They take the explicit form
\begin{align}
  \partial_t h_{ij}^\mathrm{TT} &=
  2 c \hat{\gamma}_{ij}^{\mathrm{TT}kl}
  [\hat{\gamma}_{km} \hat{\gamma}_{ln}
  \left( \pi^{mn}_\mathrm{LL} + \pi^{mn}_\mathrm{TT} \right)]
  \nonumber \\
  & + \mathcal{O} \left( \frac{1}{c^5} \right) +
  \mathcal{O} \left( \frac{h}{c} \right)
  \nonumber \\
  & = 2 c \hat{\gamma}_{ik} \hat{\gamma}_{jl}
  \pi_\mathrm{TT}^{kl} + \mathcal{O} \left( \frac{1}{c^5} \right) +
  \mathcal{O} \left( \frac{h}{c} \right),
  \label{eq:EOM_hTT} \\
  \partial_t \pi^{ij}_\mathrm{TT} &=
  - c\hat{\gamma}^{ij}_{\mathrm{TT}kl}
  \Biggl[- \frac{2}{c^4} \left( \hat{\nabla}_m U \hat{\nabla}_n U +
  4\pi G \frac{S^*_m S^*_n}{D^*} \right)
  \hat{\gamma}^{km} \hat{\gamma}^{ln}
  \nonumber \\
  & \qquad \qquad ~~\, - \frac{1}{2} \hat{\Delta}
  \left( \hat{\gamma}^{km} \hat{\gamma}^{ln}
  h_{mn}^\mathrm{TT} \right) \Biggr]
  \nonumber \\ 
  & + \mathcal{O} \left( \frac{1}{c^5} \right) +
  \mathcal{O} \left( \frac{h}{c} \right).
  \label{eq:EOM_piTT}
\end{align}%
The non-conformally flat part of the 3-metric appears first at the
second post-Newtonian approximation. Its leading order is obtained by
inserting the above expression for $ \partial_t \pi^{ij}_\mathrm{TT} $
into the time derivative of Eq.~(\ref{eq:EOM_hTT}). The resulting
equation is of wave type. In the near zone, all terms of order
$ 1 / c^6 $ may be neglected, in particular the time derivative
contribution to the d'Alembertian operator. Hence the non-conformally
flat part of the 3-metric satisfies
\begin{align}
  \hat{\Delta} h_{ij}^\mathrm{TT} & =
  - \frac{\hat{\gamma}_{ij}^{\mathrm{TT}kl}}{c^4}
  \left( 4 \hat{\nabla}_k U \hat{\nabla}_l U +
  16\pi G \frac{S^*_k S^*_l}{D^*} \right)
  \nonumber \\
  & + \mathcal{O} \left( \frac{1}{c^6} \right),
  \label{eq:htt_poisson_equation}
\end{align}%
which is identical up to second post-Newtonian order to
Eq.~(\ref{eq:equation_hTT}).

The equation for the lapse function $ \alpha $ is derived from the
gauge conditions~(\ref{eq:condition_gamma}, \ref{eq:condition_pi})
combined with the evolution equation~(\ref{eq:adm_metric_equation_2})
for the extrinsic curvature $ K_{ij} $. From the identity $ \pi^{ij}
\hat{\gamma}_{ij} = 0 $, we deduce that the trace is negligible at
this level of approximation:
\begin{align}
  \hat{\gamma}^{ij} K_{ij} & =
  \frac{1}{2 \sqrt{\bar{\gamma}}} (2 \sqrt{\bar{\gamma}} \hat{\gamma}_{ij}
  K^{ij}) + 
  \mathcal{O} \left( \frac{h}{c^3} \right)
  \nonumber \\
  & = \frac{1}{2 \sqrt{\bar{\gamma}}} \pi^{ij} \hat{\gamma}_{ij} +
  \mathcal{O} \left( \frac{h}{c^3} \right)
  \nonumber \\
  & = \mathcal{O} \left( \frac{1}{c^7} \right).
\end{align}%
By contracting Eq.~(\ref{eq:adm_metric_equation_2}) with the inverse
3-metric $\gamma^{ij}$, we arrive at
\begin{equation}
  \frac{\gamma^{ij}}{c} \partial_t K_{ij} =  
  \frac{\phi^{-4}}{c} \partial_t
  \left( \hat{\gamma}^{ij} K_{ij} \right) +
  \mathcal{O} \left( \frac{1}{c^8} \right) =
  \mathcal{O} \left( \frac{1}{c^8} \right),
\end{equation}
from which follows that
\begin{align}
  & - \nabla_i \nabla^i \alpha + \alpha R + K^{ik}
  \left( - 2 \alpha K_{ik} + 2 \nabla_i \beta_k \right)
  \nonumber \\
  & \quad - \frac{4\pi G}{c^4} (- S + 3 E) =
  \mathcal{O} \left( \frac{1}{c^8} \right).
  \label{eq:lapse_equation}
\end{align}%
Due to the fact that $ K^{ik} $ is symmetric and trace-free modulo
$ \mathcal{O}(h / c^3) $ corrections, the product
$ K^{ik} (- 2 \alpha K_{ik} + 2 \nabla_i \beta_k) $ can be written as
$ K^{ik} (- 2 \alpha K_{<ik>} + 2 \nabla_{<i} \beta_{k>}) +
\mathcal{O}(h / c^6) $. The terms inside the parentheses vanish
according to the symmetric trace-free version of
Eq.~(\ref{eq:adm_metric_equation_1}),
\begin{equation}
  - 2 \alpha K_{<ij>} + 2 \nabla_{<i} \beta_{j>} =
  \frac{1}{c} \partial_t \gamma_{<ij>} =
  \mathcal{O} \left( \frac{1}{c^5} \right),
\end{equation}
so that $ K^{ik} (- 2 \alpha K_{ik} + 2 \nabla_i \beta_k) = \mathcal{O}(1/c^8)
$. Next, we see from the Hamiltonian constraint equation that the
interaction and field parts of the scalar curvature $ R $ appearing in
Eq.~(\ref{eq:lapse_equation}) are actually of order $ \mathcal{O}(h / c^4) =
\mathcal{O}(1 / c^8) $. On the other hand, we know that $ E^* = \mathcal{O}(h
/ c^4) + \mathrm{pure\ matter\ terms} $. We have similar equalities for $ E $
and $ S = S^i_i $. Therefore, we find
\begin{align}
  \frac{1}{\sqrt{\bar{\gamma}}} \hat{\nabla}_i
  \left( \sqrt{\bar{\gamma}} \gamma^{ij} \hat{\nabla}_j \alpha \right) & =
  \mathcal{O} \left( \frac{1}{c^8} \right) +
  \mathrm{pure\ matter\ part}
  \nonumber \\
  & = \hat{\Delta} \alpha - \hat{\nabla}_i \hat{\nabla}_j \alpha \,
  \hat{\gamma}^{ik} \hat{\gamma}^{jl} h^\mathrm{TT}_{kl}.
\end{align}%
At the lowest approximation, we may replace $ \alpha $ by
$ (-g_{00} + \beta_i \beta^i)^{1/2} = 1 - U / c^2 + \mathcal{O}(1 / c^4) $.
In the end, the elliptic equation for the lapse in the presence of
$ h_{ij}^\mathrm{TT} $ is modified to
\begin{equation}
  \hat{\Delta} \alpha =
  \left(\! \hat{\Delta} \alpha \!\right)_{\!h_{ij}^\mathrm{TT} = 0}
  \! -
  \frac{1}{c^2} \hat{\gamma}^{ik} \hat{\gamma}^{jl} \,
  h^\mathrm{TT}_{ij} \hat{\nabla}_k \hat{\nabla}_l U +
  \mathcal{O} \left( \frac{1}{c^8} \right)\!.
\end{equation}
This is the desired CFC+ metric equation for the lapse function
$ \alpha $. It can easily be transformed into
Eq.~(\ref{eq:alpha_phi}) in Sect.~\ref{subsec:cfc+}, remembering that $ \phi $
coincides with the CFC conformal factor at this level. Since $ U / c^2 $
is Newtonian, $ \hat{\gamma}^{ik} \hat{\gamma}^{jl} h^\mathrm{TT}_{ij}
\hat{\nabla}_k \hat{\nabla}_l U/ c^2 $ corresponds to the second
post-Newtonian order for the dynamics.

The equation for the shift can be obtained in principle by contracting
the 3-metric evolution with the help of the Euclidean metric
$ \hat{\gamma}^{ij} $. However, the new terms, by reference to the
conformally flat case, are proportional to a product of the type $ K $
(or $ \beta $) times $ h^\mathrm{TT}_{ij} $. They are therefore
negligible at the second post-Newtonian level, and will not be
computed here. As already pointed out, the equation for the conformal
factor remains unaffected at that level as well. Thus in general, all
CFC equations except the one for the lapse function remain unaltered.


\section{Inversion of the equation for the 3-metric
  correction in CFC+}
\label{app:inversion_htt}

To express the second post-Newtonian correction $ h^\mathrm{TT}_{ij} $
in CFC+ with the help of intermediate potentials and thus to invert
Eq.~(\ref{eq:equation_hTT}) into Eq.~(\ref{eq:hTT}) in
Sect.~\ref{subsec:cfc+}, we proceed in three stages:

(i) We make the action of
$ \hat{\gamma}_{ij}^{\mathrm{TT}kl} $ explicit in
Eq.~(\ref{eq:equation_hTT}). The result is integrated formally by
means of the Poisson integral operator $ \hat{\Delta}^{-1} $. By
virtue of its linearity property, we obtain a weighted sum of Poisson
potentials of generic type $ \hat{\Delta}^{-1} F_{mn} $ (up to
possible index contractions), or super-potentials of the form
$ \hat{\Delta}^{-2} \hat{\nabla}_i \hat{\nabla}_j F_{mn} =
\hat{\Delta}^{-1} (\hat{\Delta}^{-1} \hat{\nabla}_i \hat{\nabla}_j F_{mn}) $
and $ \hat{\Delta}^{-3} \hat{\nabla}_i \hat{\nabla}_j \hat{\nabla}_k F_{mn} $.
(ii) We transform the super-potentials into simple Poisson potentials
in order to get rid of all derivatives acting directly on the
sources. (iii) We insert the resulting quantities into the
transverse traceless tensor
$ \hat{\Delta}^{-1} \hat{\gamma}_{ij}^{\mathrm{TT}kl} F_{kl} $, and
perform some additional manipulations that lead to the final
expression. These steps are performed in detail in the following.

It is straightforward to expand the operator
$ \hat{\gamma}_{ij}^{\mathrm{TT}kl} $ defined in
Eq.~(\ref{eq:tt_projector}) and to apply it to the source
$ F_{kl} $ given in Eq.~(\ref{eq:Fij_definition}). Taking into account
the symmetry in the two indices $ k $ and $ l $, inverting the
tensor Poisson equation~(\ref{eq:equation_hTT}) then yields
\begin{align}
  h^\mathrm{TT}_{ij} & = \hat{\Delta}^{-1}
  \left( \hat{\gamma}_{ij}^{\mathrm{TT}kl} F_{kl} \right)
  \nonumber \\
  & = \hat{\Delta}^{-1} F_{ij} -
  \frac{1}{2} \hat{\gamma}_{ij} \hat{\Delta}^{-1}
  \left( \hat{\gamma}^{kl} F_{kl} \right)
  \nonumber \\
  & - 2 \hat{\Delta}^{-2}
  \left( \hat{\gamma}^{kl} \hat{\nabla}_k \hat{\nabla}_{(i} F_{j)l} \right) +
  \frac{1}{2} \hat{\Delta}^{-2} \hat{\nabla}_i \hat{\nabla}_j
  \left( \hat{\gamma}^{kl} F_{kl} \right)
  \nonumber \\
  & + \frac{1}{2} \hat{\gamma}_{ij} \hat{\Delta}^{-2}
  \hat{\nabla}_k \hat{\nabla}_l
  \left( \hat{\gamma}^{mk} \hat{\gamma}^{nl} F_{mn} \right)
  \nonumber \\
  & + \frac{1}{2} \hat{\Delta}^{-3} \hat{\nabla}_i \hat{\nabla}_j
  \hat{\nabla}_k \hat{\nabla}_l
  \left( \hat{\gamma}^{mk} \hat{\gamma}^{nl} F_{mn} \right).
  \label{eq:hTT_generic}
\end{align}%
As the Poisson integral $ \hat{\Delta}^{-1} F_{kl} $ converges, all
other (super-)potentials entering Eq.~(\ref{eq:hTT_generic}) are also
well defined. However, they cannot be easily handled. For instance,
quantities such as
$ \hat{\Delta}^{-2} F_{mn} $ or $ \hat{\Delta}^{-3} F_{mn} $ are a
priori meaningless, which shows that the derivatives cannot commute
with the integrals. In order to operate on the sources without meeting
any serious restrictions, it is convenient to resort to the tool of
Hadamard finite part regularization.

It is out of our scope to review all properties of the Hadamard finite
part. However, as it will be used intensively in the following, we
recall its definition for completeness, as well as those of its features we
need in our derivation. When a function $ f(\mb{x}) $,
$ \mb{x} \in \mathbb{R}^3 $, is smooth outside a finite number of
singularities, locally integrable, but not integrable on
$ \mathbb{R}^3 $, we can consider instead the new integrand
$ (|\mb{x} - \mb{x}_0| / r_0)^B f(\mb{x}) $, where $ B $ is a complex
number, $ r_0 $ a positive number, and where $ |\mb{x} - \mb{x}_0| $
denotes the Euclidean norm of $ \mb{x} - \mb{x}_0 $, $ \mb{x}_0 $
being an arbitrary vector of $ \mathbb{R}^3 $. The integral
$ \int_{| \mbsm{x} - \mbsm{x}_0 | > r_0} \mathrm{d}^3 \mb{x} \, (|\mb{x} -
\mb{x}_0|/r_0)^B f(\mb{x}) $ defined by means of the natural measure
$ \mathrm{d}^3\mb{x} = \mathrm{d}x^1 \, \mathrm{d}x^2 \, \mathrm{d}x^3 $
in Cartesian coordinates converges for $ B $ belonging to an appropriate
domain $ D $ of the complex plane. It can be regarded as a holomorphic
function on $ D $. It is then possible to extend
$ I_f(B) = \int \mathrm{d}^3\mb{x} \, (|\mb{x} - \mb{x}_0|/r_0)^B f(\mb{x}) $
by analytic continuation as close to the point $ B = 0 $ as desired,
and to obtain its Laurent expansion
$ \sum_{k \in \mathbb{Z}} (I_f)_k B^k $ there, as explained by
\citet{blanchet_86_a}. The zeroth order coefficient $ (I_f)_0 $ is often
referred to as the finite part integral of $ f(\mb{x}) $, which in our
notation reads
\begin{equation}
  \fpart_{B = 0} \int \mathrm{d}^3\mb{x} \, \sqrt{\hat{\gamma}}
  \left( \frac{|\mb{x} - \mb{x}_0 |}{r_0} \right)^B \!\!\! f (\mb{x}) =
  (I_f)_0.
\end{equation}
The finite part integral of $ f $ may depend on the arbitrary radius
$ r_0 $; in fact, this will typically happen when the result contains
logarithms. Nonetheless, as long as $ f $ is integrable, its finite
part integral coincides with $ \int \mathrm{d}^3\mb{x} \, f $ and does
not show any dependence on $ r_0 $.

An important property of the Hadamard regularization is that the
finite part Poisson integral of a smooth function always exists,
whereas the simple Poisson integral may not. The covariant expression
of this finite part reads
\begin{equation}
  \fpart \hat{\Delta}^{-1}_{\mbsm{x}_0} f =
  \fpart_{B = 0} \int \frac{\mathrm{d}^3\mb{x}' \, \sqrt{\hat{\gamma}}}{- 4
  \pi} 
  \left( \frac{|\mb{x}' - \mb{x}_0|}{r_0} \right)^B \!\!
  \frac{f (\mb{x}')}{|\mb{x} - \mb{x}'|},
\end{equation}
where the Euclidean volume element has been written as
$ \mathrm{d}^3\mb{x} \, \sqrt{\hat{\gamma}} $ in an arbitrary
coordinate system. When $ \hat{\Delta}^{-1} f $ exists, it satisfies
$ \fpart \hat{\Delta}^{-1}_{\mbsm{x}_0} f = \hat{\Delta}^{-1} f $.
In any case, the regularized Poisson integral is the particular
solution of a Poisson equation of the type $ \hat{\Delta} g = f $.
Thus, the operator $ \fpart \hat{\Delta}^{-1} $ constitutes a genuine
generalization of the ordinary Poisson operator $ \hat{\Delta}^{-1} $
and will be denoted as $ \hat{\Delta}^{-1}_{\mbsm{x}_0} $
henceforth. It has the important property that it commutes with the
spatial derivatives $ \hat{\nabla}_i $, which allows us to work on the
form of the elementary (super-)potentials by applying simple and
systematic rules.

An important formula related to the Hadamard regularization is the one
giving the generalized Poisson integral of the distance to the field
point $ \mb{x} $,
\begin{equation}
  \hat{\Delta}_{\mbsm{x}}^{-1} |\mb{x} - \mb{x}'|^A =
  \frac{|\mb{x} - \mb{x}'|^{A +2}}{(A + 2) (A + 3)},
  \label{eq:Matthew_0}
\end{equation}
for an arbitrary complex exponent $ A \neq -2, -3 $. Notably, it can
be used to evaluate the action of the operator
$ \hat{\Delta}^{-1}_{\mbsm{x}} = \fpart_{C=0} \int \mathrm{d}^3\mb{x}'' \,
\sqrt{\hat{\gamma}} / (-4 \pi r_0^C |\mb{x} - \mb{x}''|) $ on the
``$r^a$''-potential $ \fpart_{B=0} \int \mathrm{d}^3\mb{x}' \,
\sqrt{\hat{\gamma}} |\mb{x} - \mb{x}'|^{a + B} f / (-4 \pi r_0^B) $ with
$a \in \mathbb{Z}$.
By permuting the two triple integrals, we obtain the relation
\begin{align}
  & \hat{\Delta}^{-1}_{\mbsm{x}} \fpart_{B=0} \int \frac{\mathrm{d}^3\mb{x}' \,
  \sqrt{\hat{\gamma}}}{-4 \pi r_0^B}
  |\mb{x} - \mb{x}'|^{a + B} f
  \nonumber \\
  & \quad = \fpart_{B=0} \frac{\int \mathrm{d}^3\mb{x}' \, \sqrt{\hat{\gamma}}
  |\mb{x} - \mb{x}'|^{a + B + 2}f}{- 4 \pi r_0^B
  (a + B + 2) (a + B + 3)}.
\end{align}%
The result has the same form as the source. If we make
$ \hat{\Delta}^{-1} $ act on it, we arrive at a quantity of the same
type. This provides a straightforward procedure to determine the
action of $ \hat{\Delta}^{-p} = (\hat{\Delta}^{-1})^p $,
$ p \in \mathbb{N} $, on the original integral iteratively. In this
way, we find: 
\begin{align}
  & \hat{\Delta}_{\mbsm{x}}^{-p} \fpart_{B=0} \int \frac{\mathrm{d}^3\mb{x}' \,
  \sqrt{\hat{\gamma}}}{-4 \pi r_0^B} |\mb{x} - \mb{x}'|^{a + B} f
  \nonumber \\
  & \quad = \fpart_{B=0} \int \frac{\mathrm{d}^3\mb{x}' \,
  \sqrt{\hat{\gamma}}}{-4 \pi r_0^B} (a + B + 1 + 2p)^{-1}
  \nonumber \\ 
  & \qquad \qquad \qquad ~ \times (a + B + 2p)^{-1} \dots
  (a + B + 2)^{-1}
  \nonumber \\
  & \qquad \qquad \qquad ~ \times |\mb{x} - \mb{x}'|^{a + B + 2p} f.
  \label{eq:Delta_-p_alpha_potential}
\end{align}%
If the usual ``$ r^a $''-potential with source $ f $ exists, the
$ p $th iterated Poisson integral $ \hat{\Delta}^{-1}_{\mbsm{x}} $ of
its $ p $th derivative also does. It is equal to
$ \hat{\nabla}_{i_1} \dots \hat{\nabla}_{i_p} $ applied to the
right-hand side of Eq.~(\ref{eq:Delta_-p_alpha_potential}). When we
put all derivatives under the integration symbol, we end up with a
convergent integral. At this stage, one does not need any
regularization anymore, so we may take $ B = 0 $ if none of the
terms $ (a + 2p + 1) $, $ (a + 2p) $, \dots,
$ (a + 2) $ vanishes.
Finally, we pull the coefficients
$ (a + 1 + 2p)^{-1} (a + 2p)^{-1} \dots (a + 1 + p)^{-1} $
out of the integration symbol and reintroduce the finite part. We
have then proved the formula giving the explicit action of $
\hat{\Delta}^{-p}_\mathbf{x}$ 
on arbitrary sources,
\begin{align}
  & \hat{\Delta}_{\mbsm{x}}^{-p} \fpart_{B=0} \int \frac{\mathrm{d}^3\mb{x}' \,
  \sqrt{\hat{\gamma}}}{-4 \pi r_0^B} |\mb{x} - \mb{x}'|^{a + B}
  \hat{\nabla}_{i_1} \dots \hat{\nabla}_{i_p} f
  \nonumber \\
  & \quad = \frac{\hat{\nabla}_{i_1} \dots \hat{\nabla}_{i_p}
  \fpart_{B=0} \int \frac{\mathrm{d}^3\mbsm{x}' \,
  \sqrt{\hat{\gamma}}}{- 4 \pi r_0^B}
  |\mb{x} - \mb{x}'|^{a + B + 2p}f}{(a + 1 + 2p) (a + 2p)
  \dots (a + 2)},
  \label{eq:alpha_potential_form}
\end{align}%
to be valid for $ a + 2p + 1 \notin \mathbb{N} $.

This allows us to write the potentials and super-potentials in a form
suitable for numerical integration. For this purpose, we can recourse
to fairly standard techniques, some of which have been used in
particular by \citet{blanchet_90_a} to deal with the derivative of the
Newtonian super-potential. With the help of
Eq.~(\ref{eq:alpha_potential_form}) and the commutation relation
$ \hat{\Delta}_{\mbsm{x}}^{-1} \hat{\nabla}_i = \hat{\nabla}_i 
\hat{\Delta}_{\mbsm{x}}^{-1} $, we
transform the expression of the (super-)potentials $ \hat{\Delta}^{-2}
\hat{\nabla}_i \hat{\nabla}_j F_{kl} $ or $ \hat{\Delta}^{-3} \hat{\nabla}_i
\hat{\nabla}_j \hat{\nabla}_k \hat{\nabla}_l F_{mn} $ entering the
non-conformal part of the 3-metric (with possible index
contractions). By doing this, we try to minimize the number of free
indices that remain inside the integral and, for numerical reasons, to
get rid of all spatial derivatives of the densities $ D^* $ or
$ S_i^* $ in the final expressions. Let us detail for instance the
transformation of
$ \hat{\Delta}^{-2} \hat{\nabla}_i \hat{\nabla}_j F_{kl}$. It is
achieved in four steps: (i) We let both derivatives commute with the
integration symbol, 
$ \hat{\Delta}^{-2} \hat{\nabla}_i \hat{\nabla}_j F_{kl} = \hat{\nabla}_i
\hat{\nabla}_j \hat{\Delta}_{\mbsm{x}}^{-2} F_{kl} $, 
(ii) we rewrite $ \hat{\Delta}_{\mbsm{x}}^{-2} F_{kl} $ according to
Eq.~(\ref{eq:Delta_-p_alpha_potential}) with $ p = 2 $, (iii) we make
one of the derivatives act on the kernel $ |\mb{x} - \mb{x}'|^{1 + B} $
so that there remains only one unevaluated spatial derivative
operating on a linear combination of simple Poisson integrals, and
(iv) we let the derivative act. This yields
\begin{align}
  \hat{\Delta}^{-2} \hat{\nabla}_i \hat{\nabla}_j F_{kl} =
  \frac{1}{2} \Bigl[ & \hat{\gamma}_{ij} \hat{\Delta}_{\mbsm{x}}^{-1} F_{kl} +
  \hat{\gamma}_{jm} x^m \hat{\nabla}_i \hat{\Delta}_{\mbsm{x}}^{-1} F_{kl}
  \nonumber \\
  & - \hat{\nabla}_i \hat{\Delta}_{\mbsm{x}}^{-1}
  (\hat{\gamma}_{jm} x^m F_{kl}) \Bigr].
\end{align}%
The respective transformation of
$ \hat{\Delta}^{-3} \hat{\nabla}_i \hat{\nabla}_j \hat{\nabla}_k F_{mn} $
is very similar:
\begin{align}
  & \hat{\Delta}^{-3} \hat{\nabla}_i \hat{\nabla}_j \hat{\nabla}_k F_{mn}
  \nonumber \\
  & \quad = \frac{1}{24} \hat{\nabla}_i \hat{\nabla}_j
  \hat{\nabla}_k \fpart_{B = 0} \int \frac{\mathrm{d}^3 \mb{x'} \,
  \sqrt{\hat{\gamma}}}{-4 \pi r_0^B} |\mb{x} - \mb{x}'|^{3 + B} F_{mn}
  \nonumber \\
  & \quad = \frac{1}{8} \hat{\nabla}_k \int \frac{\mathrm{d}^3 \mb{x'} \,
  \sqrt{\hat{\gamma}}}{-4 \pi} \biggl[ \hat{\gamma}_{ip}
  \hat{\gamma}_{jq} \frac{(x^p - x'^p) (x^q - x'^q)}{|\mb{x} - \mb{x}'|}
  \nonumber \\
  & \qquad \qquad \qquad \qquad \qquad
  + \hat{\gamma}_{ij} |\mb{x} - \mb{x}'| \biggr] F_{mn}
  \nonumber \\
  & \quad = \frac{1}{8} \Bigl[ 3 \hat{\gamma}_{(ij} \hat{\gamma}_{k)p} x^p
  \hat{\Delta}_{\mbsm{x}}^{-1} F_{mn} - 3 \hat{\gamma}_{(ij}
  \hat{\gamma}_{k)p} \hat{\Delta}_{\mbsm{x}}^{-1} (x^p F_{mn})
  \nonumber \\
  & \qquad \quad ~ - 2 \hat{\gamma}_{p(i} \hat{\gamma}_{j)q} x^p
  \hat{\nabla}_k \hat{\Delta}_{\mbsm{x}}^{-1} (x^q F_{mn})
  \nonumber \\
  & \qquad \quad ~ + \hat{\gamma}_{ip} \hat{\gamma}_{jq} x^p x^q
  \hat{\nabla}_k \hat{\Delta}_{\mbsm{x}}^{-1} F_{mn}
  \nonumber \\
  & \qquad \quad ~ + \hat{\gamma}_{ip} \hat{\gamma}_{jq} \hat{\nabla}_k
  \hat{\Delta}_{\mbsm{x}}^{-1} (x^p x^q F_{mn}) \Bigr].
\end{align}%
We are now able to deduce the expression for $ h^\mathrm{TT}_{ij} $
up to second post-Newtonian order, parametrized by means of the
intermediate elementary potentials $ \mathcal{S} $, $ \mathcal{S}_i $,
$ \mathcal{T}_i $, $ \mathcal{R}_i $, and $ \mathcal{S}_{ij} $ as
\begin{align}
 h^\mathrm{TT}_{ij} & = \frac{1}{2} \mathcal{S}_{ij} -
  3 x^k \hat{\nabla}_{(i} \mathcal{S}_{j)k} + \frac{5}{4} \hat{\gamma}_{jm}
  x^m \hat{\nabla}_i 
  \left( \hat{\gamma}^{kl} \mathcal{S}_{kl} \right) 
  \nonumber \\
  & +
  \frac{1}{4} x^k x^l \hat{\nabla}_i \hat{\nabla}_j
  \mathcal{S}_{kl} + 3 \hat{\nabla}_{(i} \mathcal{S}_{j)} - \frac{1}{2} x^k
  \hat{\nabla}_i \hat{\nabla}_j \mathcal{S}_k 
  \nonumber \\
  & + \frac{1}{4} \hat{\nabla}_i \hat{\nabla}_j \mathcal{S} -
  \frac{5}{4} \hat{\nabla}_i \mathcal{T}_j -
  \frac{1}{4} \hat{\nabla}_i \mathcal{R}_j \nonumber \\
  & + \hat{\gamma}_{ij} \left[ \frac{1}{4} \hat{\gamma}^{kl} \mathcal{S}_{kl} 
  + x^k \hat{\gamma}^{lm} \hat{\nabla}_m \mathcal{S}_{kl} - \hat{\gamma}^{kl}
  \hat{\nabla}_k \mathcal{S}_l  \right] \nonumber \\
  & +\mathcal{O} \left( \frac{1}{c^6} \right).
  \label{eq:app_hTT}
\end{align}%
This is the expression we use in Eq.~(\ref{eq:hTT}) in
Sect.~\ref{subsec:cfc+} with the potentials determined by the
scalar/vector/tensor Poisson
equations~(\ref{eq:lap_s}--\ref{eq:lap_sij}).


\section{Multipole expansion of the intermediate potentials}
\label{app:multipole_expansion}

Here we derive the multipole expansion of the intermediate potentials
$ \mathcal{S} $, $ \mathcal{S}_i $, $ \mathcal{T}_i $,
$ \mathcal{R}_i $, and $ \mathcal{S}_{ij} $ needed for obtaining
boundary conditions for these potentials. This is done by means of the
formula~(C.9) of \cite{blanchet_01_a} specializing the matching relation
first established in \citet{blanchet_98_a} for retarded quantities.

For any generic source $ f $ which admits outside the system a multipole
expansion of the form
$ \mathcal{M}(f) = \sum_{p = -\infty}^{p_0} f_p (\mb{n}) r^p $ with
$ \mb{n} = \mb{x} / r $ and $ p_0 < -2  $, the multipole expansion of
the Poisson integral  $ \hat{\Delta}^{-1} f $ is given by
\begin{align}
  \mathcal{M} \left( \hat{\Delta}^{-1} f \right) & =
  \sum_{l = 0}^{+\infty} \frac{(-1)^l}{l!} \hat{\nabla}_{i_1}
  \dots \hat{\nabla}_{i_l} \left( \frac{1}{r} \right)
  \nonumber \\
  & \times \fpart_{B = 0} \int \frac{\mathrm{d}^3 \mb{x}' \,
  \sqrt{\hat{\gamma}}}{- 4 \pi} \left( \frac{r'}{r_0} \right)^B \!\!
  x'^{i_1} \dots \, x'^{i_l} f
  \nonumber \\
  & + \hat{\Delta}^{-1}_{\mbsm{0}} \mathcal{M}(f).
  \label{eq:multipole}
\end{align}%
In the special case where the source $ f $ has compact support,
$ \mathcal{M}(f) $ is identically zero and thus the last term above
vanishes. We recover the standard multipole formula used in
electrostatics for spatially limited systems.

At this stage, the multipole expansions of all our elementary
potentials may be derived by application of
Eq.~(\ref{eq:multipole}). We start with $ \mathcal{S}_{ij} $ which
goes to zero at the highest order. It involves the monopole integral
with non-compact-supported source
$ \int \mathrm{d}^3\mb{x}\, \sqrt{\hat{\gamma}} \hat{\nabla}_i U
\hat{\nabla}_j U / (-4 \pi) $. Remarkably, this integral admits an
alternative expression whose source has compact support, which is
a very useful feature for
the numerical calculations. To perform the transformation, we first replace
the second potential $ U $ in the integrand
$ \hat{\nabla}_i U \hat{\nabla}_j U / ( -4 \pi) $ by
$ \hat{\Delta} \chi/2 $, where $ \chi $ denotes the Newtonian
super-potential
$ \chi = \int \mathrm{d}^3\mb{x}' \, |\mb{x} - \mb{x}'| D^* $. Next, we
integrate this Laplacian by parts, being careful of keeping all
contributions from the derivatives of $ r^B $. The result is
\begin{align}
  & \fpart_{B = 0} \int \frac{\mathrm{d}^3 \mb{x} \,
  \sqrt{\hat{\gamma}}}{-4 \pi r_0^B} r^B \, \hat{\nabla}_i
  \left( \frac{\chi}{2} \right) \hat{\nabla}_j \hat{\Delta} U
  \nonumber \\
  & \quad + \fpart_{B = 0} \int \frac{\mathrm{d}^3 \mb{x} \,
  \sqrt{\hat{\gamma}}}{-4 \pi r_0^B} \Bigl[ B (B+1) r^{B-2} \,
  \hat{\nabla}_j U
  \nonumber \\
  & \qquad \qquad \qquad \qquad \qquad \; + 2 B r^{B - 1} n^k \,
  \hat{\nabla}_j \hat{\nabla}_k U \Bigr] \hat{\nabla}_i
  \left( \frac{\chi}{2} \right).
\end{align}%
We remark here that the second finite part on the right-hand side
vanishes. Indeed, the integration does not generate any pole in
$ B $ able to compensate the cancellation of the pre-factor $ B $
itself. Consequently, after a last integration by parts, we arrive at
the equality
\begin{equation}
  \int \frac{\mathrm{d}^3\mb{x} \, \sqrt{\hat{\gamma}}}{-4 \pi}
  \hat{\nabla}_i U \hat{\nabla}_j U = - \frac{1}{2} \int \mathrm{d}^3 \mb{x} \,
  \sqrt{\hat{\gamma}} D^* \hat{\nabla}_i \hat{\nabla}_j \chi.
\end{equation}

It is not difficult to check that
$ \hat{\nabla}_j \chi = \hat{\gamma}_{jk} x^k \hat{\Delta}^{-1}
(- 4 \pi D^*) - \hat{\Delta}^{-1} (- 4 \pi \hat{\gamma}_{jk} x^k D^*) $ by
virtue of 
relation~(\ref{eq:alpha_potential_form}). Using an integration by
parts of the type
\begin{equation}
  \int \mathrm{d}^3\mb{x} \, \sqrt{\hat{\gamma}} f \, \hat{\Delta}^{-1} g =
  \int \mathrm{d}^3\mb{x} \, \sqrt{\hat{\gamma}} g \, \hat{\Delta}^{-1} f,
  \label{eq:Delta_by_part}
\end{equation}
we are finally able to show that
\begin{align}
  & \int \frac{\mathrm{d}^3\mb{x} \, \sqrt{\hat{\gamma}}}{-4 \pi}
  \hat{\nabla}_i U \hat{\nabla}_j U
  \nonumber \\
  & \quad = - \int \mathrm{d}^3\mb{x} \, \sqrt{\hat{\gamma}} D^*
  \left(\hat{\gamma}_{jk} x^k \hat{\nabla}_i U + \frac{1}{2} \hat{\gamma}_{ij}
  U \right).  
\label{eq:integral_dUdU}
\end{align}%
Inserting the latter relation into the general
formula~(\ref{eq:multipole}) specialized for
$ f = -4 \pi S_i^* S_j^*/D^* - \hat{\nabla}_i U \hat{\nabla}_j U $,
we get
\begin{align}
  \mathcal{M}(\mathcal{S}_{ij}) & = \frac{1}{r}
  \int \mathrm{d}^3\mb{x} \, \sqrt{\hat{\gamma}} D^*
  \left( \frac{S_i^* S_j^*}{D^{*\,2}} + \hat{\gamma}_{jk} x^k \hat{\nabla}_i U
  + 
  \frac{1}{2} \hat{\gamma}_{ij} U \right)
  \nonumber \\
  & + \mathcal{M} \left( - 4 \pi \frac{S^*_i S^*_j}{D^*} -
  \hat{\nabla}_i U \hat{\nabla}_j U \right).
  \label{eq:M_Sij}
\end{align}%

When examining the second term in the above equation, one immediately
observes that $ \mathcal{M}(S^*_i S^*_j / D^*) = 0 $ as the source
$ (S^*_i S^*_j / D^{*\,2}) D^* $ has compact support. Furthermore we
notice that
$ \mathcal{M}(\hat{\nabla}_i U \hat{\nabla}_j U) = \mathcal{O} (1/r^4)$,
hence
$ \hat{\Delta}_{\mbsm{0}}^{-1} \mathcal{M}(\hat{\nabla}_i U
\hat{\nabla}_j U) = \mathcal{O} (1/r^2) $ according to dimensional
analysis. It may also be checked directly with the help of the
Matthew formula $ \hat{\Delta}^{-1}_\mathbf{0} (n^{<i_1} \ldots
n^{i_l>} r^a) = n^{<i_1} \ldots n^{i_l>} r^{a+2}/[(a+2-l)(a+3+l)] $.

We continue with the calculation of $ \mathcal{M}(\mathcal{S}_i) $. The
part of its source with non-compact support amounts essentially to $ x^k 
\hat{\nabla}_i U \hat{\nabla}_k U $. The integral of the latter quantity
may be transformed with the same technique as the one used to establish
Eq.~(\ref{eq:M_Sij}). We find that
\begin{align}
  & \fpart_{B=0} \int \frac{\mathrm{d}^3\mb{x} \, \sqrt{\hat{\gamma}}}{-4 \pi
  r_0^B} r^B
  x^k \, \hat{\nabla}_i U \hat{\nabla}_k U
  \nonumber \\
  & \quad = \frac{1}{2} \fpart_{B=0} \int \frac{\mathrm{d}^3\mb{x} \,
  \sqrt{\hat{\gamma}}}{-4 \pi r_0^B} r^B \Bigl( x^k \hat{\nabla}_k \chi
  \hat{\nabla}_i (-4 \pi D^*)
  \nonumber \\
  & \qquad \qquad \qquad \qquad \qquad \qquad ~\,
  + 2 \hat{\gamma^{kl}} \hat{\nabla}_k \chi
  \hat{\nabla}_l \hat{\nabla}_i U \Bigr).
\end{align}%
We may then perform an integration by parts affecting the derivative
$ \hat{\nabla}_i D^* $ of the first term and the derivative
$ \hat{\nabla}_k \chi $ of the second term. The contribution of
$ \hat{\nabla}_k r^B $ is proportional to $ B $. It can give rise to a
definite non-zero result only when it is multiplied by the factor $ 1 / B $
coming from the radial integration of $ 1 / r^3 $. As the
corresponding angular integral vanishes, so does it. Noticing that
$ \hat{\Delta} \chi \hat{\nabla}_i U = 2 U \hat{\nabla}_i U =
\hat{\nabla}_i U^2 $, we see that the resulting integral reads
\begin{align}
  & \frac{1}{2} \int \mathrm{d}^3\mb{x} \, \sqrt{\hat{\gamma}} D^*
  \left( x^k \hat{\nabla}_i \hat{\nabla}_k \chi + \hat{\nabla}_i \chi \right)
  \nonumber \\
  & \quad -  \fpart_{B=0} \int \mathrm{d}^3\mb{x} \,
  \sqrt{\hat{\gamma}} \, \left( \frac{r}{r_0} \right)^B \, \hat{\nabla}_i U^2. 
\end{align}%
After integrating the second term by parts, we get an integral whose
source is proportional to $ B $. Following the same argument as
before, it can be shown to be simply zero. Expressing again the
super-potential as a combination of Poisson integrals, and performing
other integrations by parts including those of the
form~(\ref{eq:Delta_by_part}), we arrive at
\begin{align}
  & \fpart_{B=0} \int \frac{\mathrm{d}^3\mb{x} \, 
\sqrt{\hat{\gamma}}}{-4 \pi r_0^B} \, r^B x^k \,
  \hat{\nabla}_i U \hat{\nabla}_k U
  \nonumber \\
  & \quad = - \int \mathrm{d}^3\mb{x} \, \sqrt{\hat{\gamma}} D^*
  \hat{\gamma}_{ij} \left( x^j U + x^j x^k \hat{\nabla}_k U \right).
\end{align}%
On the other hand, the source $ x^k \hat{\nabla}_i U \hat{\nabla}_k U $
behaves as $M^2 \hat{\gamma}_{ik} n^k / r^3 = - M^2 \hat{\nabla}_i (r^{-2}/2)
$, where $ M $ is the baryonic rest mass density defined in
Sect.~\ref{subsec:cfc+}. A direct integration leads to the relation $
\hat{\Delta}_{\mbsm{0}}^{-1} (1 / r^2) = \ln r + \mathrm{const.} $, which can
also be seen from a dimensional argument, so that $
\hat{\Delta}_{\mbsm{0}}^{-1} \mathcal{M}(x^k \hat{\nabla}_i U \hat{\nabla}_k U
) $ decreases asymptotically as $ - M^2 \hat{\nabla}_i (\ln r)/2$. Therefore,
we have proved the approximate equality
\begin{align}
   \mathcal{M}(\mathcal{S}_i)
  &= \frac{1}{r} \int \mathrm{d}^3\mb{x} \, \sqrt{\hat{\gamma}} D^*
  \left[ x^k S_i S_k + \hat{\gamma}_{ij} x^j (U + x^k \hat{\nabla}_k U) \right]
  \nonumber \\
  & + \frac{M^2}{2r} \hat{\gamma}_{ik} n^k +
  \mathcal{O} \left( \frac{1}{r^2} \right).
\end{align}%

In the multipole expansion of $ \mathcal{T}_i $, the only
term entering its composition that is generated by a quantity with non-compact
support is the Poisson potential  
$ \hat{\Delta}^{-1} (\hat{\gamma}_{ij} \hat{\gamma}^{kl} x^j \hat{\nabla}_k U 
\hat{\nabla}_l U) $. 
Its monopole part is
composed of an integral over the source, which simplifies by virtue of
the relation $ 2 \hat{\gamma}^{kl} \hat{\nabla}_k U \hat{\nabla}_l U= 
\hat{\Delta} U^2 - 2 U \hat{\Delta} U $ to
\begin{align}
&  \fpart_{B=0} \int \frac{\mathrm{d}^3\mb{x}}{-4 \pi} \sqrt{\hat{\gamma}}
\left( \frac{r}{r_0} \right)^B
  \left(  \hat{\gamma}_{ij} \hat{\gamma}^{kl} x^j \hat{\nabla}_k U
  \hat{\nabla}_l U  \right) \nonumber \\ &
\quad = -  \int  \mathrm{d}^3\mb{x} \, \sqrt{\hat{\gamma}} D^*
  \hat{\gamma}_{ik} x^k U, 
\end{align}
and of a contribution originating from the Poisson inverse operator
applied to
$ \mathcal{M}(\hat{\gamma}_{ij} \hat{\gamma}^{kl} x^j \hat{\nabla}_k U
\hat{\nabla}_l U) = 
- M^2 \hat{\nabla}_i (r^{-2}/2) + \mathcal{O}(1/r^4) $. We have thus
\begin{equation}
  \hat{\Delta}_{\mbsm{0}}^{-1} \mathcal{M} (\hat{\gamma}_{ij} \hat{\gamma}^{kl}
  x^j \hat{\nabla}_k U \hat{\nabla}_l U) = - \frac{M^2}{2 r}
  \hat{\gamma}_{ij} n^j +
  \mathcal{O} \left(\frac{1}{r^2} \right).
\end{equation}

The treatment of the leading term in the multipole expansion of
$ \mathcal{R}_i $ is similar. The integral of
$ \hat{\nabla}_i (x^k x^l\hat{\nabla}_k U \hat{\nabla}_l U) $ is simply
zero, by virtue of the Gauss theorem, and the monopole part of
$ - \hat{\Delta}^{-1} \hat{\nabla}_i (x^k x^l \hat{\nabla}_k U
\hat{\nabla}_l U) $ is the same as that of
$ 2 \hat{\gamma}_{ij} \hat{\gamma}^{kl} x^j \hat{\nabla}_k U \hat{\nabla}_l U
$.

We conclude with the multipole expansion for the potential
$ \mathcal{S} $. Since its source is already of compact support, it
can simply be expanded with the help of the standard multipole formula
mentioned after Eq.~(\ref{eq:multipole}).

In the end this yields the desired multipole expansions of the
elementary potentials in the second second post-Newtonian expansion,
which correspond to
Eqs.~(\ref{eq:mult_expansion_S}--\ref{eq:mult_expansion_Sij}) in
Sect.~\ref{subsec:cfc+}.

\section{Post-Newtonian and post-Minkowskian link for gravitational waves}
\label{app:GW}

The equation for null outgoing geodesics in the ADM gauge reads
$ t - r / c - 2 M_\mathrm{ADM} G \ln [r /(c \eta)] / c^3 + \mathcal{O}(G^2) =
\mathrm{const.} $, with $ M_\mathrm{ADM} = M + \mathcal{O}(1/c^2)$ being the
ADM mass and $ \eta $ an 
arbitrary positive constant with the dimension of time. For $ r $
large enough compared to the typical wavelength, this also gives the
link between the time $ t $ and the distance $ r $ of a field point
$ \mb{x} $ located near $ \mathcal{I}^+ $. In the post-Newtonian
framework, $ r / c $ is regarded as a small quantity (with respect to
$ t $), which precisely reflects the fact that the post-Newtonian
metric is not accurate in the wave zone. Therefore, the
expression~(\ref{eq:hTT}) for $ h_{ij}^\mathrm{TT} $ cannot be used
there. Instead, we should consider the post-Minkowskian version of
Eq.~(\ref{eq:htt_poisson_equation}), where all quantities are expanded
in powers of the gravitational constant $ G $. Higher order terms in
$ 1 / c^2 $ must not be a priori neglected as long as they appear at a
level of approximation below the one we want to reach.

In general, this approach is not guaranteed to lead to a well-defined
perturbative scheme in the ADM gauge in the sense that the
contribution of order $ G^{n+1} $ may become greater than the
contribution of order $ G^n $ in the neighborhood of
$ \mathcal{I}^+ $. It is worth to briefly point at the origin of the
possible problem. Once the field $ {h_n}_{ij}^\mathrm{TT} $ is known
at the $ n $th post-Minkowskian ($ n $PM) level, the next
approximation $ {h_{n + 1}}_{ij}^\mathrm{TT} $ is determined by the
wave-like equation of the type $ \hat{\Box} {h_{n+1}}_{ij}^\mathrm{TT} =
\hat{\gamma}_{ij}^{\mathrm{TT}kl} {\Lambda_{n+1}}_{kl} (h^\mathrm{TT}) +
\mathcal{O} (G^{n+2}) $ obtained by combing Eqs.~(\ref{eq:h_dot})
and~(\ref{eq:pi_dot}) as explained in
Appendix~\ref{app:calc_htt}. The source $ {\Lambda_{n+1}}_{ij} $
depends non-linearly on
$ h_{ij}^\mathrm{TT} $, which can thus be replaced by its $ n $PM
value $ {h_n}_{kl}^\mathrm{TT} $. In the end, the solution of the
previous wave equation is the retarded integral of
$ \hat{\gamma}_{ij}^{\mathrm{TT}kl} {\Lambda_{n+1}}_{kl}
(h^\mathrm{TT}) $, which we denote in short as
$ \hat{\Box}^{-1} \hat{\gamma}_{ij}^{\mathrm{TT}kl} 
{\Lambda_{n+1}}_{kl} $. The transverse traceless projector can itself
be written as a converging integral with the help of a modified
version of Eq.~(\ref{eq:alpha_potential_form}), in which the
derivatives are applied to the kernel
$ |\mathbf{x} - \mathbf{x}'|^{\alpha + B + 2p} $ with $ \alpha = -1 $.
Even if the retarded integral converges, it may be conveniently
replaced by its Hadamard finite part. The numerical result is not
affected, but the integral appearing under the operator
$ \mathrm{FP}_{B = 0} $ can now be shown to commute with the integral
defining $ {h_{n+1}}_{ij}^\mathrm{TT} $ after an appropriate analytic
continuation. As a consequence, the finite part of
$ \hat{\gamma}_{ij}^{\mathrm{TT}kl} $ regarded as an integral
operator, which may be referred to as
$ \hat{\gamma}_{ij}^{\mathrm{TT}kl} $ without any ambiguity, can be
pulled out from the d'Alembertian inverse symbol:
$ \hat{\Box}^{-1} \hat{\gamma}_{ij}^{\mathrm{TT}kl} {\Lambda_{n+1}}_{kl} = 
\hat{\gamma}_{ij}^{\mathrm{TT}kl} \hat{\Box}^{-1} {\Lambda_{n+1}}_{kl} $.
The behavior of $ {\Lambda_{n+1}}_{ij}({h_n}^\mathrm{TT}) $ when
$ r \to + \infty $ follows from a formula extending
Eq.~(\ref{eq:multipole}) to the case where the quantity to be
calculated is a retarded integral. In particular, the multipole
expansion of $ h_{ij}^\mathrm{TT} $ involves the retarded integral of
the multipole expansion of the source, regularized by means of the
Hadamard finite part
$ \hat{\Box}_\mathbf{0}^{-1} \mathcal{M} ({\Lambda_{n+1}}_{ij}) $.
This contribution turns out to generate terms tending towards 
zero as $ (\ln r)^2 / r $ at the 2PM order, which are thus out
of control. The structure of the field has not been investigated yet
at higher PM order in ADM gauge, but a similar phenomenon is likely to
happen in that case too.

Fortunately, the previous problem can be cured at the post-linear
level either by moving to some radiative gauge where
$ u = t - r / c $ is a conformal coordinate, or by absorbing the
logarithms into the arguments of some multipole moments. In harmonic
gauge, the logarithms occurring in the post-Minkowskian iteration have
been shown \citep{blanchet_87_a} to be removable to all orders. In ADM
gauge, the possibility of such an elimination has been checked
explicitly at the post-linear level by \citet{schaefer_90_a} from a
formula adapting the derivation of Appendix~\ref{app:calc_htt} in
order to include in the Hamiltonian the relative Newtonian part of the
2PM corrections to all terms quadratic in
$ h_{ij}^\mathrm{TT} $ (or $ \pi^{ij}_\mathrm{TT} $). The new wave
equation for the field variable reads
\begin{align}
  \hat{\Box} h_{ij}^\mathrm{TT} &= \gamma^{\mathrm{TT}kl}_{ij }
  \biggl[ F_{kl} + \frac{4}{c^2} \Bigl( -2 \pi G D^* h_{kl}^\mathrm{TT}
  \nonumber \\
  & \qquad \qquad \qquad \qquad ~ + \hat{\gamma}^{mn} \hat{\nabla}_m U
  \hat{\nabla}_n h_{kl}^\mathrm{TT} \! +
  U \hat{\Delta} h_{kl}^\mathrm{TT} \Bigr) \biggr]
  \nonumber \\
  & + \mathcal{O} \! \left( \! \frac{ \left( h^\mathrm{TT}
  \right)^0}{c^6} \right) \! + 
  \mathcal{O} \! \left( \! \frac{ \left( h^\mathrm{TT} \right)^1}{c^4}
  \right) \! +
  \mathcal{O} \left( \! \left( h^\mathrm{TT} \right)^2 \right)\!. \!\!
  \label{eq:htt_tail}
\end{align}%
The Green function method provides the required solution,
but the computation is rather delicate. The
result is a retarded integral, on which it remains to apply the
transverse trace-free projector $ \hat{\gamma}^{\mathrm{TT}kl}_{ij} $.
The action of the second spatial derivatives
$ \hat{\nabla}_k \hat{\nabla}_l $ on a quantity of the form
$ f (t - r / c) / r $ leads to the ``monopole'' term
$ \hat{\gamma}_{kp} \hat{\gamma}_{lq} n^p n^q \ddot{f} (t - r / c) / (r c^2)
$, each dot denoting a time derivative, plus $ 1 / r^2 $ corrections. It
follows that 
$ \hat{\Delta}^{-1} \hat{\nabla}_k \hat{\nabla}_l [f (t - r / c) / r] =
\hat{\gamma}_{kp} \hat{\gamma}_{lq} n^p n^q f (t - r / c) / r +
\mathcal{O} (1/r^2) $. Similarly, it is not difficult to check that
$ \hat{\Delta}^{-2} \hat{\nabla}_{i} \hat{\nabla}_{j}
\hat{\nabla}_{k} \hat{\nabla}_{l} [f(t-r/c) / r] = \hat{\gamma}_{ip}
\hat{\gamma}_{jq} \hat{\gamma}_{kr} \hat{\gamma}_{ls} n^p n^q n^r n^s
f (t - r / c) / r + \mathcal{O} (1 / r^2) $. 
This entails that the
operator~(\ref{eq:tt_projector}) applied to a term $ f(t - r / c) / r $ 
reduces to the projector $ P_{ij}^{\mathrm{TT}kl} $. Using the latter
property, we recover the well-known relation
\begin{align}
  h^\mathrm{TT}_{ij} & = \frac{1}{r} P^{\mathrm{TT}kl}_{ij} \nonumber \\ &
  \times \left\{ \! M_{kl} \! \left[ t \!-\! \frac{r}{c} \!-\! \frac{2 G}{c^3}
  \left( M + \mathcal{O}\left(\frac{1}{c^2} \right) \right) \ln
  \left( \! \frac{r}{c \eta} \! \right) \right. \right. \nonumber \\ & \left.
 \qquad \qquad + \mathcal{O}(G^2) \bigg] + \mathcal{O}
  \left( \frac{1}{c^2} \right) \! \right\}  + \mathcal{O} \left(
  \frac{1}{r^2} \right), \label{eq:hTT_monopole}
\end{align}
with $ M_{ij}(t) = \int \mathrm{d}^3 \mb{x} \, \sqrt{\hat{\gamma}} F_{ij}/(-4 \pi) $
being the Newtonian monopole moment associated to the source $F_{ij}$. 

Because of the algebraic transverse traceless projection, we may
substitute $ M_{<ij>} $ by $ M_{ij} $ without affecting the outcome in
the above relation. Since we still work at the 2PN level, we do not
need to include any higher order corrections to the expression of
$ M_{ij} $. The
retarded argument
$ t - r / c - 2 G M_\mathrm{ADM} / c^3 \ln [r / (c \eta)] + \mathcal{O}(G^2) $
coincides with the null radiative coordinate. It can be viewed as 
the actual retarded time $ t_\mathrm{ret} $ of the field variable at
$\mathcal{I}^+$ up to the 1PM order. Taking the preceding into
account, we arrive at
\begin{equation}
  h^\mathrm{TT}_{ij} \approx P^{\mathrm{TT}kl}_{ij}
  \frac{M_{<kl>}(t_\mathrm{ret})}{r} +
  \mathcal{O}\left(\frac{1}{r^2} \right).
\end{equation}
We now improve the formula for $ M_{ij} $ by making use of
Eq.~(\ref{eq:integral_dUdU}). This leads to
\begin{equation}
  M_{<ij>} = \frac{4 G}{c^4} \int \mathrm{d}^3 \mb{x} \, \sqrt{\hat{\gamma}}
  D^* \left( \frac{S^*_{<i} S^*_{j>}}{{D^*}^2} + x^k
  \hat{\gamma}_{k<i} \hat{\nabla}_{j>} U \right).
\end{equation}
Modulo 1PN errors, this integral equals
$ 4 G \int \mathrm{d}^3\mb{x} \, \sqrt{\hat{\gamma}} D
(\hat{\gamma}_{ki} \hat{\gamma}_{lj} v^{<k} v^{l>} +
x^k \hat{\gamma}_{k<i} \hat{\nabla}_{j>} U)/c^4 $. This is proportional to
(twice) the second time derivative of the Newtonian quadrupole
$ 2 \ddot{Q}_{ij} $ at this level, as can be checked with
the help of the relation 
\begin{align}
  & \frac{\mathrm{d}}{\mathrm{d}t} \int \mathrm{d}^3\mb{x} \,
  \sqrt{\hat{\gamma}} D^* f(\mb{x},t)
  \nonumber \\
  & \quad = \int \mathrm{d}^3\mb{x} \, \sqrt{\hat{\gamma}} D^*
  \left( v^i \hat{\nabla}_i + \partial_t \right) f(\mb{x}, t),
\end{align}
which is valid for any $ C^1 $ function $ f $. We have thus derived the
Newtonian quadrupole formula
\begin{equation}
  h^\mathrm{TT}_{ij} \sim P^{\mathrm{TT}kl}_{ij} \frac{2 G \ddot{Q}_{kl}}{r
  c^4}
\end{equation}
in the neighborhood of future null infinity.

It now is possible to investigate the asymptotic behavior of $
h^\mathrm{2PN}_{ij} $ in a similar way as we did for the intermediate
potentials. The computation does not present any particular subtleties in
contrast to that of the full field variable dominant contribution. The
operator $ \hat{\gamma}_{ij}^{\mathrm{TT} kl} $ commutes with the Poisson
integral so that $ \mathcal{M} (\hat{\Delta}^{-1}
\hat{\gamma}_{ij}^{\mathrm{TT}kl} F_{kl}) = \hat{\gamma}_{ij}^{\mathrm{TT}kl}
\mathcal{M} (\hat{\Delta}^{-1} F_{kl}) =
\hat{\gamma}_{ij}^{\mathrm{TT} kl} (M_{kl} / r) + \mathcal{O} (1/r^2) $.
The action of the transverse trace-free projector is computed with the
help of Eq.~(\ref{eq:Matthew_0}). This gives
\begin{align}
  h^\mathrm{2PN}_{ij} (\mb{x}, t) & =
  \biggl[ \frac{3}{16}
  \hat{\gamma}_{(ij} \hat{\gamma}^{kl)} +
  \frac{3}{4}  \delta^k_{(i} \hat{\gamma}_{j)p} n^l n^p \nonumber \\
  & \quad ~~ +
  \frac{3}{16} \hat{\gamma}_{ip} \hat{\gamma}_{jq} n^p n^q n^k n^l 
  \nonumber \\
  & \quad ~~ - \frac{5}{16} \left( \hat{\gamma}_{ij} n^k n^l +
  \hat{\gamma}^{kl} \hat{\gamma}_{ip} \hat{\gamma}_{jq}  n^p n^q \right)
  \biggr] 
  \nonumber \\
  & \times \frac{M_{kl} (t)}{r} + \mathcal{O} \left(\frac{1}{r^2} \right),
  \label{eq:h_ij_2pn}
\end{align}
with $ M_{<ij>} = 2 G \ddot{Q}_{ij} / c^4 $. By applying the operator
$ P^{\mathrm{TT}kl}_{ij} $ on both sides of Eq.~(\ref{eq:h_ij_2pn}),
we find that $ P^{\mathrm{TT}kl}_{ij} h^\mathrm{2PN}_{kl} =
P^{\mathrm{TT}kl}_{ij} G \ddot{Q}_{kl}/(4r c^4)$. As a conclusion,
with the notations of
Sect.~\ref{subsec:gravitational_wave_extraction}, we have (cf.\
Eqs.~(\ref{eq:h+2pn}, \ref{eq:hx2pn})):
\begin{equation}
  P^{\mathrm{TT}kl}_{ij} h^\mathrm{2PN}_{kl}(\mb{x},t_\mathrm{ret}) \sim
  \frac{1}{8} h^\mathrm{quad}_{ij}(\mb{x}, t).
\end{equation}

\bibliographystyle{aa}

\end{document}